\documentclass[12pt]{article}

\usepackage{setspace}
\usepackage[utf8]{inputenc}
\usepackage[margin=1in]{geometry}
\usepackage[titletoc,title]{appendix}
\usepackage[svgnames]{xcolor}
\usepackage{listings}
\usepackage{changepage}
\usepackage{amsmath,amsfonts,amssymb,mathtools}
\usepackage{graphicx,float}
\usepackage{indentfirst}
\usepackage{multirow}
\usepackage{graphics}
\usepackage{caption}
\usepackage{subcaption}
\usepackage{tikz}
\usepackage{float}
\usepackage{longtable}
\usepackage[round]{natbib}
\usepackage{hyperref}
 
\allowdisplaybreaks

\title{Bayesian Hierarchical Modeling for Bivariate Multiscale Spatial Data with Application to Blood Test Monitoring}

\bigskip
\author
{\vspace{-1em}\textbf{Shijie Zhou}\footnote{Corresponding author.} \vspace{1em}\\
   Department of Statistics \\
   Florida State University \\
   117 N. Woodward Ave., Tallahassee, Florida, U.S.A. \\
   Email: sz20bh@fsu.edu \vspace{2em}\\
	\and
	\vspace{-1em}\textbf{Jonathan R. Bradley} \vspace{1em}\\
     Department of Statistics\\
     Florida State University\\ 
     117 N. Woodward Ave., Tallahassee, Florida, U.S.A.\\
    Email: jrbradley@fsu.edu}

\date{}

\begin{document}
	\maketitle
        \clearpage

	\begin{abstract}
		\linespread{1.3}\selectfont
		In public health applications, spatial data collected are often recorded at different spatial scales (or geographic regions/divisions) and over different correlated variables. We are motivated by one such data from the Dartmouth Atlas Project, consisting of average annual percentage of diabetic Medicare enrollees who have taken the hemoglobin A1c test observed on hospital service areas (HSA) and the average annual percentage of diabetic Medicare enrollees who have taken blood lipid test recorded on counties. Counties are not perfectly nested within HSAs and so it is not immediate how one can capitalize on the bivariate relationships between these two scales. It is well known in spatial statistics that one can improve predictions by taking into account correlations between variables and scales. The established strategy for multiscale spatial data is referred to as spatial change of support (COS), which is ubiquitous in univariate settings. However, there are very few methods available that simultaneously use multivariate and multiscale correlations to improve predictions. There are several existing multivariate spatial models that can be easily combined with multiscale spatial methods to improve predictions when analyzing multivariate multiscale spatial data. In this paper, we propose three new models from such combinations for bivariate multiscale spatial data in a Bayesian context, where we model average percentages as Gaussian distributed. In particular, we extend spatial random effects models, multivariate conditional autoregressive (MCAR) models, and ordered hierarchical models through a multiscale spatial approach. To investigate the relative performance of these new methods, we simulate data from each of the three models and compare the corresponding predictions. We found that the multiscale MCAR model's predictions are more robust to model misspecification, but is computationally more intensive. In our analysis of 2015 Texas annual average percentage receiving blood tests from the Dartmouth Atlas Project, we were able to produce finer resolution predictions (partitioning of HSAs and counties) than univariate analyses, determined that the novel multiscale MCAR was preferable among the three methods according to out-of-sample metrics, and determined the HSA with the highest within-HSA variability of hemoglobin A1c blood testing. Additionally, we compare the univariate multiscale models to the bivariate multiscale models and see clear improvements in prediction over univariate analyses for both simulations and real data analyses.
		
		
	\end{abstract}

	\section{Introduction}
	
	Multiscale spatial analysis has gained attention in spatial, spatio-temporal statistics, and many subject matter domains over the past decades (see \citealp{WG04} for a standard reference). The term ``multiscale'' refers to the case when more than one, possibly misaligned (i.e., non-nested), spatial resolution is involved in the analysis. By multivariate multiscale spatial data, we mean we observe more than one spatial response at multiple spatial scales. For example, the Dartmouth Atlas dataset provides data on hospital service areas (HSA), counties, and other pre-defined regions for several variables, including the average annual percentage of diabetic Medicare enrollees aged 65-75 having hemoglobin A1c test and the average annual percentage of diabetic Medicare enrollees aged 65-75 having blood lipids (LDL-C) test (\citealp{Data22}). 
	
	We are motivated by this data on the annual average percentage of individuals receiving certain blood tests since blood tests play a crucial role during the treatment of diabetes. Certain blood tests, such as the hemoglobin A1c test, are considered to be the standard measures of diabetes management \citep{Delamater06} and regular measurements are shown to be beneficial to diabetes control \citep{LHM90}. Considering the importance of blood tests for diabetic patients, our goal is to jointly analyze the average annual percentage of medicare 65-75 enrollees who have taken the hemoglobin A1c test and the average annual percentage of medicare 65-75 enrollees who have taken the blood lipids test, obtained from the Dartmouth Atlas Project.

 \begin{figure}
		\centering
		\begin{subfigure}[t]{0.45\textwidth}
			\centering
			\includegraphics[width=\linewidth]{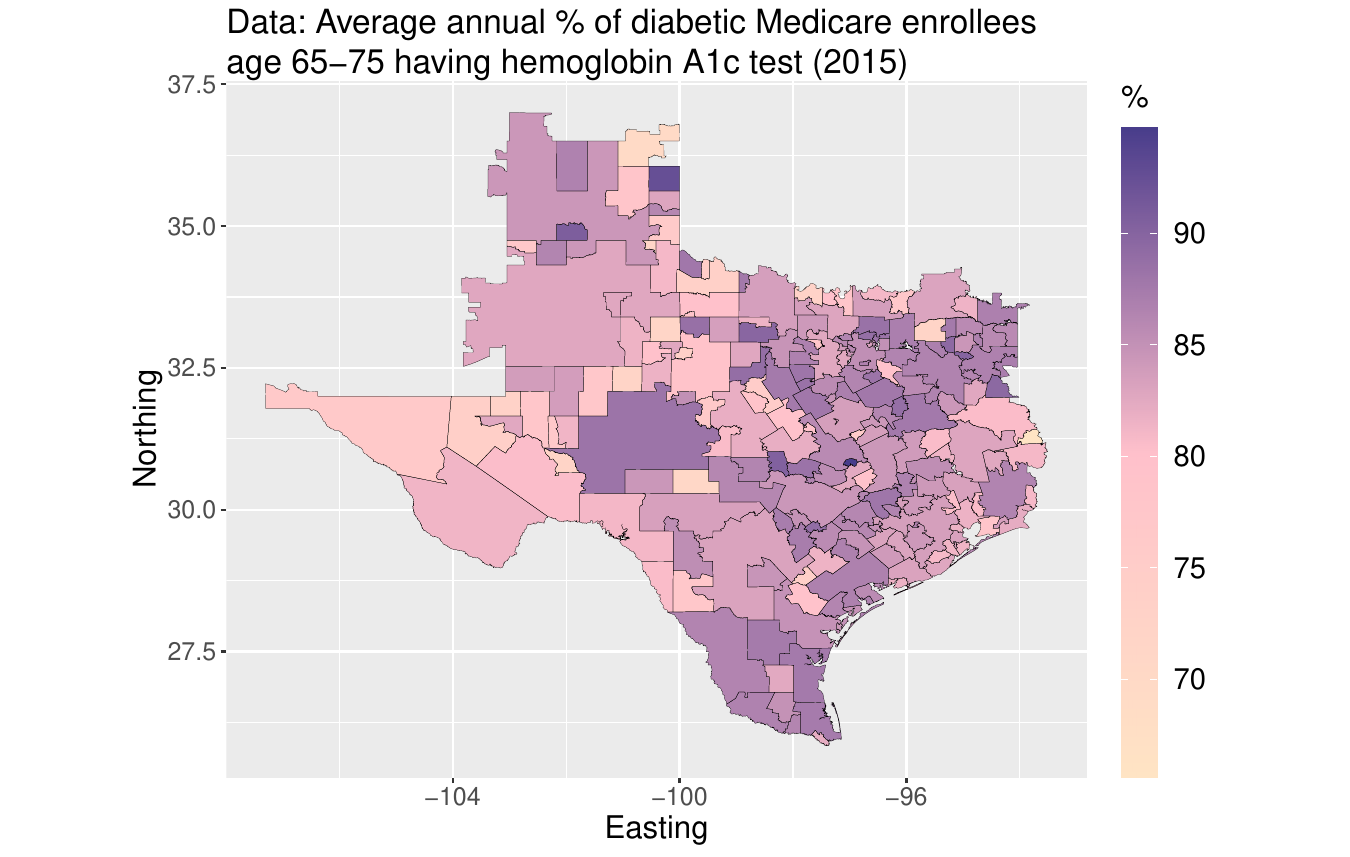} 
			\caption{}
			\label{A1c_hsa_obs}
		\end{subfigure}
		\hspace{0.001cm}
		\begin{subfigure}[t]{0.45\textwidth}
			\centering
			\includegraphics[width=\linewidth]{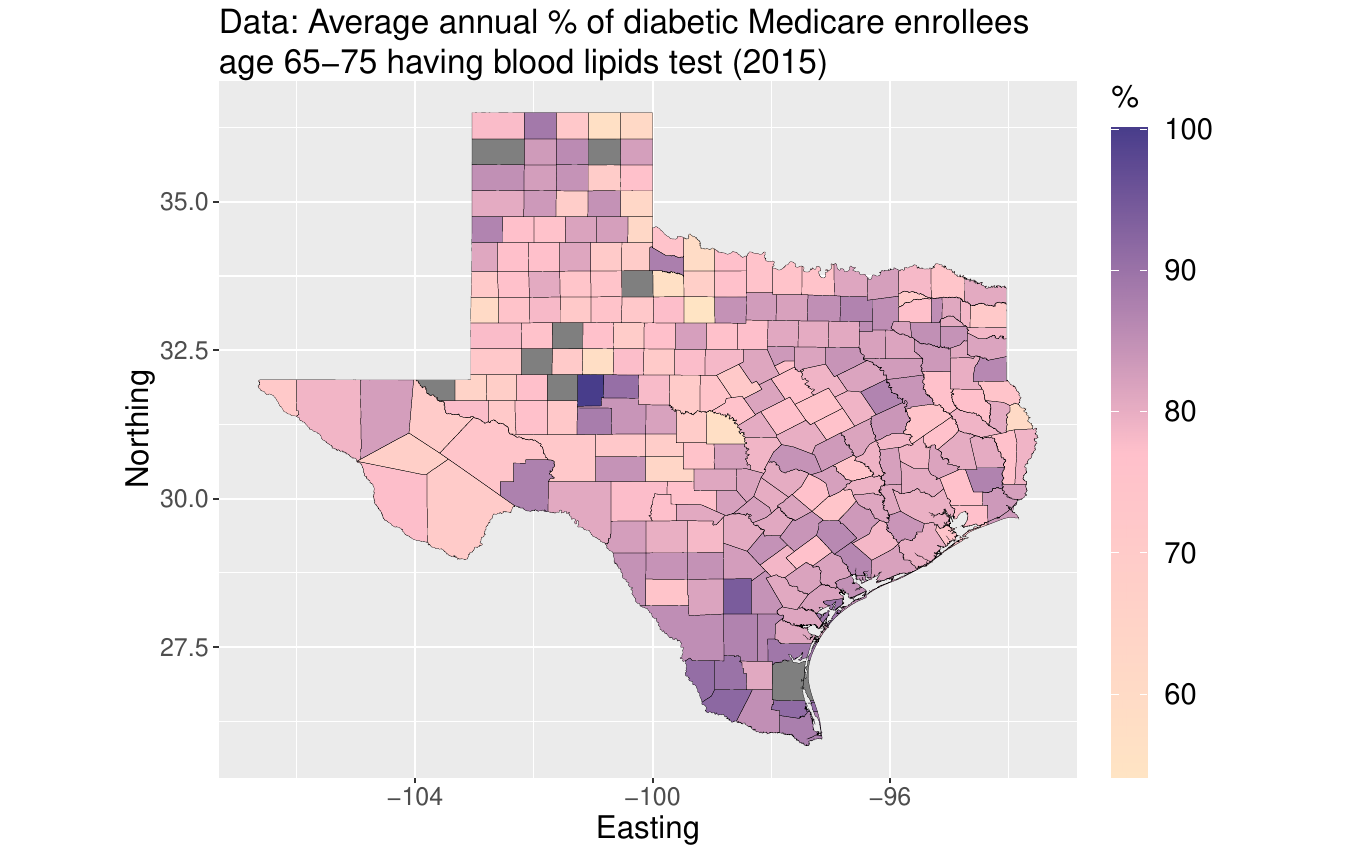} 
			\caption{}
			\label{lipid_county_obs}
		\end{subfigure}
            \hspace{0.001cm}
		\begin{subfigure}[t]{0.5\textwidth}
			\centering
			\includegraphics[width=\linewidth]{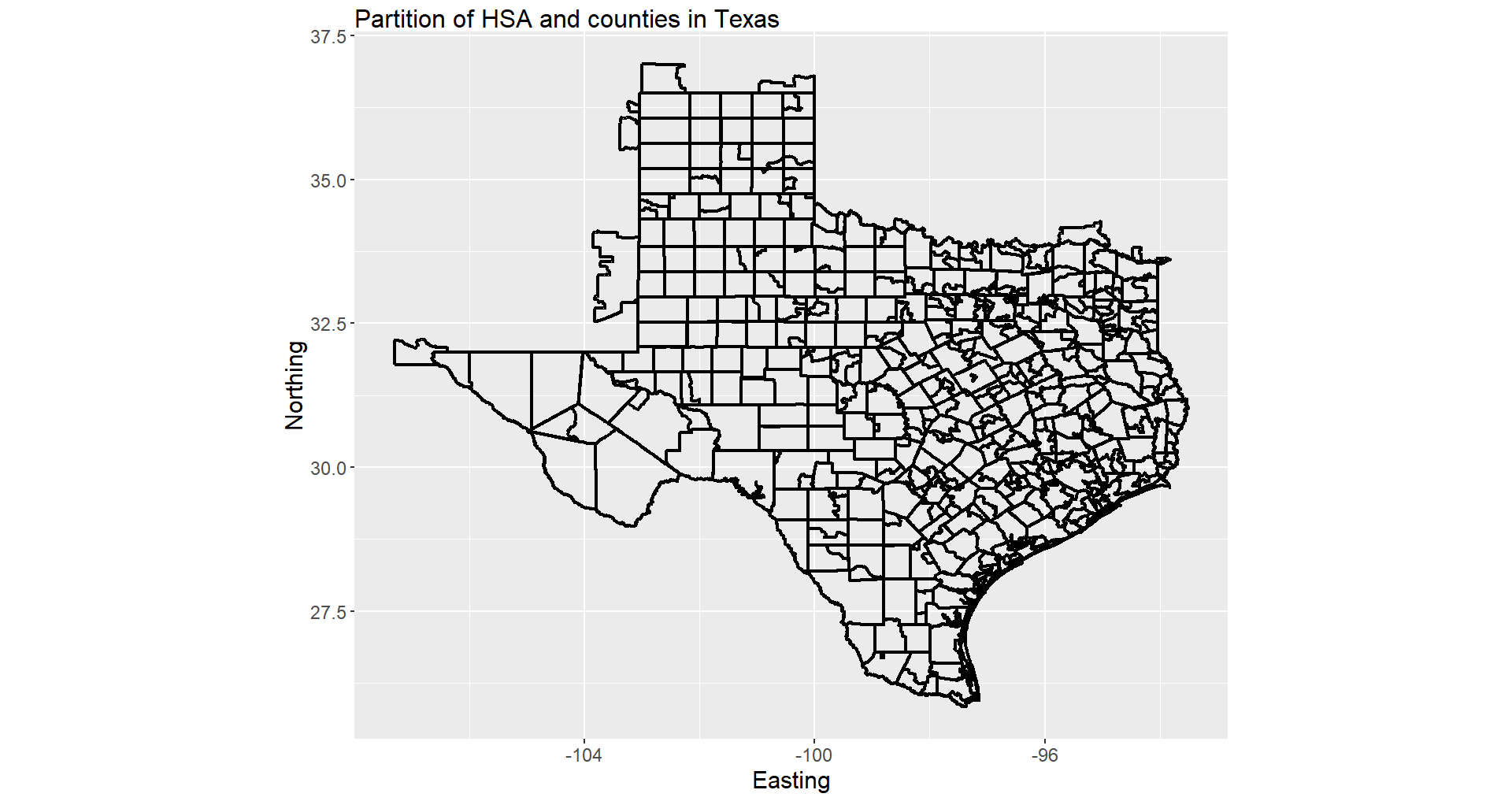} 
			\caption{}
			\label{Texas partition}
		\end{subfigure}
		\caption{Texas blood test monitoring data in 2015. (a) Average annual percentage of diabetic Medicare enrollees age 65-75 having hemoglobin A1c test --- observed on Texas hospital service areas (HSA). (b) Average annual percentage of diabetic Medicare enrollees age 65-75 having blood lipids test --- observed on Texas counties. (c) Partitioning of HSAs and counties in Texas. }
		\label{fig:analysisobs}
	\end{figure}

 In Figure \ref{fig:analysisobs}, we plot the annual average percentage of diabetic Medicare enrollees who have taken the hemoglobin A1c test recorded on Texas hospital service areas (HSA) with annual average percentage of diabetic Medicare enrollees who have taken the blood lipid test recorded on Texas counties. HSAs and counties are the observed supports for this data, and they are not perfectly nested within each other. Ideally, one would be interested in predicting on a more detailed fine resolution (e.g., the partitioning of HSAs and counties given in the third panel of Figure \ref{fig:analysisobs}). Predictions on aggregate scales, such as HSAs and counties, are more susceptible to the modifiable areal unit problem \citep{Banerjee15}, which the notation that inferences on different spatial scales can produce different conclusions. To achieve this goal, one needs to combine existing methodologies in a novel way. In particular, we consider combining several multivariate spatial models with a multiscale spatial approach to analyze such bivariate multiscale spatial data.

 In addition to produce predictions on the finer resolution (i.e., partition of counties and HSAs), we are also interested in assessing the within HSA heterogeneity of hemoglobin A1c testing (originally observed on the HSA scale). In general, HSAs are geographic divisions that represent regions where a healthcare provider (e.g., a hospital) operates and offers medical services. As such, the amount of variability of average percent blood testing within an HSA describes how consistent blood testing is offered and administered by healthcare providers. Consequently, we are interested in identifying the HSA/healthcare providers that are the least consistent throughout the region. Finally, we are also interested in determining which of the newly proposed methods is preferable in terms of their predictive performance (e.g., Watanabe-Akaike information criterion).
	
	In univariate settings, a typical multiscale problem occurs when we observe data at one scale but are instead interested in conducting inference on another scale. This is commonly known as the spatial change of support (COS) problem (see \citealp{GY02} for a comprehensive review). One type of COS is to use data collected at one areal scale to predict on another areal scale. These two scales are often misaligned in practice, and COS is achieved through a partitioning of the available supports (i.e., source supports) and the support for inference (i.e., target support), and assuming the process is piece-wise constant over this partitioning (\citealp{MC98}; \citealp{MC99}). There are other types of COS and multiscale approaches such as points-to-regions prediction where COS can be achieved by integrating the point process up to an areal support (\citealp{Benedetti22}; \citealp{bradley17}; \citealp{Gelfand01}; \citealp{Qu21}; \citealp{WB2005}); however, this type of multiscale features are not present in the Dartmouth Atlas Data. Our goal is to develop innovative bivariate multiscale spatial models through combining standard multivariate spatial models with the appropriate multiscale approach for correlated spatial responses observed over spatially misaligned areal scales.
	
	We mainly consider three multivariate spatial models: the spatial random effects model (\citealp{CJ08}), the multivariate conditional autoregressive (MCAR) model (\citealp{Gelfand03}), and ordered (by variable) hierarchical model (\citealp{RB99}). In the spatial random effects model, dependence between variables is induced through a mean-zero random vector shared across all locations and variables. This model uses areal-referenced basis function expansions. One areal-referenced basis function to use is the Moran's I basis functions (\citealp{bradley15}; \citealp{bradley18}) or the Obled-Creutin basis functions (\citealp{bradley17}; \citealp{OC86}). In the bivariate case, we consider an ordered hierarchical model and use basis functions to model one of the variables. Ordering variables may be very difficult in such models, but is straightforward for bivariate data (\citealp{Banerjee15}, pg. 274). We also consider the MCAR model (\citealp{CB03}; \mbox{\citealp{Gelfand03}}). As an extension of the CAR model, the MCAR model is naturally defined for areal data and adopts a separable-type structure using the Kronecker product in the covariance specification for computational gains \citep{MG93}.
	
	In this paper, we propose three bivariate multiscale spatial models by combining the multivariate spatial random effects model, MCAR model, and ordered hierarchical model with the multiscale spatial approach within a Bayesian hierarchical modeling framework. Bayesian modeling is fairly natural for our application, since our data is particularly high-dimensional and Bayesian modeling allows for uncertainty quantification. We call these models the multiscale spatial random effects (MS-SRE) model, the multiscale MCAR (MS-MCAR) model, and the multiscale ordered hierarchical (MS-OH) model. All three strategies, namely, MS-SRE, MS-MCAR, and MS-OH are new (i.e., to our knowledge have not been formally introduced in the literature), and they each represent a contribution. 

 In each of these new models, our responses are assumed conditionally Gaussian distributed. This is reasonable, since the responses are averages, and averages are often roughly normally distributed (e.g., via central limit theorem). In general, this assumption is reasonable when one observes averages or large count-valued responses. However, in practice, one might observe zero-valued counts in their dataset and as a result, a Poisson distribution may be more reasonable. As such one straightforward adjustment to our new methods, for different datasets,  would be to replace the Gaussian data models with, say, a Poisson distribution with a log-link.
	
Our work not only presents multiple new bivariate multiscale spatial models but also provides a comparison of these new models based on their predictive performance (i.e., root mean squared error (RMSE), Watanabe-Akaike information criterion (WAIC), continuous rank probability score (CRPS), and coverage of credible intervals) and computational efficiency. We conduct an extensive simulation study in which we simulate data from each of our bivariate multiscale spatial models and fit all models to compare the predictions. This is done in an effort to assess how robust each model is to model misspecification. We also demonstrate our models' capability of predicting at a finer resolution through the Texas blood tests monitoring data. The remainder of this paper will be organized as follows. In Section 2, we will review the three single-scale bivariate spatial models along with the multiscale approach. Section 3 gives the methodologies for combining the bivariate spatial models with the multiscale approach, and presents Bayesian hierarchical model specifications for all the three combinations we investigate. Section 4 consists of a simulation study that compares the three proposed models for bivariate multiscale data. We analyze data obtained from the Dartmouth Atlas Study in Section 5. A discussion in Section 6 then follows to conclude this paper. Derivations of full-conditional distributions and additional discussions are provided in the appendices.

	\section{Preliminaries}
	
	In this section, we review single-scale bivariate spatial models (Section 2.1) and an univariate multiscale approach (Section 2.2). Details on prior specifications are provided in Section 2.3.
	
	\subsection{Review: Single-scale bivariate spatial models}
	In this review, we consider a bivariate and intercept-only setting. Suppose we observe $\{Y_1(A_1),...,Y_1(A_n),Y_2(A_1),...,Y_2(A_n)\}$, where $Y_j(A_i)$ denotes the $j$-th variable observed on $i$-th areal unit for variables/processes $j=1,2$ and $i=1,...,n$. Here $A_i \subset D$ where $D\subset \mathbb{R}^d$ is the spatial domain of interest. For ease of exposition, we specify types of spatial random effects (SRE), multivariate conditional autoregressive (MCAR), and ordered hierarchical (OH) models in Table \ref{tab:singlescalemultimodels}.
	
	\begin{table}[h]
		\centering
		\resizebox{\columnwidth}{!}{
			\begin{tabular}{ l  l  l  l }
				\hline
				\hline
				& SRE model & MCAR model & OH model \\
				\hline
				DM 1 & $Y_1(A_i)|\beta_1,\boldsymbol{\eta},\sigma_1^2 \stackrel{ind}{\sim} N(\beta_1+\boldsymbol{g}_1(A_i)'\boldsymbol{\eta},\sigma^2_1)$
				& $Y_1(A_i)|\beta_1,\boldsymbol{\psi},\sigma_1^{2} \stackrel{ind}{\sim} N(\beta_1+\psi_{1i},\sigma_1^{2})$ & $Y_1(A_i)|\beta_1,\boldsymbol{\eta},\sigma_1^2 \stackrel{ind}{\sim} N(\beta_1+\boldsymbol{g}_1(A_i)'\boldsymbol{\eta},\sigma^2_1)$ \\
				DM 2 &  $Y_2(A_i)|\beta_2,\boldsymbol{\eta},\sigma_2^2 \stackrel{ind}{\sim} N(\beta_2+\boldsymbol{g}_2(A_i)'\boldsymbol{\eta},\sigma^2_2)$ & $Y_2(A_i)|\beta_2,\boldsymbol{\psi},\sigma_2^{2} \stackrel{ind}{\sim} N(\beta_2+\psi_{2i},\sigma_2^{2})$ & $Y_2(A_i)|\beta_0,\beta_1,\beta_2,\boldsymbol{\eta},\sigma_2^2 \stackrel{ind}{\sim} N(\beta_0+\beta_2(\beta_1+\boldsymbol{g}_1(A_i)'\boldsymbol{\eta}),\sigma_2^2)$ \\
				PrM & $\boldsymbol{\eta}|\sigma^2_{\eta},\phi \sim MVN(\boldsymbol{0},\boldsymbol{K})$ & $\boldsymbol{\psi}|\rho,\tau,\nu^2 \sim MVN(\boldsymbol{0},\boldsymbol{\Sigma}\otimes (\boldsymbol{D}-\rho\boldsymbol{W})^{-1})$ & $\boldsymbol{\eta}|\sigma^2_{\eta},\phi \sim MVN(\boldsymbol{0},\boldsymbol{K})$\\
				& \quad where \,\, $\boldsymbol{K}=\{\sigma^2_{\eta}\exp(-\phi\|\boldsymbol{c}_i-\boldsymbol{c}_j\|)\}$ & \quad where \,\, $\boldsymbol{\Sigma}=\nu^2\boldsymbol{T}(\tau)$ & \quad where \,\,$\boldsymbol{K}=\{\sigma^2_{\eta}\exp(-\phi\|\boldsymbol{c}_i-\boldsymbol{c}_j\|)\}$\\
				PM 1 & $\beta_k \stackrel{i.i.d}{\sim} N(0,\sigma^2_{\beta}) \quad for \,\,k=1,2$ & $\beta_k \stackrel{i.i.d}{\sim} N(0,\sigma^2_{\beta}) \quad for \,\,k=1,2$ &$\beta_k \stackrel{i.i.d}{\sim} N(0,\sigma^2_{\beta}) \quad for \,\,k=0,1,2$\\
				PM 2 & $\sigma^2_{\eta} \sim IG(a_{\eta},b_{\eta})$ & $\nu^2 \sim IG(a_{\nu},b_{\nu})$ &$\sigma^2_{\eta} \sim IG(a_{\eta},b_{\eta})$\\
				PM 3 & $\sigma_k^2 \stackrel{i.i.d}{\sim} IG(a_{\sigma},b_{\sigma}) \quad for \,\,k=1,2$ &$\sigma_k^2 \stackrel{i.i.d}{\sim} IG(a_{\sigma},b_{\sigma}) \quad for \,\,k=1,2$ &$\sigma_k^2 \stackrel{i.i.d}{\sim} IG(a_{\sigma},b_{\sigma}) \quad for \,\,k=1,2$ \\
				PM 4 & $\phi \sim Unif(a_{\phi},b_{\phi})$ & $\rho \sim Unif(a_{\rho},b_{\rho})$ &$\phi \sim Unif(a_{\phi},b_{\phi})$\\
				PM 5 & \textemdash & $\tau \sim Unif(a_{\tau},b_{\tau})$ & \textemdash \\
				\hline
				\hline
			\end{tabular}
		}
		\caption{\label{tab:singlescalemultimodels}Single-scale bivariate spatial models. Here DM, PrM, and PM are abbreviations for the data model, process model, and parameter model, respectively. Let $N(\mu,\sigma^2)$ be the univariate normal distribution with mean $\mu$ and variance $\sigma^2$, and $MVN(\boldsymbol{\mu},\boldsymbol{\Sigma})$ be the multivariate normal distribution with vector mean $\boldsymbol{\mu}$ and covariance matrix $\boldsymbol{\Sigma}$. Let $IG$ and $Unif$ be the inverse gamma and uniform distributions, respectively. Let $\boldsymbol{g}_k:\{A_1,...,A_n\}\rightarrow \mathbb{R}^r$ be an $r$-dimensional vector of basis functions for $k=1,2$.}
	\end{table}

	In the SRE model, the basis functions $\boldsymbol{g}_k(A_i)$ ($k=1,2$) are chosen to be the $r$-dimensional Moran's I basis functions as defined in \citet{bradley15} and \citet{HH13}. In particular, we set $\boldsymbol{g}_k(A_i)$ equal to the first $r$ eigenvectors of the Moran's I operator (\citealp{Moran50}). The Moran's I operator, denoted $\boldsymbol{M}(\boldsymbol{W})$, has the following form
	\begin{equation*}
		\label{eqn:moransi}
		\boldsymbol{M}(\boldsymbol{W})=(\boldsymbol{I}_n-\boldsymbol{1}_n(\boldsymbol{1}_n'\boldsymbol{1}_n)^{-1}\boldsymbol{1}_n')\boldsymbol{W}(\boldsymbol{I}_n-\boldsymbol{1}_n(\boldsymbol{1}_n'\boldsymbol{1}_n)^{-1}\boldsymbol{1}_n')
	\end{equation*}
	where $\boldsymbol{I}_n$ is the $n\times n$ identity matrix, $\boldsymbol{1}_n$ is the $n$-dimensional vector of all ones for the intercept-only setting, and $\boldsymbol{W}$ is the $n\times n$ adjacency matrix defined by areal units $\{A_1,...,A_n\}$. The $(i,j)$-th entry of the adjacency matrix $\boldsymbol{W}$ is $1$ if $A_i$ is a neighbor of $A_j$, and is $0$ if it is not. Using the spectral decomposition $\boldsymbol{M}(\boldsymbol{W})=\boldsymbol{\Phi}\boldsymbol{\Lambda}\boldsymbol{\Phi}'$, we can specify $\boldsymbol{G}=(\boldsymbol{g}{_k}(A_1)',...,\boldsymbol{g}_k(A_n)')'$ to be the first $r$ columns of $\boldsymbol{\Phi}$. It has been discussed recently \citep{KC22} that basis matrices that deconfound, like the Moran's I basis, produce regression coefficients whose posterior mean is equivalent to the ordinary least squares estimate. This was one of the original motivations for spatial deconfounding \citep{hodges}, as the ordinary least squares estimator is invariant to misspecifications in the spatial covariance. \citet{hanks} showed that predictions are robust to this choice of basis functions. Since COS is a prediction problem, the Moran's I basis is a particularly apt choice. 
 
 In the multiscale space-time COS literature, it has been demonstrated (\citealp{Raim21}) that predictions are fairly robust to the choice of $\boldsymbol{K}=cov(\boldsymbol{\eta})$. As such, we choose a standard structure for $\boldsymbol{K}$. In particular, the $r\times r$ covariance matrix $\boldsymbol{K}=\{\sigma^2_{\eta}\exp(-\phi\|\boldsymbol{c}_i-\boldsymbol{c}_j\|)\}$, where $\{\boldsymbol{c}_1,\boldsymbol{c}_2,...,\boldsymbol{c}_r\}\in D$ are the prespecified knots \citep{Banerjee08}. We specify the knots to be equally spaced throughout the spatial domain. This is a very common choice in the basis function literature \citep{Banerjee08,Banerjee15,Simpson12}, as there is often very little prior knowledge on the placement of these knots, and placing a prior on these parameters is extremely inefficient \citep{Guhaniyogi11}. Of course, there are several alternative options that have been used in the past, including but not limited to Obled-Creutin basis functions for $\boldsymbol{g}_k(\cdot)$ (\citealp{bradley17}; \citealp{OC86}), inverse Wishart distribution for $\boldsymbol{K}$, and diagonal $\boldsymbol{K} = \sigma_{K}^{2}\boldsymbol{I}_{r}$ with $\sigma_{K}^{2}$ given an inverse gamma prior, among others. We do not adopt the simplistic diagonal structure $\boldsymbol{K} = \sigma_{K}^{2}\boldsymbol{I}_{r}$ as this specification is less robust to basis function misspecification because the spatial covariance between the Gaussian process at regions $A$ and $B$ is completely determined by the choice of basis functions up to a proportionality constant, since $cov(\boldsymbol{g}(A)^{\prime}\boldsymbol{\eta},\boldsymbol{g}(B)^{\prime}\boldsymbol{\eta}) = \boldsymbol{g}(A)^{\prime}\boldsymbol{K}\boldsymbol{g}(B)  = \sigma_{K}^{2}\boldsymbol{g}(A)^{\prime}\boldsymbol{g}(B)$.
	
	In the MCAR model, $\boldsymbol{\psi}=(\psi_{11},...,\psi_{1n},\psi_{21},...,\psi_{2n})'$ is assumed a multivariate normal process with mean zero and covariance $\boldsymbol{\Sigma}\otimes (\boldsymbol{D}-\rho\boldsymbol{W})^{-1}$, where $\boldsymbol{W}$ is the adjacency matrix for areal units $\{A_1,...,A_n\}$, $\boldsymbol{D}=diag(w_{i+}:i=1,...,n)$ where $w_{i+}$ is the sum of all the elements in row $i$ of $\boldsymbol{W}$, $\rho$ is an unknown scalar, and $\boldsymbol{\Sigma}$ is a $2\times 2$ covariance matrix. The symbol ``$\otimes$" denotes the Kronecker product that combines the spatial dependence and the nonspatial correlation between the two variables. The matrix $\boldsymbol{\Sigma}$ is usually given an inverse Wishart prior (\citealp{Banerjee15}). However, this can be simplified in bivariate setting by rescaling the responses so that it is reasonable to assume $\boldsymbol{\Sigma}=\nu^2\boldsymbol{T}(\tau)$, where the scalar $\nu^2$ gives the variance and $\boldsymbol{T}(\tau)$ is a $2\times 2$ matrix with one on the diagonal and correlation parameter $\tau$ everywhere else. In our motivating application, the range of values of the two responses are re-scaled to both be from zero to one, suggesting that it is reasonable to assume $\boldsymbol{\Sigma}=\nu^2\boldsymbol{T}(\tau)$. This specification greatly speeds up the computation since it now only requires updating two unknowns instead of four unknowns.

    We also use basis functions for the first response variable in the OH model \citep{MHH17}. The choice of basis functions and priors for the parameters are identical to those in the SRE model. The key difference is that the dependence between variables is not induced through a shared random vector like in the SRE model but through a hierarchy structure in the data models. Specifically, $Y_2(A_i)$ in the OH model is assumed a mean that is a scaled version of the mean of $Y_1(A_i)$ (i.e., $\beta_0+\beta_2(\beta_1+\boldsymbol{g}_1(A_i)'\boldsymbol{\eta})$).

    In multivariate settings, one needs to be mindful on whether the multivariate responses are on the same scale (e.g., $Y_{1}$ may fall in the range -3 to 3, and $Y_{2}$ may fall in the range -300 to 300). For the SRE model, the basis functions $\textbf{g}_{1}(\cdot)$ and $\textbf{g}_{2}(\cdot)$ are allowed to be different, and this includes by a scaling factor (e.g., one can set $\textbf{g}_{2}(\cdot) = 100 \hspace{2pt}\textbf{g}_{1}(\cdot)$), which one can use to appropriately scale the Gaussian process. The OH model naturally re-scales the latent Gaussian process through $\beta_{2}$, and MCAR rescales the latent process $\boldsymbol{\psi}$ via the variance parameters in $\boldsymbol{\Sigma}$. In bivariate context (as is in our case) responses on different orders of magnitude are less of a concern, since one can simply rescale (or transform) one/both of the responses so that both responses are of the same order of magnitude. Moreover, in our motivating application the responses are average percentages and are consequently, restricted to the same range of zero to one. As such, we assume $\textbf{g}_{1} = \textbf{g}_{2}$ and assume constant variance in $\boldsymbol{\Sigma}$.

	\subsection{A review of a multiscale approach}

 \sloppy 
 Spatial COS is an established strategy in spatial statistics \citep{WB2005}, which is consistently covered in standard spatial statistics textbooks \citep{cressie1993,WG04,cressie2011statistics,Banerjee15}. This approach re-weights estimates based on geographic regions. One motivation for this is that it is often the case that nearby regions are often more similar (i.e., a positive spatial correlation is present empirically). In the field of Geography, this observation is sometimes referred to as ``Tobler's First Law of Geography'' \citep {cressie1993}. In our blood test monitoring data, we observe spatial correlations empirically (Moran's I hypothesis test suggests this), and as a result, can leverage this spatial dependence to improve estimates via COS.
 
\begin{figure}[h]
		\centering
		\includegraphics[width=1\linewidth]{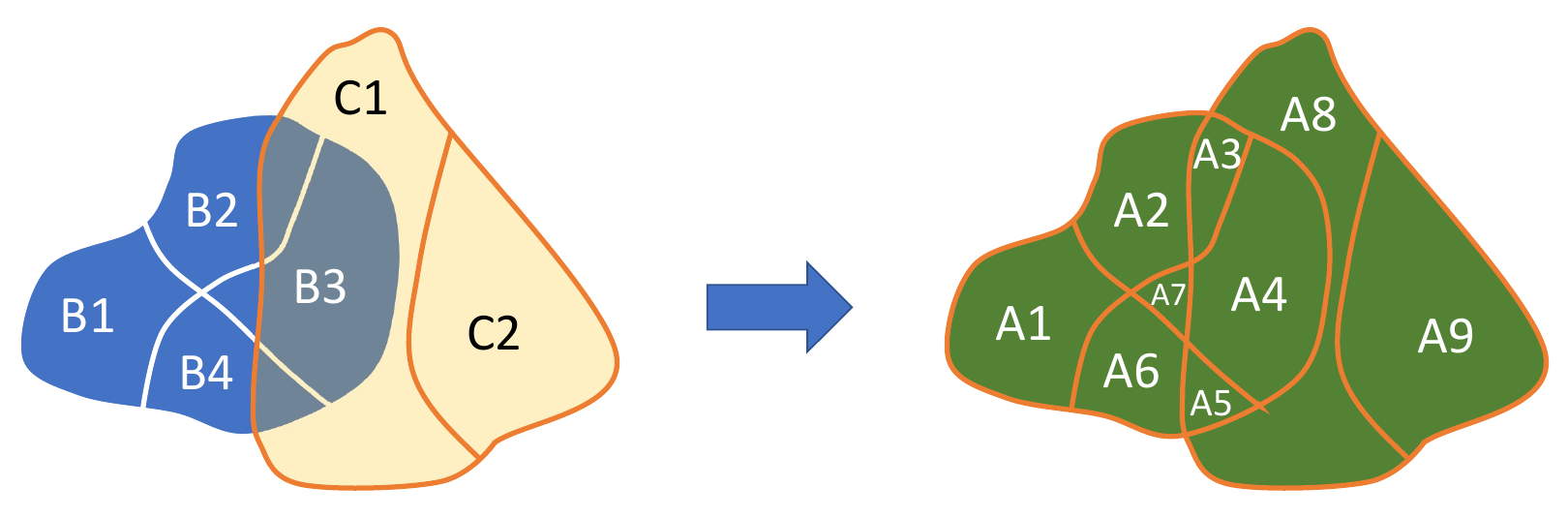}
		\caption{Partitioning plot of multiscale areal units. In the left panel the blue regions represent $D_B$ and the yellow regions represent $D_C$. The green regions represent $D_A$, which consists of the regions formed by partitioning $D_B$ and $D_C$.}
		\label{fig:partition-plot}
	\end{figure}

 We consider a multiscale approach for spatial data observed at more than one areal scale. Suppose we observe data at two sets of regions, $\{B_1,...,B_{n_b}\}$ and $\{C_1,...,C_{n_c}\}$. We call these two areal supports $D_B$ and $D_C$, and denote the data from these two supports as $Y(B_i)$ and $Y(C_j)$, respectively. Suppose that the source supports $D_B$ and $D_C$ are spatially misaligned, our inferential goal is to obtain predictions at a finer resolution with regions $\{A_1,...,A_{n_a}\}$. This target support, denoted $D_A$, is formed by the partition of $D_B$ and $D_C$. We call $D_A$ the ``partition scale." Figure~\ref{fig:partition-plot} demonstrates the partitioning for such sets of regions. In this demonstration, we let $n_b=4$ and $n_c=2$. The resulting partition scale $D_A$ has $n_a=9$ areal units. The COS strategy expresses the observed data on the source supports by geographically re-weighting the response on the target support. As an example, we can write based on Figure~\ref{fig:partition-plot} that
\begin{align}
    \label{topdownintro}
    Y(B_1)&=Y(A_1), \quad
    Y(B_2)=\frac{
			|A_2|Y(A_2)+|A_3|Y(A_3)}{|A_2\cup A_3|},\nonumber\\
   Y(B_3)&=\frac{|A_4|Y(A_4)+|A_7|Y(A_7)}{|A_4\cup A_7|},\quad
   Y(B_4)=\frac{|A_5|Y(A_5)+|A_6|Y(A_6)}{|A_5\cup A_6|}  \\	Y(C_1)&=\frac{|A_3|Y(A_3)+|A_4|Y(A_4)+|A_5|Y(A_5)+|A_8|Y(A_8)}{|A_3\cup A_4\cup A_5 \cup A_8|},\quad
  Y(C_2)=Y(A_9),\nonumber
\end{align}
where $|\cdot|$ denotes the area of the region. By assuming a spatial model on $Y(A_l)$, we can define data models for $Y(B_i)$ and $Y(C_j)$ based on (\ref{topdownintro}). 

If the population of individuals (i.e., in our case healthcare provider level average percentage of diabetic Medicare enrollees aged 65 to 75 receiving a blood test) is known by the partitioning then the geographic area $|A_l|$ could be replaced by this value as opposed to the geographic area. However, in our case, these values are not known and instead, we use the geographic area. This is reasonable considering that spatial correlations are clearly present in this data as the Moran's I \citep{MI} p-values for each variable are small (less than $10^{-4}$). Thus, standard strategies to adopt spatial dependence, such as COS and the models in Section 2.1, are expected to aid with prediction. However, this standard COS strategy is known to implicitly make assumptions on the unit-level responses \citep[e.g., see][]{GY02,bradley2015spatio,bradley16,bradley17}. We provide these details in Appendix A.

	\subsection{Prior Specifications for Bayesian Analysis} The priors we choose for SRE, MCAR, and OH models specified in Table 1 are default ``flat" priors. Generally, $\beta_k$ in all three models are given normal priors with mean zero and an extremely large variance that flattens out the distribution. All variance parameters are assumed an inverse gamma prior, which is a conjugate prior for a normal likelihood. The hyperparameters (i.e., shape and scale) in the inverse gamma prior are typically chosen for the distribution to be relatively diffuse, placing minimal prior constraints on the value of the variances. The range parameter $\phi$ in the SRE and OH models are commonly assumed to have a uniform prior with bounds sufficiently wide to accommodate both short-distance and long-distance correlations. Additionally, the correlation parameter $\tau$ in the MCAR model is also given a uniform prior with bounds that cover all possible values of correlation (i.e., -1 to 1). The hyperparameters are again so that the prior can be considered relatively ``flat" for a noninformative specification. Specifically, we have $\sigma^2_{\beta}=10^6$ and $a_{\sigma}=b_{\sigma}=a_{\eta}=b_{\eta}=a_{\nu}=b_{\nu}=1$, which gives a prior variance of infinity. The parameter $\phi$ in the MS-SRE and MS-OH models is given a uniform prior with $a_{\phi}=0$ and $b_{\phi}=10$. The parameters $\rho$ and $\tau$ in the MS-MCAR model are given uniform priors as well with $a_{\rho}=0$, $b_{\rho}=1$, $a_{\tau}=-1$, and $b_{\tau}=1$, since in the bivariate case $|\rho|<1$ (\citealp{Banerjee15}, pg. 307). We continue to adopt these priors in our bivariate multiscale spatial models introduced in Section 3.
	
	\section{Methodologies}
	
	In this section, we introduce the three bivariate multiscale spatial models. Section 3.1 describes the way to combine bivariate spatial models with the multiscale approach, and Section 3.2 provides detailed model specifications for all three bivariate multiscale spatial models.
	
	\subsection{Combining SRE, MCAR, and OH models with the multiscale approach}
	
	For bivariate spatial data observed at misaligned areal supports, we consider combining the SRE, MCAR, and OH models with the multiscale approach. We refer to the combined models as the multiscale spatial random effects (MS-SRE) model, the multiscale MCAR (MS-MCAR) model, and the multiscale ordered hierarchical (MS-OH) model. Suppose the first response is observed in regions $\{B_1,...,B_{n_1}\}$, while second response is observed in regions $\{C_1,...,C_{n_2}\}$. We call these areal scales $D_1$ and $D_2$, respectively. We denote the first response as $Y_1(B_i)$ and the second response as $Y_2(C_j)$. Our inferential goal is to predict both variables at another areal scale that is the partition of $D_1$ and $D_2$. We denote this partitioned areal scale as $D_A$ with areal units $\{A_l: l=1,...,n_3\}$. Analogous to (\ref{topdownintro}), we can write $Y_1(B_i)$ and $Y_2(C_j)$ in terms of $Y_1(A_l)$ and $Y_2(A_l)$ respectively, so that assuming a bivariate spatial model on $Y_1(A_l)$ and $Y_2(A_l)$ would naturally produce data models for $Y_1(B_i)$ and $Y_2(C_j)$.
	
	We arrange the redefined $Y_1(B_i)$ and $Y_2(C_j)$ in matrix form. Let us define $\boldsymbol{Y}_1=(Y_1(B_1),...,Y_1(B_{n_1}))'$, $\boldsymbol{Y}_2=(Y_2(C_1),...,Y_2(C_{n_2}))'$, $\boldsymbol{Y}_{A1}=(Y_1(A_1),...,Y_1(A_{n_3}))'$, and $\boldsymbol{Y}_{A2}=(Y_2(A_1),...,Y_2(A_{n_3}))'$. We can write $\boldsymbol{Y}_1=\boldsymbol{P}_1\boldsymbol{Y}_{A1}$ and $\boldsymbol{Y}_2=\boldsymbol{P}_2\boldsymbol{Y}_{A2}$ based on the partition. We refer to the $n_1\times n_3$ matrix $\boldsymbol{P}_1$ and $n_2\times n_3$ matrix $\boldsymbol{P}_2$ as partition matrices. The partition matrices $\boldsymbol{P}_1$ and $\boldsymbol{P}_2$ can be explicitly defined as
	\begin{align*}
		\boldsymbol{P}_1 &=\begin{pmatrix}
			\frac{|A_1\cap B_1|}{|B_1|} & \frac{|A_2\cap B_1|}{|B_1|}  & \cdots & \frac{|A_{n_3}\cap B_1|}{|B_1|} \\
			\frac{|A_1\cap B_2|}{|B_2|} & \frac{|A_2\cap B_2|}{|B_2|} & \cdots & \frac{|A_{n_3}\cap B_2|}{|B_2|} \\
			\vdots & \ddots &  & \vdots\\
			\frac{|A_1\cap B_{n_1}|}{|B_{n_1}|} & \frac{|A_2\cap B_{n_1}|}{|B_{n_1}|} & \cdots & \frac{|A_{n_3}\cap B_{n_1}|}{|B_{n_1}|}
		\end{pmatrix}\\
		\boldsymbol{P}_2 &=\begin{pmatrix}
			\frac{|A_1\cap C_1|}{|C_1|} & \frac{|A_2\cap C_1|}{|C_1|}  & \cdots & \frac{|A_{n_3}\cap C_1|}{|C_1|} \\
			\frac{|A_1\cap C_2|}{|C_2|} & \frac{|A_2\cap C_2|}{|C_2|} & \cdots & \frac{|A_{n_3}\cap C_2|}{|C_2|}\\
			\vdots & \ddots &  & \vdots\\
			\frac{|A_1\cap C_{n_2}|}{|C_{n_2}|} & \frac{|A_2\cap C_{n_2}|}{|C_{n_2}|} & \cdots & \frac{|A_{n_3}\cap C_{n_2}|}{|C_{n_2}|} 
		\end{pmatrix},
	\end{align*}
	respectively. Putting together $\boldsymbol{Y}_1$ and $\boldsymbol{Y}_2$, we have
	
	\begin{equation}
		\label{eqn:topdown}
		\boldsymbol{Y}=\begin{pmatrix}
			\boldsymbol{Y}_1 \\
			\boldsymbol{Y}_2 
		\end{pmatrix}=\begin{pmatrix}
			\boldsymbol{P}_1\boldsymbol{Y}_{A1} \\
			\boldsymbol{P}_2\boldsymbol{Y}_{A2}
		\end{pmatrix}=\begin{pmatrix}
			\boldsymbol{P}_1 & \boldsymbol{0} \\
			\boldsymbol{0} & \boldsymbol{P}_2 
		\end{pmatrix}\begin{pmatrix}
			\boldsymbol{Y}_{A1} \\
			\boldsymbol{Y}_{A2} 	\end{pmatrix}=\boldsymbol{P}\boldsymbol{Y}_A
	\end{equation}
	where $\boldsymbol{P}$ has dimension $(n_1+n_2)\times 2n_3$. Assuming bivariate spatial SRE, MCAR, and OH models on $\boldsymbol{Y}_A$, we can build corresponding data models (DM) for $\boldsymbol{Y}$ through the predictive distribution (PD) for $\boldsymbol{Y}_A$ in (\ref{eqn:topdown}). 
	
	For illustration, we provide the PD for $\boldsymbol{Y}_{A1}$ and $\boldsymbol{Y}_{A2}$ in the case of MS-SRE, and show how the PD induces a DM (i.e., the DM is a consequence of averaging the PD over the partitioning). A summary of the remaining multiscale models can be found in Table \ref{tab:topdownmodels}. The PD for $\boldsymbol{Y}_{A}$ assuming a SRE is, 
	\begin{align}
		\begin{split}
			\label{eqn:singlescalematrix}
			\text{MS-SRE PD1} &: \boldsymbol{Y}_{A1}|\beta_1,\boldsymbol{\eta},\sigma^2_1 \sim MVN(\beta_1\boldsymbol{1}_{n_3}+\boldsymbol{G}_1\boldsymbol{\eta},\sigma_1^2\boldsymbol{I}_{n_3})\\
			\text{MS-SRE PD2} &: \boldsymbol{Y}_{A2}|\beta_2,\boldsymbol{\eta},\sigma^2_2 \sim MVN(\beta_2\boldsymbol{1}_{n_3}+\boldsymbol{G}_2\boldsymbol{\eta},\sigma_2^2\boldsymbol{I}_{n_3}),
		\end{split}
	\end{align}
	where $\boldsymbol{1}_{n_3}$ is an $n_3$-dimensional vector of all ones, $\boldsymbol{I}_{n_3}$ is an $n_3\times n_3$ identity matrix, and the $n_3\times r$ dimensional matrices $\boldsymbol{G}_1=(\boldsymbol{g}_1(A_1)',...,\boldsymbol{g}_1(A_{n_3})')'$ and $\boldsymbol{G}_2=(\boldsymbol{g}_2(A_1)',...,\boldsymbol{g}_2(A_{n_3})')'$ are the same (i.e., $\boldsymbol{G}_1=\boldsymbol{G}_2$) basis functions generated by the Moran'I operator with adjacency matrix defined by $\{A_1,...,A_{n_3}\}$. Equation (\ref{eqn:singlescalematrix}) is simply the DM1 and DM2 from a bivariate SRE model in Table \ref{tab:singlescalemultimodels}, applied to the predictive data $\boldsymbol{Y}_{A1}$ and $\boldsymbol{Y}_{A2}$. In spatial statistics, it is common to decompose a process into a ``large scale variability term'' (in this case $\beta_{j}\boldsymbol{1}_{n_{3}}$ for $j = 1,2$), ``small-scale variability'' term (in this case $\boldsymbol{G}_{j}\boldsymbol{\eta})$ for $j = 1,2$), and ``fine-scale variability'' term (modeled via the variance $\sigma_{j}^{2}$). These terms are interpreted from a purely mathematical/statistical perspective, where the goal of each term is to model different levels of smoothness in the data. The fine-scale variability is often not included in the conditional mean, which is the desired term to predict. This is because there is no cross-covariance in (\ref{eqn:singlescalematrix}) for one to leverage to improve predictions. For a standard reference for this decomposition of spatial models see \citet{cressie2011statistics}. The $r$-dimensional vector $\boldsymbol{\eta}$ is interpreted as a random effect so that, upon marginalizing $\boldsymbol{\eta}$, $\boldsymbol{Y}_{Aj}\vert \beta_{j},\boldsymbol{K},\sigma_{j}^{2}\sim MVN(\beta_{j}\boldsymbol{1}_{n_{3}},\boldsymbol{G}_{j}\boldsymbol{K}\boldsymbol{G}_{j}^{\prime}+\sigma_{j}^{2}\boldsymbol{I}_{n_{3}})$, where recall $\boldsymbol{K} = cov(\boldsymbol{\eta})$. Thus, our use of basis functions induces heterogeneous variances in the vector $\boldsymbol{Y}_{Aj}$ as the diagonal entries in the low-rank matrix $\boldsymbol{G}_{j}\boldsymbol{K}\boldsymbol{G}_{j}^{\prime}$ are not constant. That is, the variance for the $i$-th region of the $j$-th response is given by $\boldsymbol{g}_{j}(A_{i})'\textbf{K}\boldsymbol{g}_{j}(A_{i}) + \sigma_{j}^{2}$. In the spatial statistics literature, the variance parameters $\sigma_{j}^{2}$ is referred to as a ``nugget'' parameter and allows for a discontinuity in the covariance function $cov(Y_{Aj}(H),Y_{Aj}(G))$ when $G = H \in \{A_{i}\}$. It is a strong assumption to assume the covariance function is right-continuous at zero (i.e., no nugget variance). Consider an example of iid Gaussian data, which has covariance zero and a non-zero variance. Removing the nugget in this degenerate iid example assumes the data has zero variance, which is not realistic. These nugget parameters are particularly important from a mathematical perspective, since if one sets $\sigma_{j}^{2} = 0$, we obtain a low-rank covariance matrix $\boldsymbol{G}_{j}\boldsymbol{K}\boldsymbol{G}_{j}^{\prime}$ and $\boldsymbol{Y}_{Aj}$ is distributed according to the improper singular normal distribution instead of a proper Gaussian distribution.

 Based on (\ref{eqn:topdown}) and (\ref{eqn:singlescalematrix}), we can achieve data models for our observed $\boldsymbol{Y}_{1}$ and $\boldsymbol{Y}_{2}$ defined on $\{B_i\}$ and $\{C_j\}$:
	\begin{align}
		\begin{split}
			\label{eqn:multiscalematrix}
			\text{MS-SRE DM1} &: \boldsymbol{Y}_{1}|\beta_1,\boldsymbol{\eta},\sigma_1^2 \sim MVN(\beta_1\boldsymbol{1}_{n_1}+\boldsymbol{P}_1\boldsymbol{G}_1\boldsymbol{\eta},\sigma_1^2\boldsymbol{P}_1\boldsymbol{P}_1') \\
			\text{MS-SRE DM2} &: \boldsymbol{Y}_{2}|\beta_2,\boldsymbol{\eta},\sigma_2^2 \sim MVN(\beta_2\boldsymbol{1}_{n_2}+\boldsymbol{P}_2\boldsymbol{G}_2\boldsymbol{\eta},\sigma_2^2\boldsymbol{P}_2\boldsymbol{P}_2').\\
		\end{split}
	\end{align}
	Notice that for disjoint set $\{B_i; i=1,...,n_1\}$ (or $\{C_j; j=1,...,n_2\}$) each column of $\boldsymbol{P}_1$ (or $\boldsymbol{P}_2$) would have at most one nonzero element since a partitioned unit cannot belong to more than one $B_i$ (or $C_j$). Therefore, $\boldsymbol{P}_1\boldsymbol{P}_1'$ and $\boldsymbol{P}_2\boldsymbol{P}_2'$ are diagonal matrices. Additional discussion on the connection between (\ref{eqn:singlescalematrix}) and (\ref{eqn:multiscalematrix}) is given in Appendix A.

	\subsection{Markov Chain Monte Carlo specifications}
	
	We summarize MS-SRE, MS-MCAR, MS-OH models in Table \ref{tab:topdownmodels}. To implement these three models, we adopt the Metropolis-within-Gibbs Markov Chain Monte Carlo (MCMC) algorithm to sample from the posterior. Relatively flat priors are given for all of the parameters in these models (see Section 2.3). The update of $\phi$ (in MS-SRE and MS-OH models), $\rho$, and $\tau$ (in MS-MCAR model) require Metropolis-Hastings steps. See Appendix B for the derivations of the full-conditional distributions and the details on the Metropolis-Hastings steps. For the simulation study in Section 4, the three new models will be compared based on prediction performance and computational efficiency. We will generate 2,000 samples and discard the first 1,000 as burn-in for simulation. The models are implemented for the analysis of the Texas blood test monitoring data obtained from the Dartmouth Atlas Study in Section 5, where 10,000 samples will be generated with the first 2,000 discarded for burn-in. 
 
 We make use of trace plots, effective sample size and Gelman Rubin diagnostics in Appendix C. In general, the length of the MCMC should be chosen based on the convergence of the MCMC and the effective sample size. Our metrics (i.e., trace plots and Gelman Rubin diagnostics) suggested convergence after 1000 replicates, and the effective sample size was consistently around 1000 when running the sampler for another 1000 iterations. We suggest running the MCMC long enough to obtain an effective sample size on the order of 1000, but in general, the larger the effective sample size the better. Thus, for our simulation study, which included 100 independent replicate datasets, we chose a reasonable length of 2,000 iterations, which took 18 days to run. Since the real-data analysis did not require the time costly choice of 100 independent runs, we chose a larger length of 10,000.
	
	\begin{table}
		\centering
		\resizebox{\columnwidth}{!}{
			\begin{tabular}{ l  l  l }
				\hline
				\hline
				& Bayesian Hierarchical Models & COS Predictive Models \\
				\hline
				MS-SRE & $ DM 1: \, \boldsymbol{Y}_{1}|\beta_1,\boldsymbol{\eta},\sigma_1^2 \sim MVN(\beta_1\boldsymbol{1}_{n_1}+\boldsymbol{P}_1\boldsymbol{G}_1\boldsymbol{\eta},\sigma_1^2\boldsymbol{P}_1\boldsymbol{P}_1')$
				& $\boldsymbol{Y}_{A1}|\beta_1,\boldsymbol{\eta},\sigma^2_1 \sim MVN(\beta_1\boldsymbol{1}_{n_3}+\boldsymbol{G}_1\boldsymbol{\eta},\sigma_1^2\boldsymbol{I}_{n_3})$ \\
				& $DM  2: \, \boldsymbol{Y}_{2}|\beta_2,\boldsymbol{\eta},\sigma_2^2 \sim MVN(\beta_2\boldsymbol{1}_{n_2}+\boldsymbol{P}_2\boldsymbol{G}_2\boldsymbol{\eta},\sigma_2^2\boldsymbol{P}_2\boldsymbol{P}_2')$ & $\boldsymbol{Y}_{A2}|\beta_2,\boldsymbol{\eta},\sigma^2_2 \sim MVN(\beta_2\boldsymbol{1}_{n_3}+\boldsymbol{G}_2\boldsymbol{\eta},\sigma_2^2\boldsymbol{I}_{n_3})$\\
				& $PrM: \, \boldsymbol{\eta}|\sigma^2_{\eta},\phi \sim MVN(\boldsymbol{0},\boldsymbol{K})$ & \\
				& $\quad \quad \quad \quad \quad \quad \quad\quad where \,\,\boldsymbol{K}=\{\sigma^2_{\eta}\exp(-\phi\|\boldsymbol{c}_i-\boldsymbol{c}_j\|)\}$ & \\
				& $PM 1: \, \beta_k \stackrel{i.i.d}{\sim} N(0,\sigma^2_{\beta}) \quad for \,\,k=1,2$ &\\
				& $PM 2: \, \sigma^2_{\eta} \sim IG(a_{\eta},b_{\eta})$ &\\
				& $PM 3: \, \sigma_k^2 \stackrel{i.i.d}{\sim} IG(a_{\sigma},b_{\sigma}) \quad for \,\,k=1,2$ & \\
				& $PM 4: \, \phi \sim Unif(a_{\phi},b_{\phi})$ & \\
				MS-MCAR & $DM 1: \, \boldsymbol{Y}_1|\beta_1,\boldsymbol{\psi},\sigma_1^2 \sim MVN(\beta_1\boldsymbol{1}_{n_1}+\boldsymbol{P}_1\boldsymbol{\psi}_1,\sigma_1^2\boldsymbol{P}_1\boldsymbol{P}_1')$ & $\boldsymbol{Y}_{A1}|\beta_1,\boldsymbol{\psi},\sigma_1^{2} \sim MVN(\beta_1\boldsymbol{1}_{n_3}+\boldsymbol{\psi}_{1},\sigma_1^{2}\boldsymbol{I}_{n_3})$ \\
				& $DM 2: \, \boldsymbol{Y}_2|\beta_2,\boldsymbol{\psi},\sigma_2^2 \sim MVN(\beta_2\boldsymbol{1}_{n_2}+\boldsymbol{P}_2\boldsymbol{\psi}_2,\sigma_2^2\boldsymbol{P}_2\boldsymbol{P}_2')$ & $\boldsymbol{Y}_{A2}|\beta_2,\boldsymbol{\psi},\sigma_2^{2} \sim MVN(\beta_2\boldsymbol{1}_{n_3}+\boldsymbol{\psi}_2,\sigma_2^{2}\boldsymbol{I}_{n_3})$\\
				& $ PrM: \, \boldsymbol{\psi}|\rho,\tau,\nu^2 \sim MVN(\boldsymbol{0},\boldsymbol{\Sigma}\otimes(\boldsymbol{D}-\rho \boldsymbol{W})^{-1})$ & \\
				&\quad \quad \quad\quad\quad\quad\quad\quad\quad\quad\quad\quad $where \,\,\boldsymbol{\Sigma}=\nu^2\boldsymbol{T}(\tau)$ &  \\
				& $PM 1: \,\beta_k \stackrel{i.i.d}{\sim} N(0,\sigma_{\beta}^2) \quad for \,\, k=1,2$ & \\
				& $PM 2: \, \rho \sim Unif\left(a_{\rho},b_{\rho}\right)$ & \\ 
				& $PM 3: \, \tau \sim Unif(a_{\tau},b_{\tau})$ & \\
				& $PM 4: \, \nu^2 \sim IG(a_{\nu},b_{\nu})$ & \\
				& $PM 5: \, \sigma^2_k \stackrel{i.i.d}{\sim} IG(a_{\sigma},b_{\sigma})\quad for \,\,k=1,2$ & \\ 
				MS-OH & $DM 1: \, \boldsymbol{Y}_1|\beta_1,\boldsymbol{\eta},\sigma_1^2 \sim MVN(\beta_1\boldsymbol{1}_{n_1}+\boldsymbol{P}_1\boldsymbol{G}_1\boldsymbol{\eta},\sigma_1^2\boldsymbol{P}_1\boldsymbol{P}_1')$ &$\boldsymbol{Y}_{A1}|\beta_1,\boldsymbol{\eta},\sigma_1^{2} \sim MVN(\beta_1\boldsymbol{1}_{n_3}+\boldsymbol{G}_1\boldsymbol{\eta},\sigma_1^{2}\boldsymbol{I}_{n_3})$ \\
				& $DM 2: \, \boldsymbol{Y}_2|\beta_0,\beta_1,\beta_2,\boldsymbol{\eta},\sigma^2_2 \sim MVN(\beta_0\boldsymbol{1}_{n_2}+\beta_2(\beta_1\boldsymbol{1}_{n_2}+\boldsymbol{P}_2\boldsymbol{G}_1\boldsymbol{\eta}),\sigma^2_2\boldsymbol{P}_2\boldsymbol{P}_2') $ & $\boldsymbol{Y}_{A2}|\beta_0,\beta_1,\beta_2,\boldsymbol{\eta},\sigma_2^{2} \sim MVN(\beta_0\boldsymbol{1}_{n_3}+\beta_2(\beta_1\boldsymbol{1}_{n_3}+\boldsymbol{G}_1\boldsymbol{\eta}),\sigma_2^{2}\boldsymbol{I}_{n_3})$ \\
				& $PrM: \, \boldsymbol{\eta}|\sigma^2_{\eta},\phi \sim MVN(\boldsymbol{0},\boldsymbol{K})$ & \\
				&$\quad \quad \quad \quad \quad \quad \quad\quad where \,\,\boldsymbol{K}=\{\sigma^2_{\eta}\exp(-\phi\|\boldsymbol{c}_i-\boldsymbol{c}_j\|)\}$ & \\
				& $PM 1: \, \beta_k \stackrel{i.i.d}{\sim} N(0,\sigma^2_{\beta}) \quad for \,\,k=0,1,2$ & \\
				& $PM 2: \, \sigma^2_{\eta} \sim IG(a_{\eta},b_{\eta})$ & \\
				& $PM 3: \, \sigma_k^2 \stackrel{i.i.d}{\sim} IG(a_{\sigma},b_{\sigma}) \quad for \,\,k=1,2$ & \\
				& $PM 4: \, \phi \sim Unif(a_{\phi},b_{\phi})$ & \\
				\hline
				\hline
			\end{tabular}
		}
		\caption{\label{tab:topdownmodels}Multiscale bivariate spatial models. Here DM, PrM, and PM are abbreviations for the data model, process model, and parameter model, respectively. Let $N(\mu,\sigma^2)$ be the univariate normal distribution with mean $\mu$ and variance $\sigma^2$, and $MVN(\boldsymbol{\mu},\boldsymbol{\Sigma})$ be the multivariate normal distribution with vector mean $\boldsymbol{\mu}$ and covariance matrix $\boldsymbol{\Sigma}$. Let $IG$ and $Unif$ be the inverse gamma and uniform distributions, respectively. Note that in the MS-MCAR model, $\boldsymbol{\psi}=(\boldsymbol{\psi}_1',\boldsymbol{\psi}_2')'$, where $\boldsymbol{\psi}_1=(\psi_{11},...,\psi_{1n_3})'$ and $\boldsymbol{\psi}_2=(\psi_{21},...,\psi_{2n_3})'$.}
	\end{table}

	\section{Simulation}
	
\begin{figure}[H]
		\centering
		\includegraphics[width=0.7\linewidth]{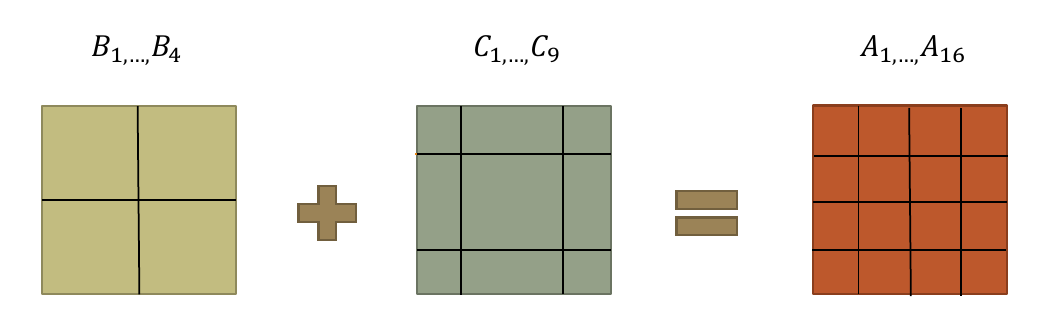}
		\caption{Illustration of the two areal supports $D_1$ and $D_2$ and their partition $D_A$ over the subregion $[0,0.2]\times [0,0.2]$. In the first panel, we plot $B_1,B_2,B_3$, and $B_4$ ($\cup_{i=1}^4 B_i=[0,0.2]\times [0,0.2]$). The remaining $B_i$ are defined according to a $10\times 10$ grid over $[0,1]\times [0,1]$. The second panel contains $C_1,...,C_9$ ($\cup_{j=1}^9 C_j=[0,0.2]\times [0,0.2]$), which is irregular. The remaining $C_j$ are defined by shifting $C_1,...,C_9$. That is, stacking $C_1,...,C_9$ over five rows and five columns produces an irregular grid over $[0,1]\times [0,1]$ with 225 areal units.}
		\label{fig:partitionsim}
	\end{figure}

	We consider bivariate multiscale spatial data observed on two misaligned areal supports. We let the first areal support $D_1$ be a regular $10\times 10$ grid on the unit square $[0,1]\times [0,1]$. We consider the second areal support $D_2$ to consist of 225 irregular regions whose union also gives the unit square. We describe the construction of $D_1$ and $D_2$ in Figure \ref{fig:partitionsim}. In the first panel of the Figure \ref{fig:partitionsim}, we plot $B_1,B_2,B_3$, and $B_4$ ($\cup_{i=1}^4 B_i=[0,0.2]\times [0,0.2]$). The remaining $B_i$ are defined according to a $10\times 10$ grid over $[0,1]\times [0,1]$. The second panel contains $C_1,...,C_9$ ($\cup_{j=1}^9 C_j=[0,0.2]\times [0,0.2]$), which is irregular. The remaining $C_j$ are defined by shifting $C_1,...,C_9$. That is, stacking $C_1,...,C_9$ over five rows and five columns produces an irregular grid over $[0,1]\times [0,1]$ with 225 areal units. We choose the areal units in $D_2$ to have this irregular structure, as shown in Figure \ref{fig:partitionsim}, such that its partition with $D_1$ would produce a $20\times 20$ regular grid (denoted $D_A$) on the unit square. For each of the three Bayesian hierarchical models specified in Table \ref{tab:topdownmodels}, we want to generate $n_1=100$ number of observations for $\boldsymbol{Y}_1$ and $n_2=225$ number of observations for $\boldsymbol{Y}_2$. Our goal is to obtain $n_3=400$ predictions for both variables on the target support $D_A$.
	
For the MS-SRE and MS-OH models, we select $r=50$ equally spaced knots using the \textit{cover.design} function in the R package \textit{fields}. We choose $\sigma^2_{\eta}=1$ and $\phi=0.1$ for the covariance $\boldsymbol{K}$ of the random vector $\boldsymbol{\eta}$. In MS-SRE and MS-MCAR, we specify the constant mean to be $\beta_1=2$ and $\beta_2=5$. In MS-OH, we choose $\beta_0=0$, $\beta_1=2$, and $\beta_2=2$. The truth of $\sigma^2_1$ and $\sigma^2_2$ in all three models are chosen such that the signal to noise ratio is 5. The signal to noise ratio is often used to specify variance parameters. In general, when the signal to noise is small (large), the conditional mean is difficult (easier) to predict. The value of the intercepts are on the same order of magnitude as what was observed in the application In MS-MCAR, a true value for $\rho$ close to one (i.e, $\rho=0.9$) is necessary to produce realistic spatial association in the latent process (\citealp{Banerjee15}, pg. 82). The variance and correlation parameters that make up the $2\times 2$ matrix $\boldsymbol{\Sigma}$ are specified as $\nu^2=1.5$ and $\tau=0.2$, respectively. This specification is particularly relevant for our motivating example, as the Moran's I test suggests significant spatial correlations. In practice, one should always check the assumption of spatial correlations (e.g., through the Moran's I test) when making use of any spatial statistical method that assumes such structure. The priors given to the parameters in all three models are considered relatively ``flat" for a noninformative specification. See Section 2.3 for details. 
	
	We generate 100 datasets from each of the MS-SRE, MS-MCAR, and MS-OH models. For each simulated dataset, we fit all of the three models to it. No lack of convergence in the MCMC is detected through visually examining the trace plots. We compute the root mean square error (RMSE) between the true latent process and the predicted latent process for both variables on both their original scales (i.e., $D_1$ or $D_2$) and the partition scale (i.e., $D_A$), and report the average and and confidence intervals (i.e., average plus or minues two standard deviations) of the RMSEs for the 100 datasets in Table \ref{tab:simresults1}a. For each of the four blocks in the table, the bold numbers on the diagonal give the smallest average RMSEs for that corresponding row. This is expected since the model where the dataset is being generated should fit the data the best. For both variables, the RMSEs calculated for the inference scale $D_A$ are always slightly larger than that for the original scale (i.e., $D_1$ and $D_2$). For the first variable, MS-OH performs significantly better than MS-MCAR when data is generated from the MS-SRE. When MS-MCAR is the truth (i.e., the model where the data is being generated from), the average RMSEs are much larger in magnitude and the misspecified models have average RMSEs that are significantly larger (more than two standard deviations away). In the case when MS-OH is the truth, the average RMSE of MS-MCAR appears to be much closer to the average RMSE of MS-OH compared to that of MS-SRE (although not true for the second variable). The other results are consistent across different variables and scales.

 In Table \ref{tab:simresults1}b, we also report the average and confidence intervals of Coverage, which is defined as the proportion of areal units whose posterior 95\% credible interval cover the true value. Comparing across different variables and scales, we find that MS-MCAR is the most robust under model misspecification. The MS-SRE model has the highest average coverage when data is generated from itself, and has the lowest average coverage when data is generated from the MS-MCAR. The result is similar for MS-OH. However, the MS-OH has average coverage for $Y_1$ significantly lower than that for $Y_2$ when MS-SRE is the truth.
 
 To show the advantage of bivariate models, we also include the average and confidence interval of RMSE for univariate multiscale models in Table \ref{tab:simresults2}. The values are obtained by fitting the univariate multiscale models to the 100 datasets simulated from their corresponding bivariate multiscale models. Comparing Table \ref{tab:simresults2} with the values bolded in Table \ref{tab:simresults1}a, we see that the univariate average RMSEs are consistently larger than the bivariate average RMSEs. We also provide a sensitivity analysis to the case where one misspecifies a bivariate relationship (i.e., no bivariate correlation exists) in Appendix D. However, this does not appear to be the case in our application, as we observe a significant cross-variable correlation parameter.
	
	\begin{table}
        \begin{subtable}[h]{1.0\textwidth}
		\centering
		\begin{tabular}{c@{\hskip 0.1in} c@{\hskip 0.1in} c@{\hskip 0.1in} c@{\hskip 0.1in} c@{\hskip 0.1in} c}
			\hline
			\hline
			Variable & Scale & Truth & Fit:MS-SRE & Fit:MS-MCAR & Fit:MS-OH \\
			\hline
			\multirow{6}{*}{$Y_1$} & \multirow{3}{*}{$D_A$} & MS-SRE & \textbf{0.015}(0.013,0.018) & 0.065(0.063,0.067) & 0.055(0.049,0.061)\\
			& & MS-MCAR & 0.477(0.469,0.485) & \textbf{0.396}(0.390,0.402) & 0.490(0.480,0.501) \\
			& & MS-OH & 0.062(0.054,0.071) & 0.043(0.040,0.046) & \textbf{0.014}(0.010,0.018) \\\cline{2-6}
			& \multirow{3}{*}{$D_1$} & MS-SRE & \textbf{0.014}(0.012,0.016) & 0.056(0.054,0.058) & 0.050(0.045,0.055) \\
			&  & MS-MCAR & 0.305(0.294,0.316) & \textbf{0.161}(0.147,0.175) & 0.325(0.309,0.341) \\
			& & MS-OH & 0.056(0.048,0.064) & 0.035(0.032,0.038) & \textbf{0.013}(0.009,0.016) \\
			\hline
			\multirow{6}{*}{$Y_2$} & \multirow{3}{*}{$D_A$} & MS-SRE & \textbf{0.015}(0.013,0.018) & 0.053(0.050,0.056) & 0.016(0.013,0.018) \\
			& & MS-MCAR & 0.495(0.485,0.505) & \textbf{0.366}(0.355,0.377) & 0.474(0.455,0.493) \\ 
			& & MS-OH & 0.045(0.038,0.052) & 0.097(0.091,0.104) & \textbf{0.026}(0.021,0.031)\\\cline{2-6}
			& \multirow{3}{*}{$D_2$}& MS-SRE & \textbf{0.014}(0.012,0.016) & 0.050(0.047,0.053) & 0.015(0.012,0.018) \\
			& & MS-MCAR & 0.415(0.403,0.427) & \textbf{0.216}(0.196,0.237) & 0.391(0.370,0.413) \\
			& & MS-OH & 0.043(0.036,0.050) & 0.091(0.085,0.098) & \textbf{0.025}(0.020,0.029)\\
			\hline
			\hline
		\end{tabular}
        \caption{RMSE}
        \label{tab:RMSE}
        \end{subtable} 

        \begin{subtable}[h]{1.0\textwidth}
        \centering
		\begin{tabular}{c@{\hskip 0.1in} c@{\hskip 0.1in} c@{\hskip 0.1in} c@{\hskip 0.1in} c@{\hskip 0.1in} c}
			\hline
			\hline
			Variable & Scale & Truth & Fit:MS-SRE & Fit:MS-MCAR & Fit:MS-OH \\
			\hline
			\multirow{6}{*}{$Y_1$} & \multirow{3}{*}{$D_A$} & MS-SRE & 0.997(0.989,1.000) & 0.981(0.977,0.985) & 0.434(0.372,0.497)\\
			& & MS-MCAR & 0.384(0.337,0.430) & 0.914(0.891,0.937) & 0.258(0.156,0.360) \\
			& & MS-OH & 0.729(0.670,0.789) & 1.000(1.000,1.000) & 0.992(0.961,1.000) \\\cline{2-6}
			& \multirow{3}{*}{$D_1$} & MS-SRE & 0.997(0.987,1.000) & 0.919(0.899,0.939) & 0.442(0.350,0.535) \\
			&  & MS-MCAR & 0.483(0.421,0.545) & 0.974(0.943,1.000) & 0.336(0.221,0.451) \\
			& & MS-OH & 0.738(0.671,0.805) & 0.995(0.984,1.000) & 0.993(0.964,1.000) \\
			\hline
			\multirow{6}{*}{$Y_2$} & \multirow{3}{*}{$D_A$} & MS-SRE & 0.996(0.986,1.000) & 0.985(0.979,0.992) & 0.999(0.995,1.000) \\
			& & MS-MCAR & 0.369(0.327,0.412) & 0.905(0.868,0.942) & 0.468(0.395,0.541) \\ 
			& & MS-OH & 0.855(0.768,0.943) & 0.972(0.960,0.984) & 0.993(0.976,1.000)\\\cline{2-6}
			& \multirow{3}{*}{$D_2$}& MS-SRE & 0.997(0.989,1.000) & 0.974(0.961,0.988) & 0.999(0.995,1.000) \\
			& & MS-MCAR & 0.426(0.362,0.490) & 0.964(0.936,0.992) & 0.567(0.463,0.672) \\
			& & MS-OH & 0.856(0.769,0.943) & 0.943(0.917,0.969) & 0.994(0.979,1.000)\\
			\hline
			\hline
		\end{tabular}
        \caption{Coverage}
        \label{tab:Coverage}
        \end{subtable}
		\caption{\label{tab:simresults1}Simulation study results. The average and confidence interval (i.e., average plus or minus two standard deviations) of RMSE and Coverage by variable for multiscale models (MS-SRE, MS-MCAR, MS-OH) when the data is generated from different truths.}
	\end{table}

\begin{figure}
		\centering
		\begin{subfigure}[t]{0.22\textwidth}
			\centering
			\includegraphics[width=\linewidth]{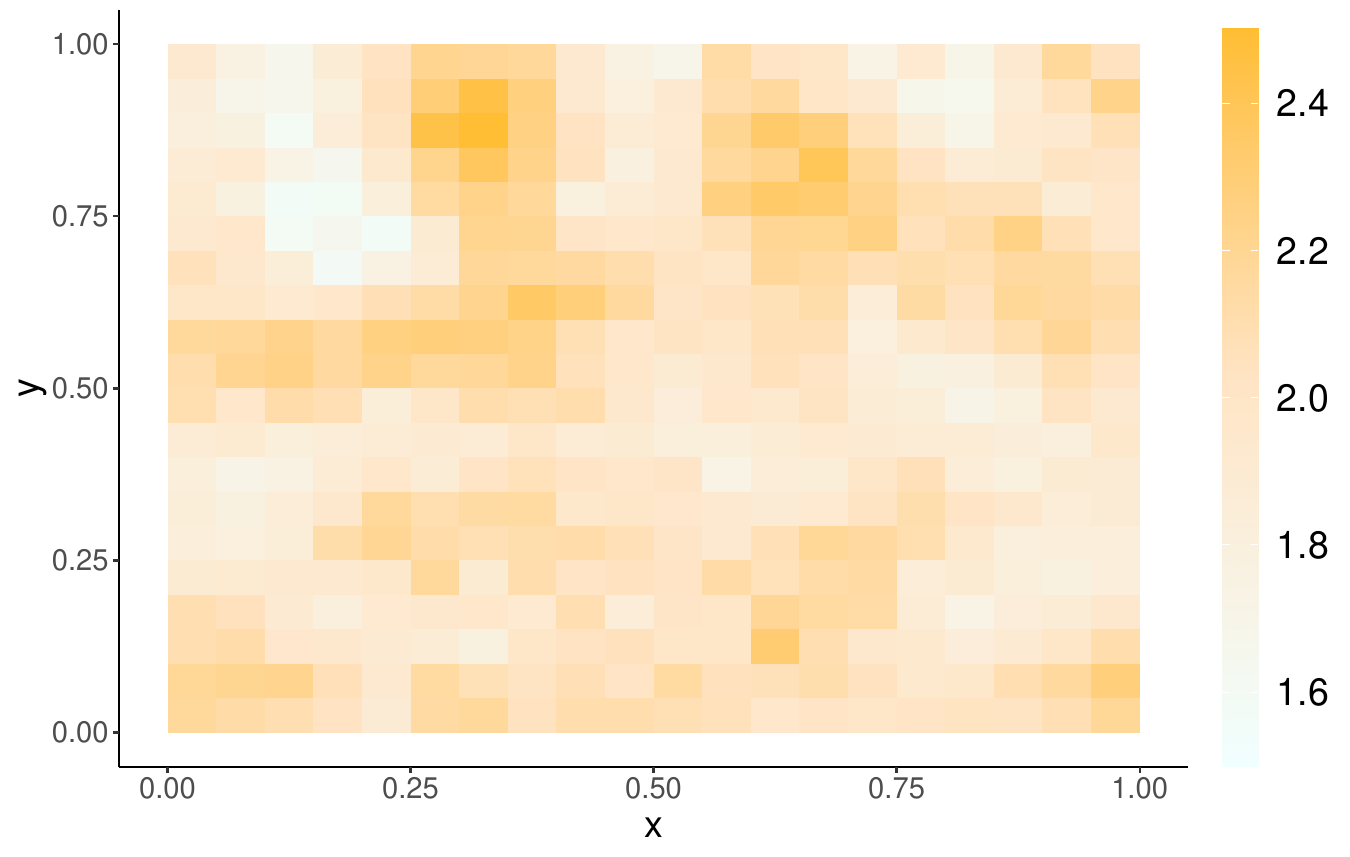} 
			\caption{MS-SRE:Truth} \label{fig:m1_y1_partition_obs}
		\end{subfigure}
		\hspace{0.05cm}
		\begin{subfigure}[t]{0.22\textwidth}
			\centering
			\includegraphics[width=\linewidth]{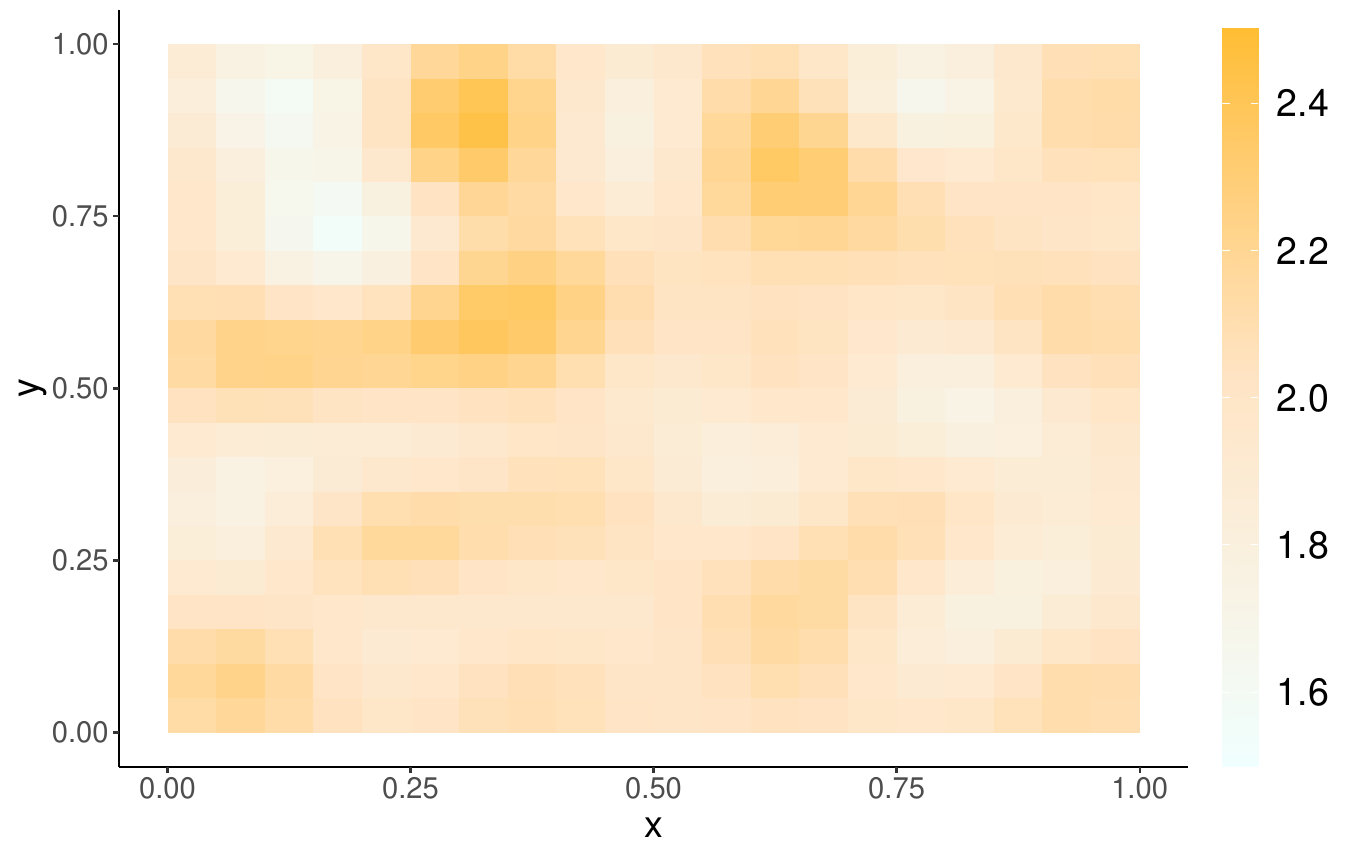} 
			\caption{MS-SRE:Fit} \label{fig:m1_m1_y1_partition_pred}
		\end{subfigure}
		\hspace{0.05cm}
		\begin{subfigure}[t]{0.22\textwidth}
			\centering
			\includegraphics[width=\linewidth]{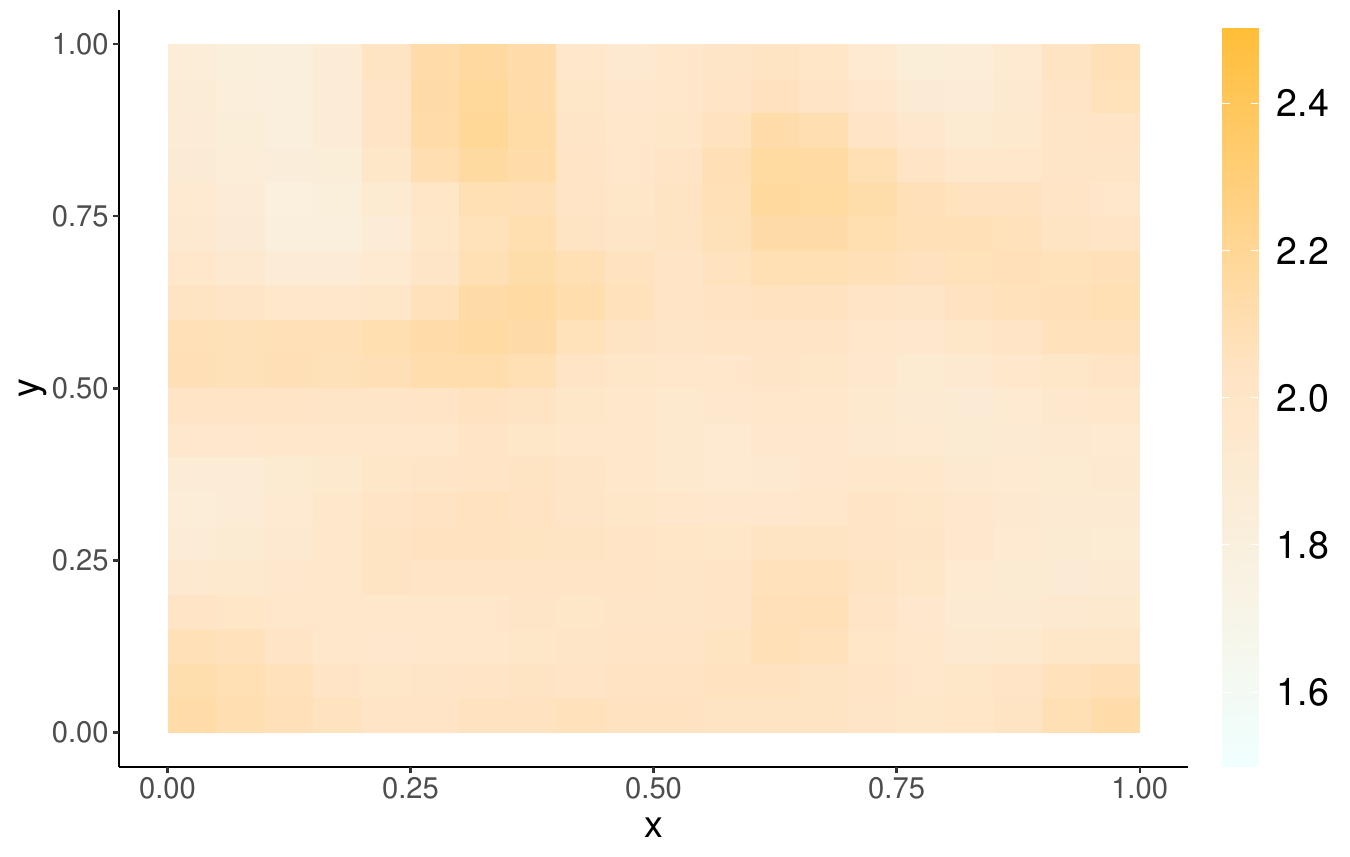} 
			\caption{MS-MCAR:Fit} \label{fig:m1_m2_y1_partition_pred}
		\end{subfigure}
		\hspace{0.05cm}
		\begin{subfigure}[t]{0.22\textwidth}
			\centering
			\includegraphics[width=\linewidth]{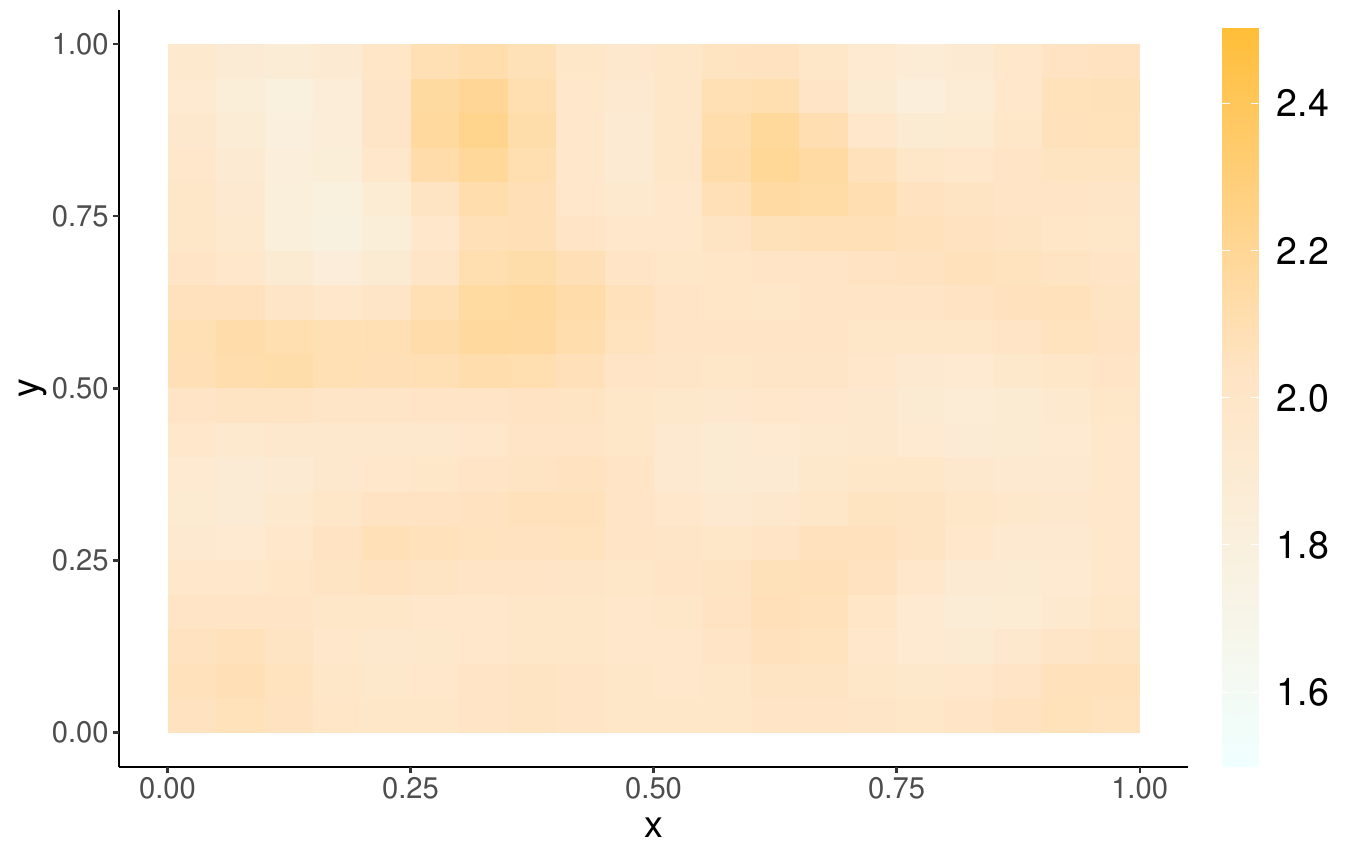} 
			\caption{MS-OH:Fit} \label{fig:m1_m3_y1_partition_pred}
		\end{subfigure}
		
		\vspace{0.1cm}
		\begin{subfigure}[t]{0.22\textwidth}
			\centering
			\includegraphics[width=\linewidth]{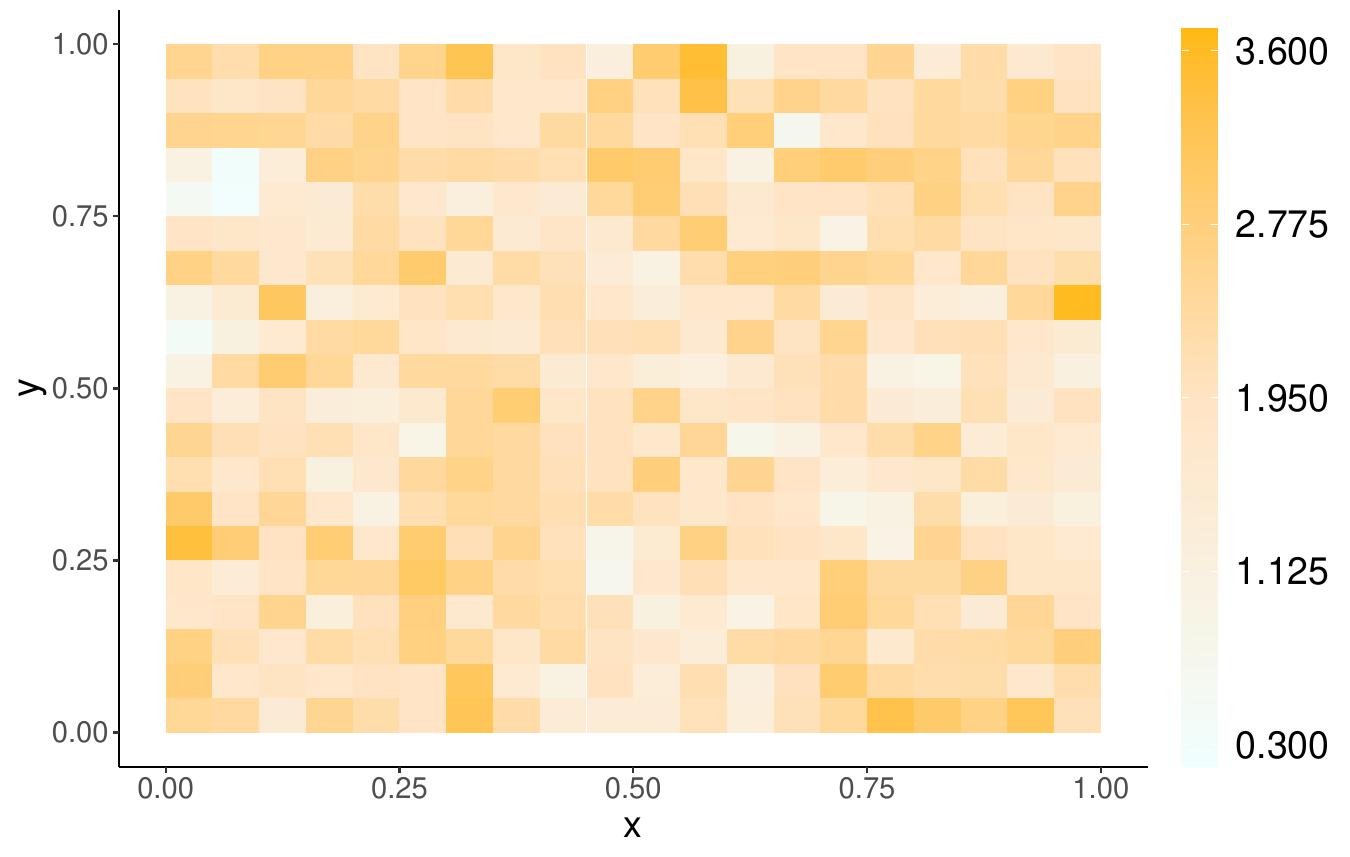} 
			\caption{MS-MCAR:Truth} \label{fig:m2_y1_partition_obs}
		\end{subfigure}
		\begin{subfigure}[t]{0.22\textwidth}
			\centering
			\includegraphics[width=\linewidth]{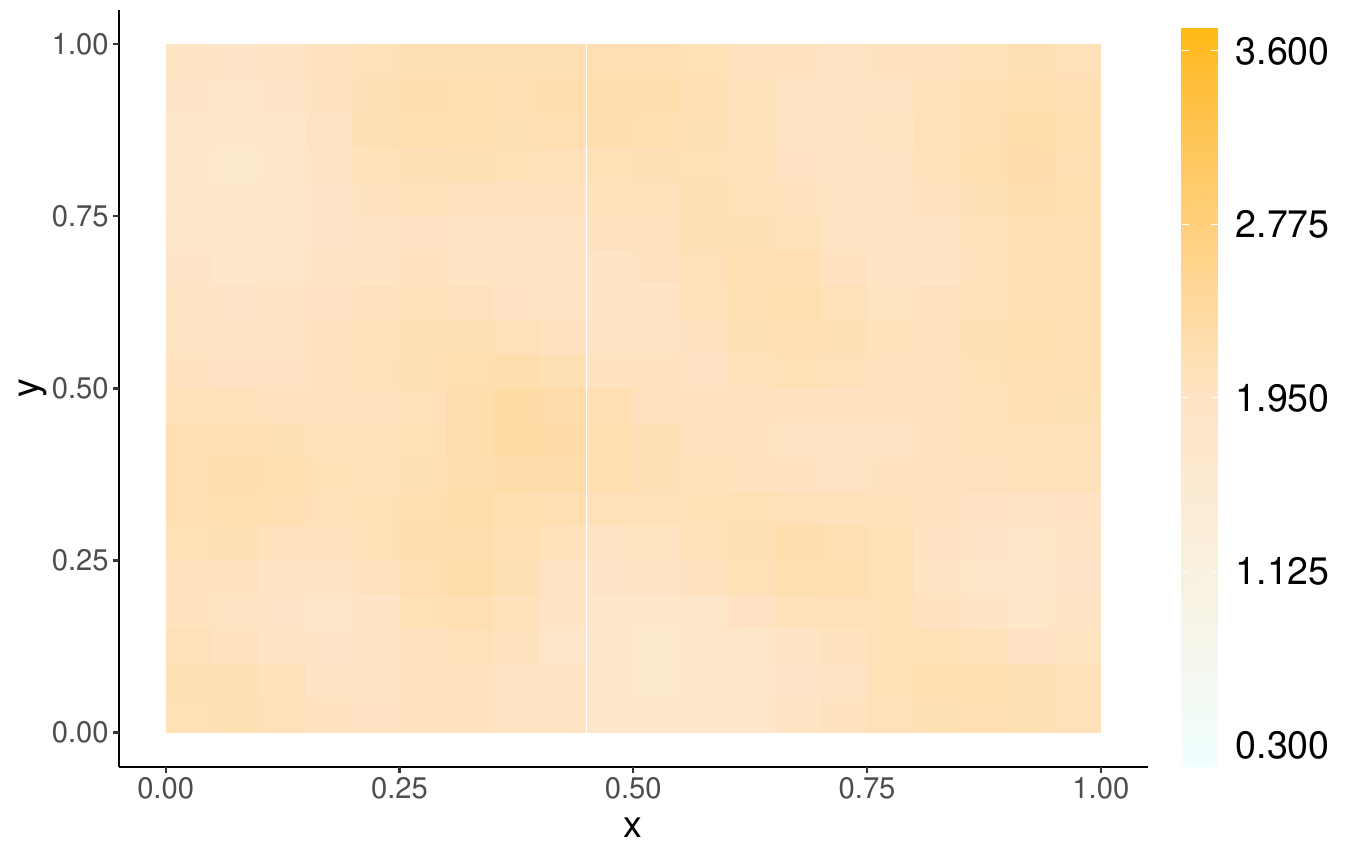} 
			\caption{MS-SRE:Fit} \label{fig:m2_m1_y1_partition_pred}
		\end{subfigure}
		\hspace{0.1cm}
		\begin{subfigure}[t]{0.22\textwidth}
			\centering
			\includegraphics[width=\linewidth]{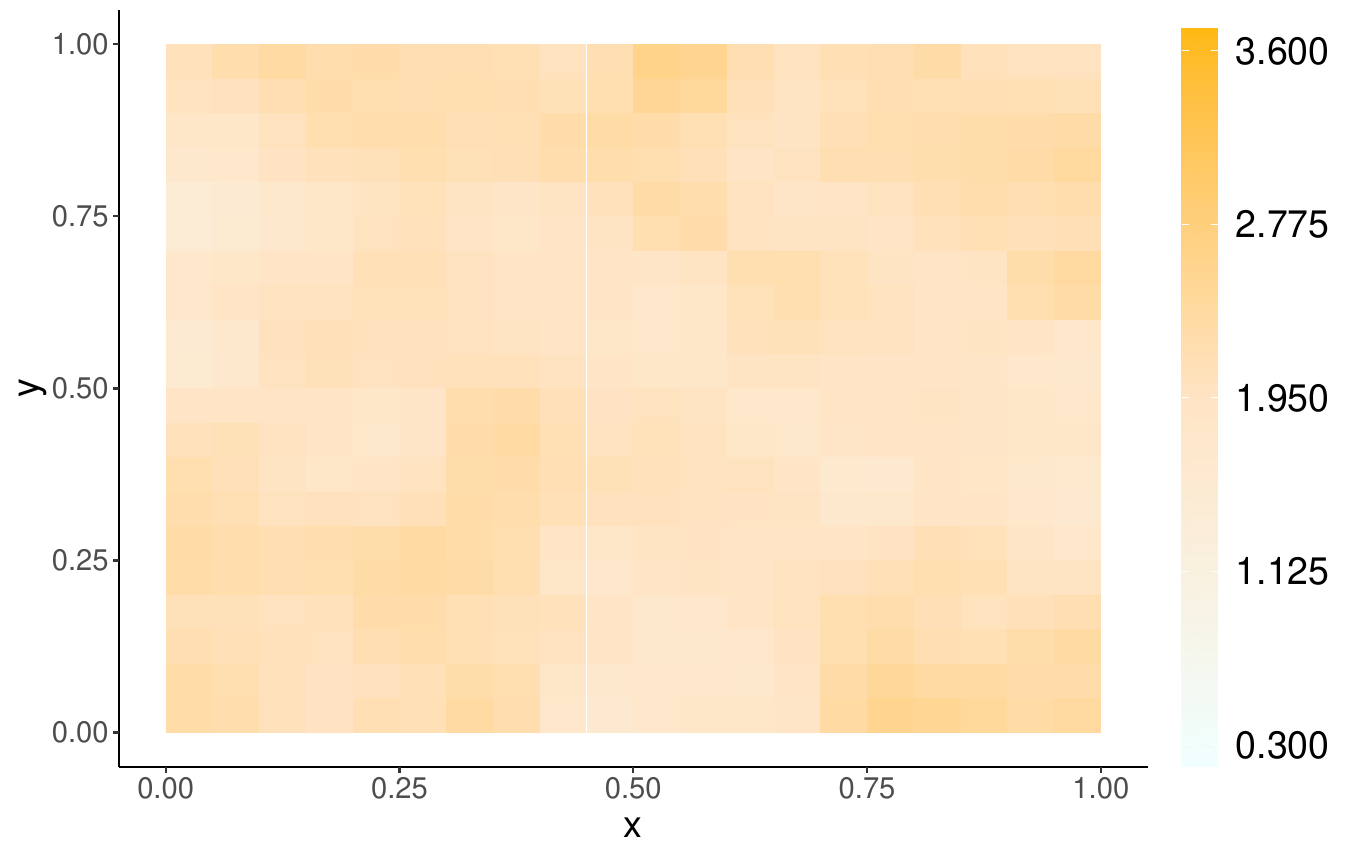} 
			\caption{MS-MCAR:Fit} \label{fig:m2_m2_y1_partition_pred}
		\end{subfigure}
		\hspace{0.1cm}
		\begin{subfigure}[t]{0.22\textwidth}
			\centering
			\includegraphics[width=\linewidth]{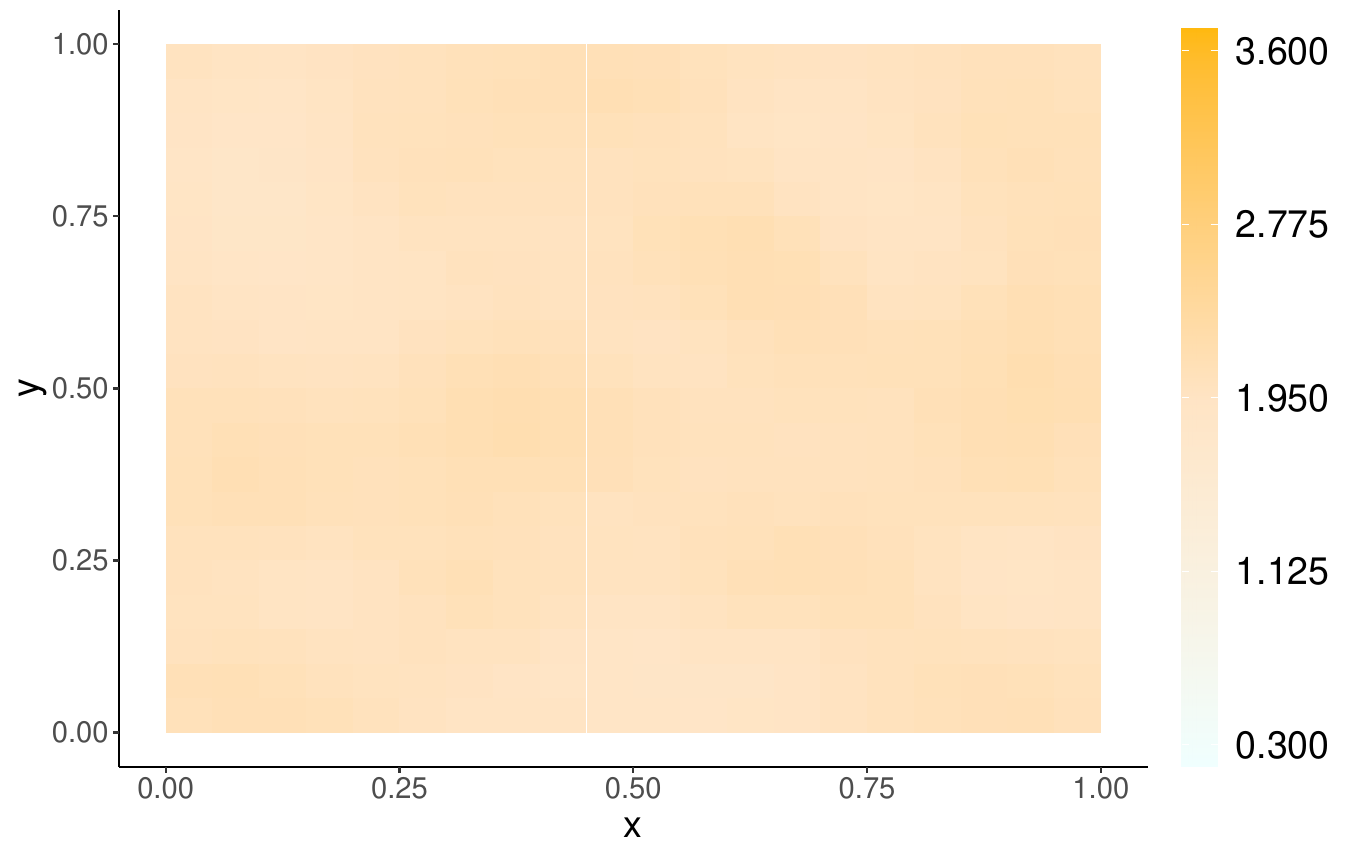} 
			\caption{MS-OH:Fit} \label{fig:m2_m3_y1_partition_pred}
		\end{subfigure}
		
		\vspace{0.1cm}
		\begin{subfigure}[t]{0.22\textwidth}
			\centering
			\includegraphics[width=\linewidth]{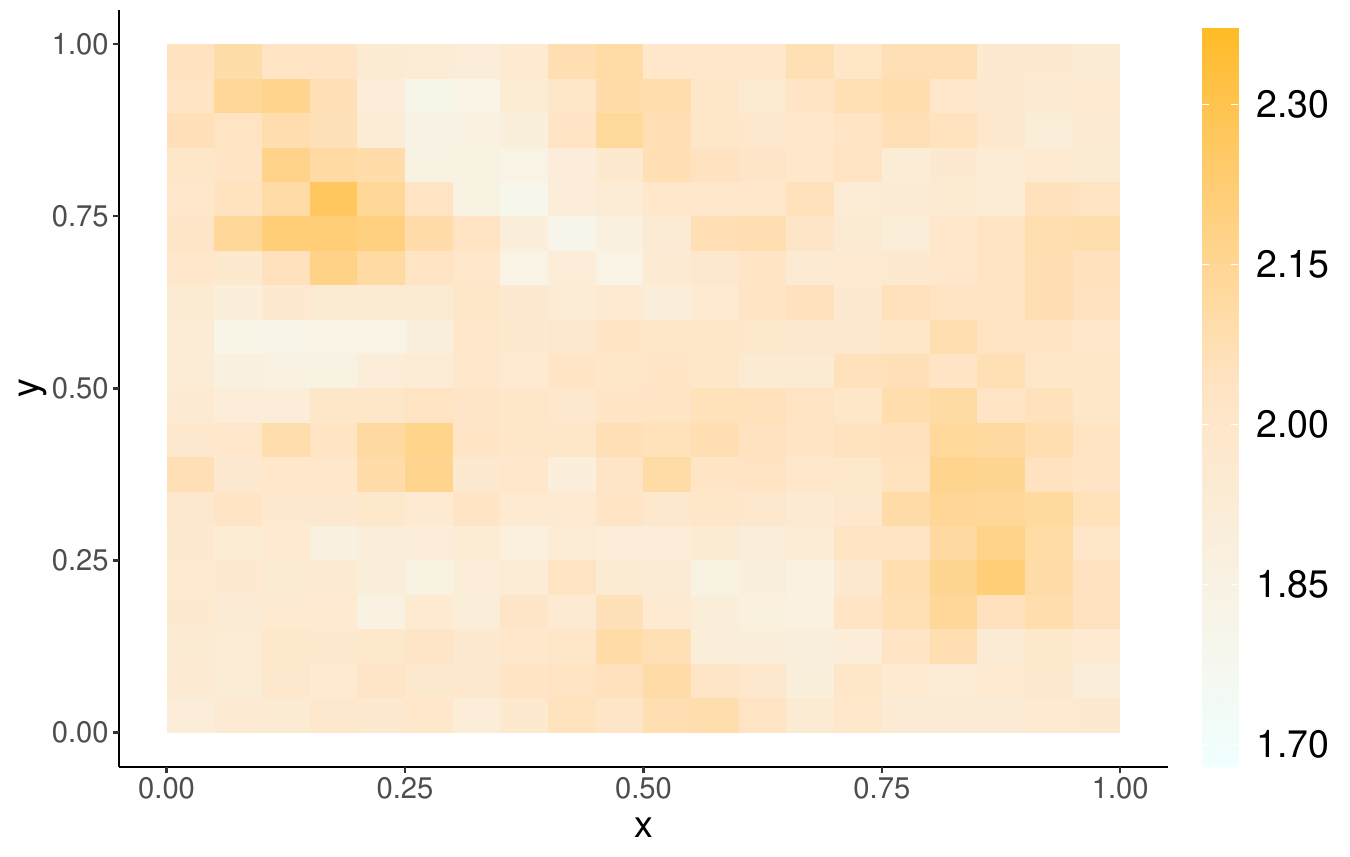} 
			\caption{MS-OH:Truth} \label{fig:m3_y1_partition_obs}
		\end{subfigure}
		\begin{subfigure}[t]{0.22\textwidth}
			\centering
			\includegraphics[width=\linewidth]{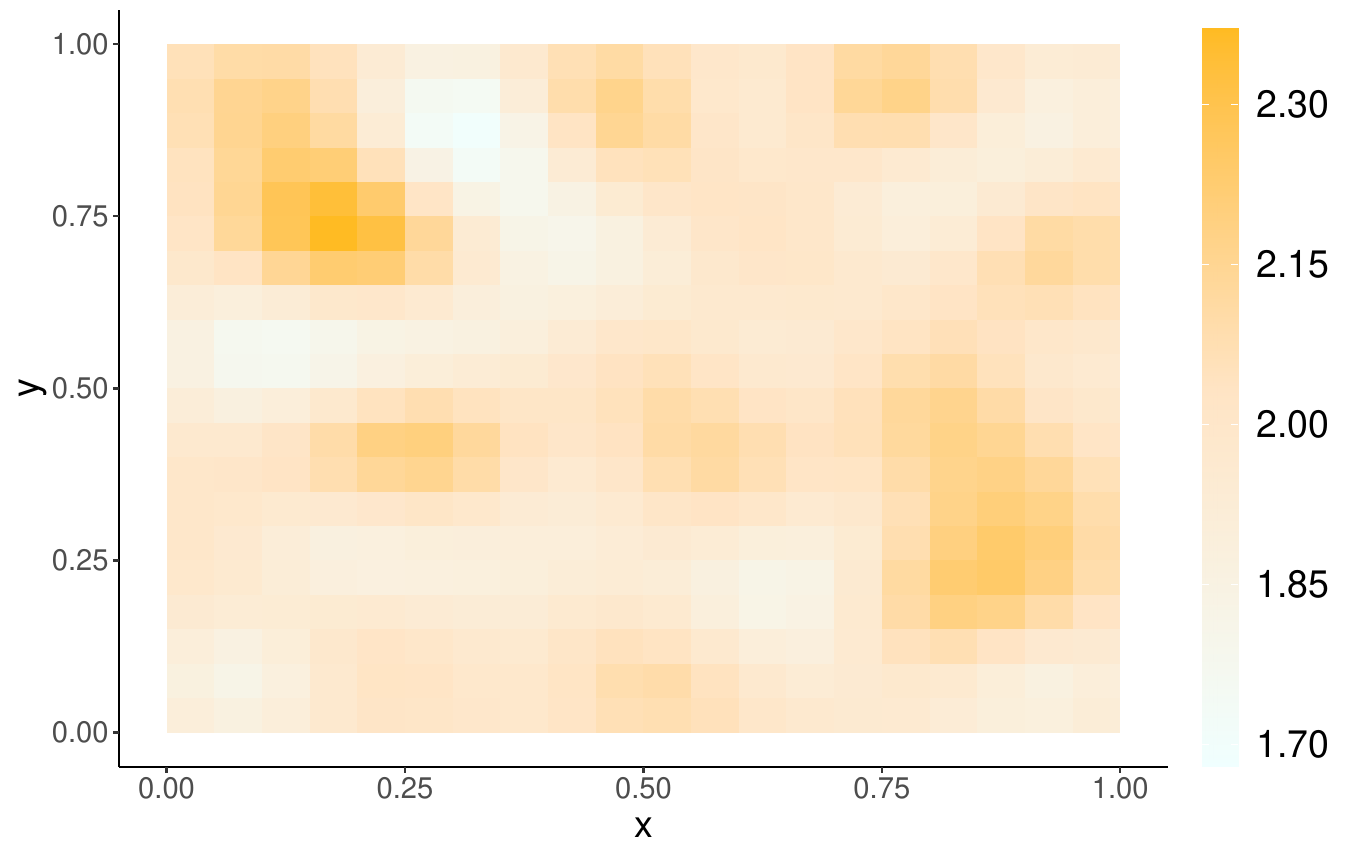} 
			\caption{MS-SRE:Fit} \label{fig:m3_m1_y1_partition_pred}
		\end{subfigure}
		\hspace{0.1cm}
		\begin{subfigure}[t]{0.22\textwidth}
			\centering
			\includegraphics[width=\linewidth]{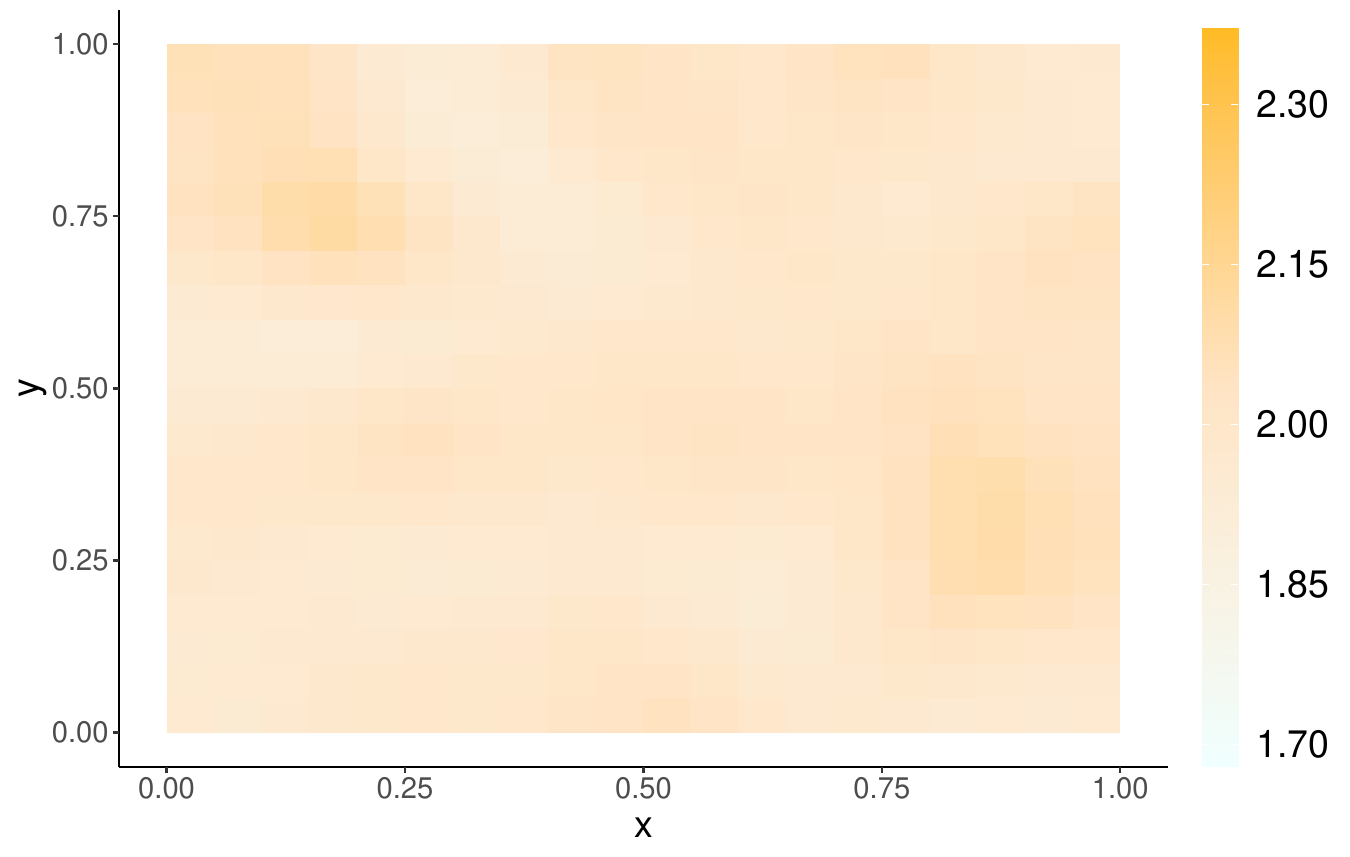} 
			\caption{MS-MCAR:Fit} \label{fig:m3_m2_y1_partition_pred}
		\end{subfigure}
		\hspace{0.1cm}
		\begin{subfigure}[t]{0.22\textwidth}
			\centering
			\includegraphics[width=\linewidth]{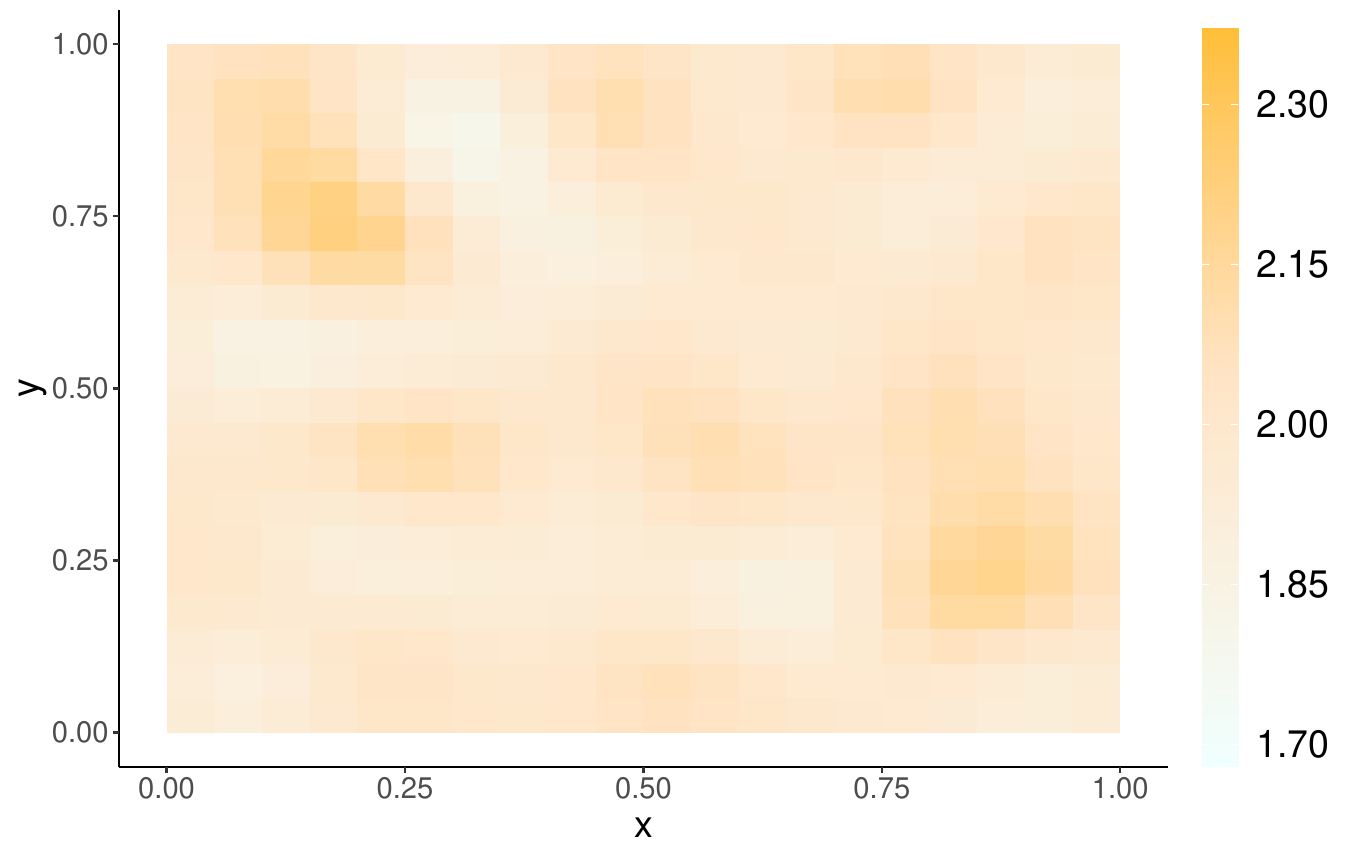} 
			\caption{MS-OH:Fit} \label{fig:m3_m3_y1_partition_pred}
		\end{subfigure}
		\caption{Data vs. prediction for $Y_1$ on $D_A$. The first column presents the simulated data on the partition scale $D_A$ where each subcaption indicates the truth (i.e., from which model the data is simulated). The subsequent three columns give the predictions obtained by fitting the model indicated in the subcaption to the data with truth specified in the first plot of the corresponding row.}
		\label{fig:simulationpred1-y1}
	\end{figure}

\begin{figure}
		\centering
		\begin{subfigure}[t]{0.22\textwidth}
			\centering
			\includegraphics[width=\linewidth]{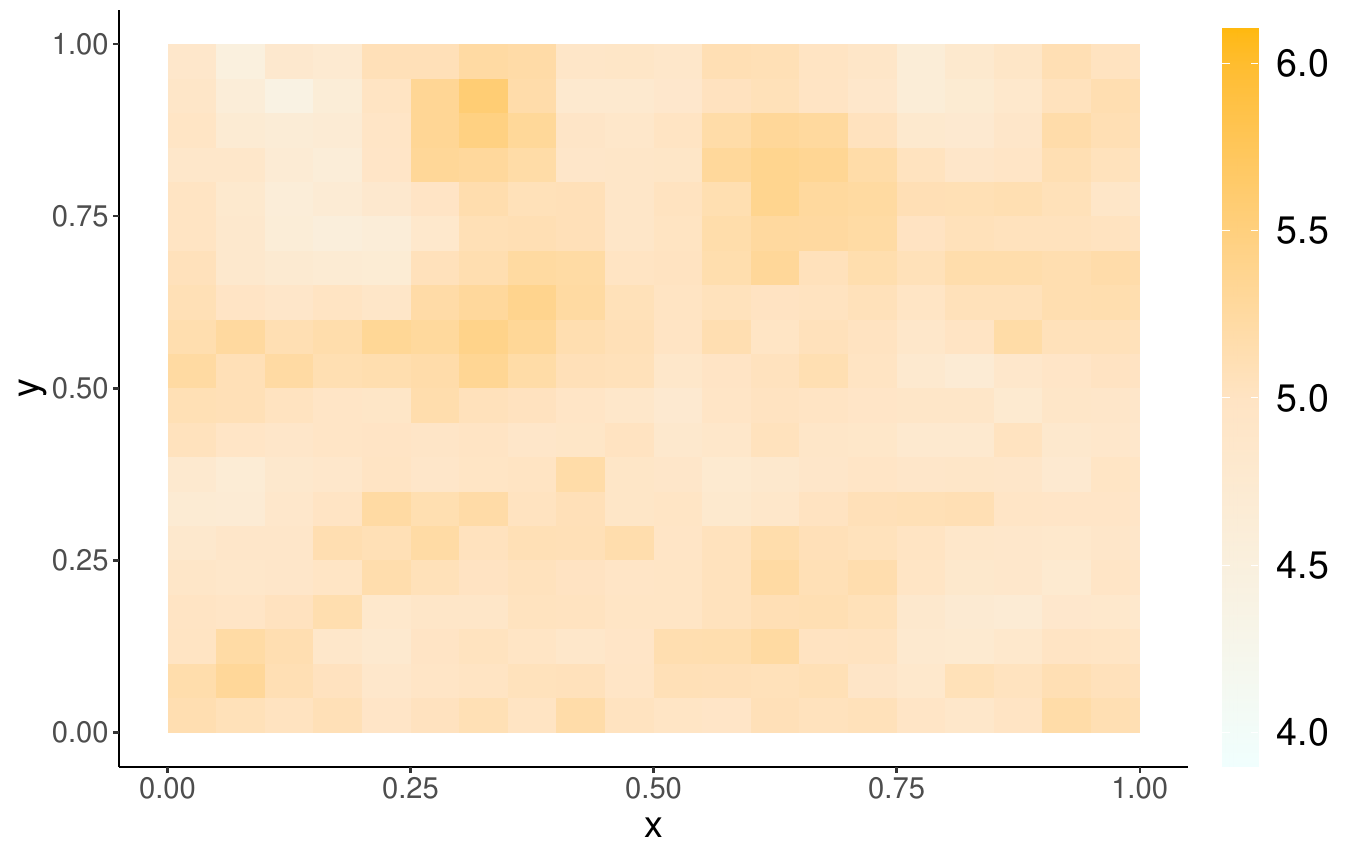} 
			\caption{MS-SRE:Truth} \label{fig:m1_y2_partition_obs}
		\end{subfigure}
		\begin{subfigure}[t]{0.22\textwidth}
			\centering
			\includegraphics[width=\linewidth]{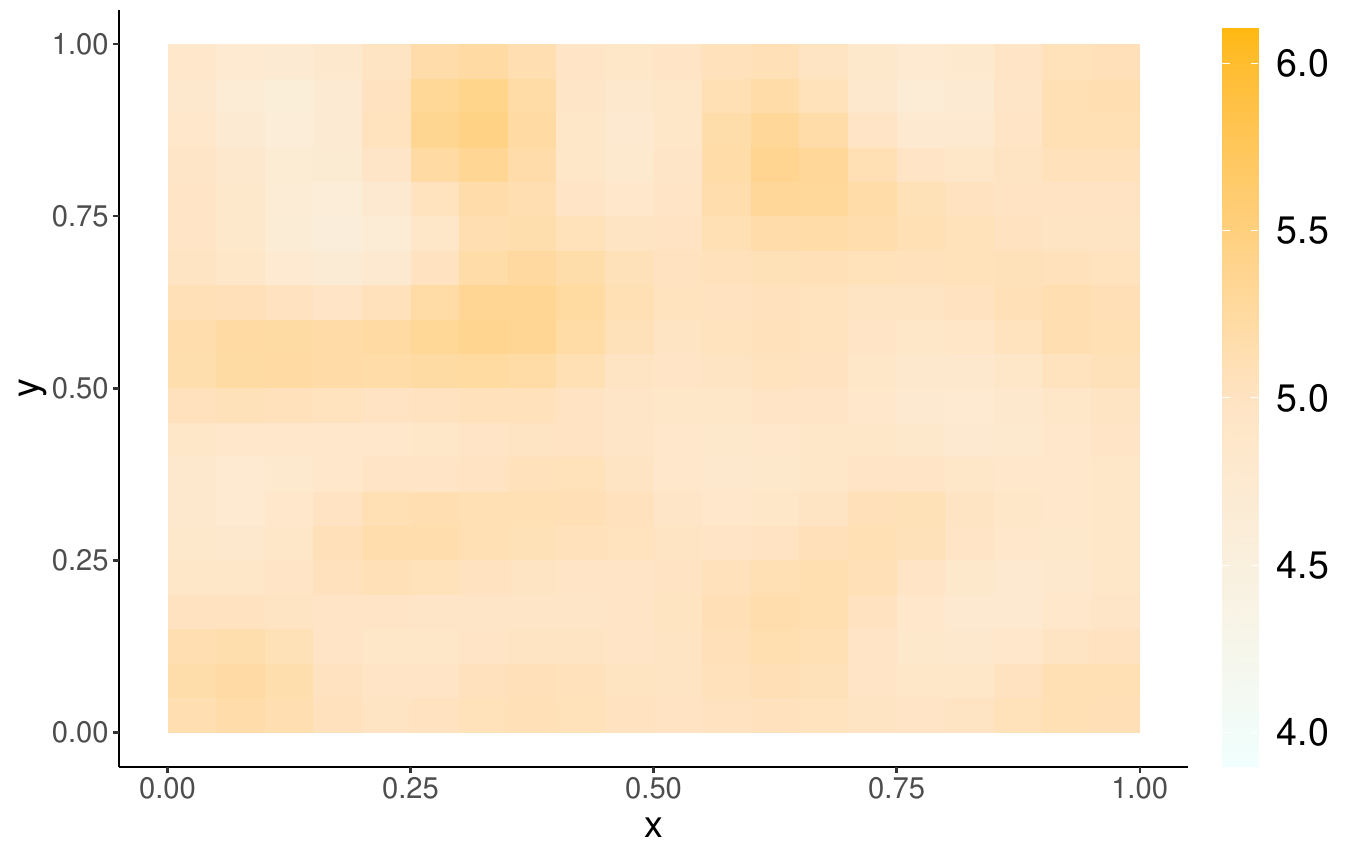} 
			\caption{MS-SRE:Fit} \label{fig:m1_m1_y2_partition_pred}
		\end{subfigure}
		\hspace{0.1cm}
		\begin{subfigure}[t]{0.22\textwidth}
			\centering
			\includegraphics[width=\linewidth]{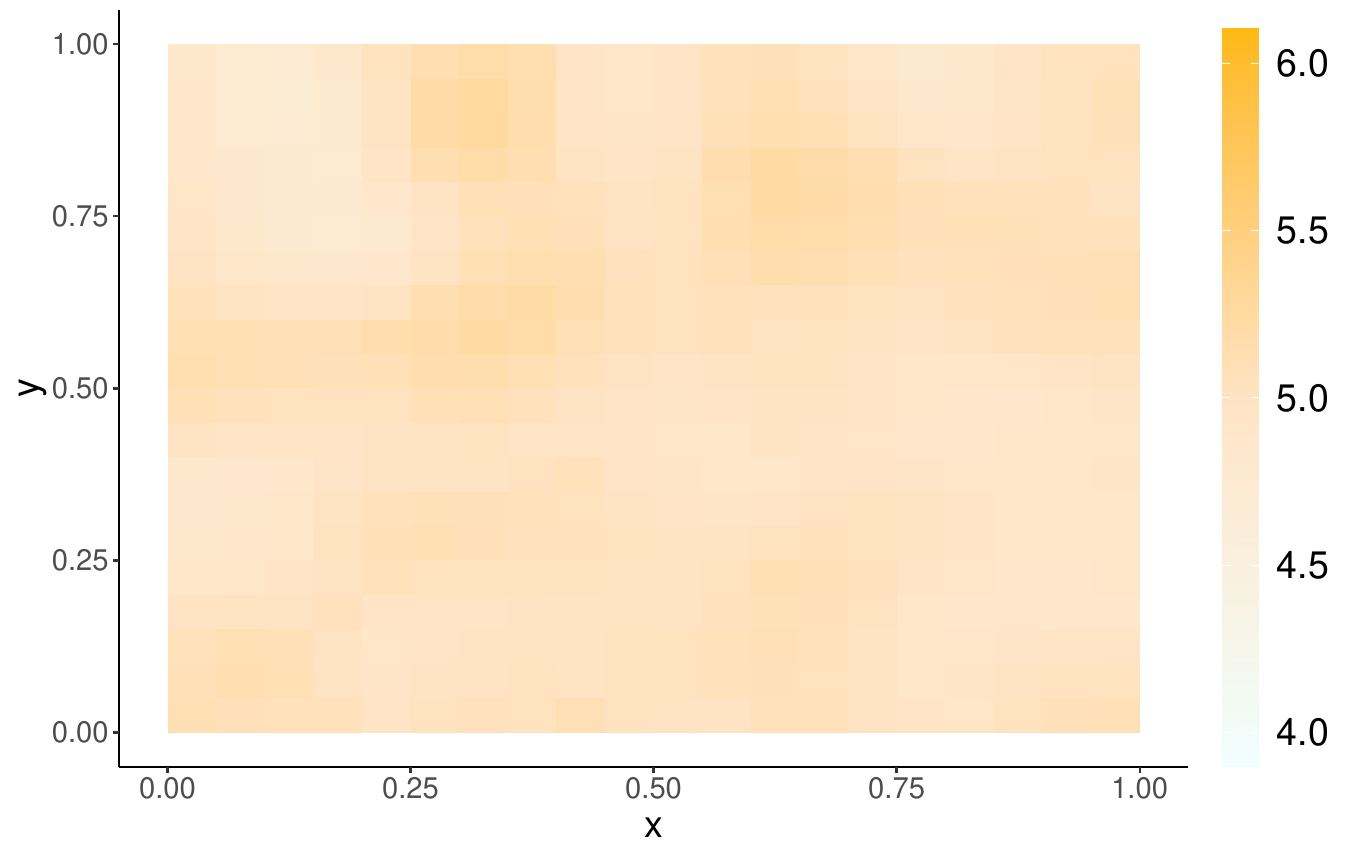} 
			\caption{MS-MCAR:Fit} \label{fig:m1_m2_y2_partition_pred}
		\end{subfigure}
		\hspace{0.1cm}
		\begin{subfigure}[t]{0.22\textwidth}
			\centering
			\includegraphics[width=\linewidth]{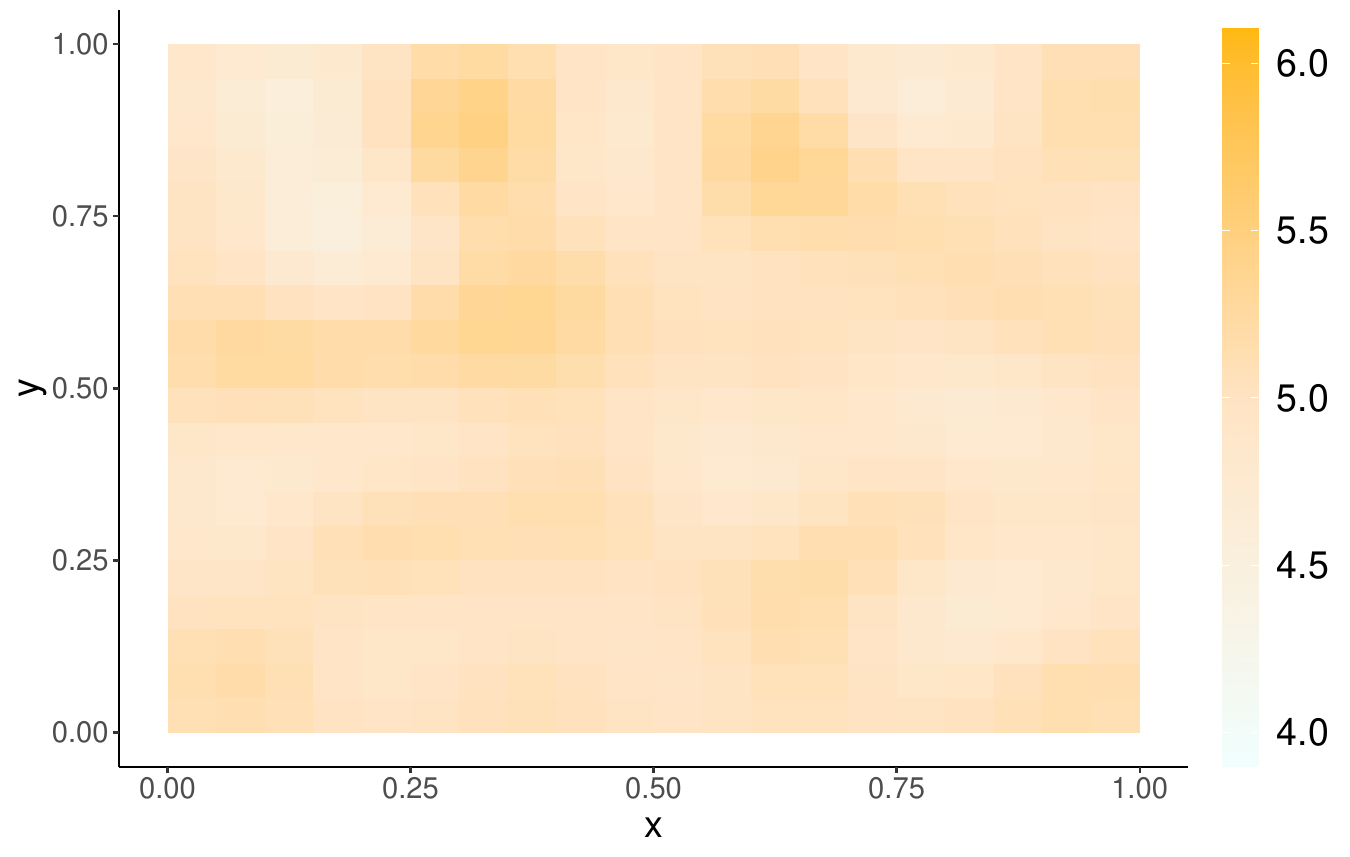} 
			\caption{MS-OH:Fit} \label{fig:m1_m3_y2_partition_pred}
		\end{subfigure}
		
		\vspace{0.1cm}
		\begin{subfigure}[t]{0.22\textwidth}
			\centering
			\includegraphics[width=\linewidth]{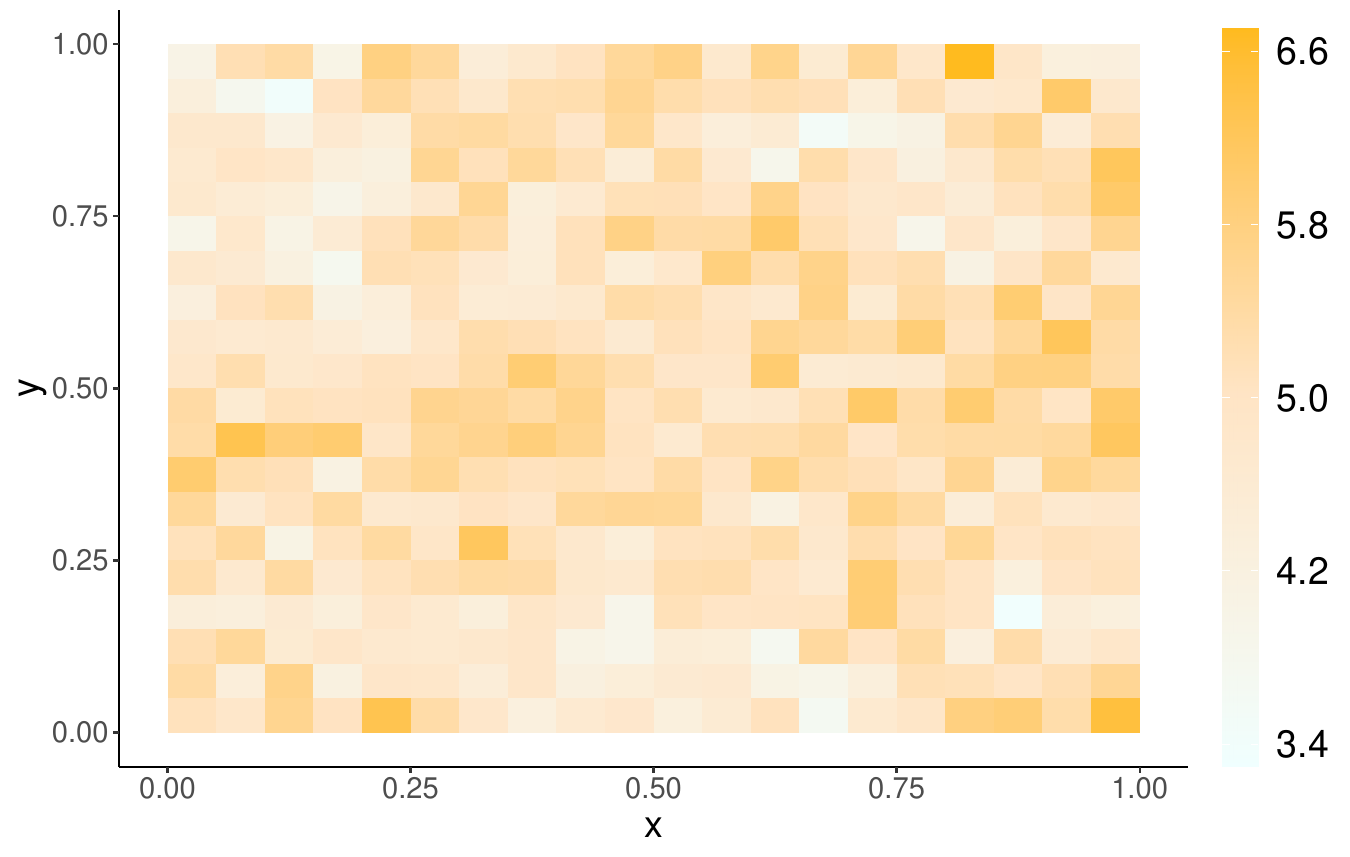} 
			\caption{MS-MCAR:Truth} \label{fig:m2_y2_partition_obs}
		\end{subfigure}
		\begin{subfigure}[t]{0.22\textwidth}
			\centering
			\includegraphics[width=\linewidth]{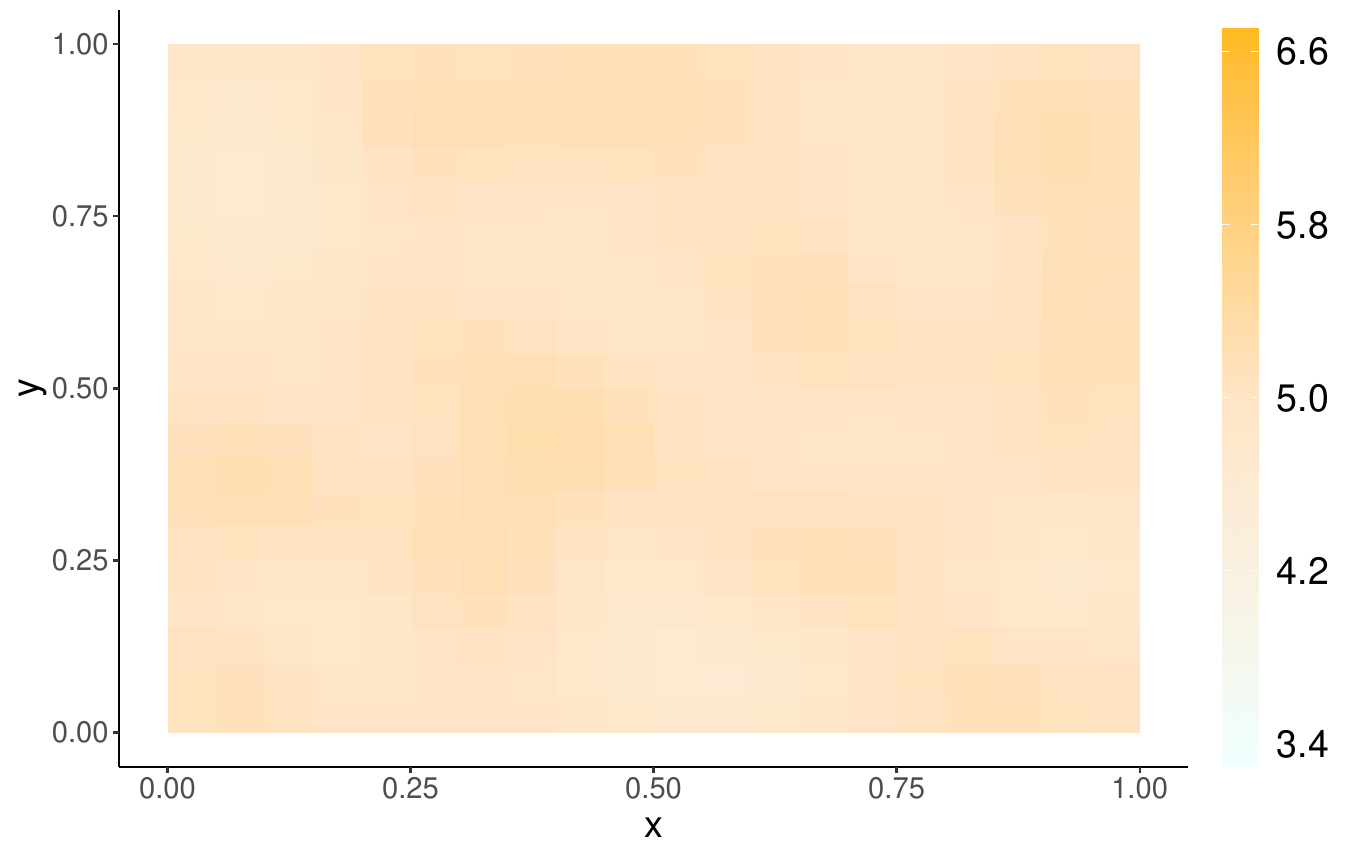} 
			\caption{MS-SRE:Fit} \label{fig:m2_m1_y2_partition_pred}
		\end{subfigure}
		\hspace{0.1cm}
		\begin{subfigure}[t]{0.22\textwidth}
			\centering
			\includegraphics[width=\linewidth]{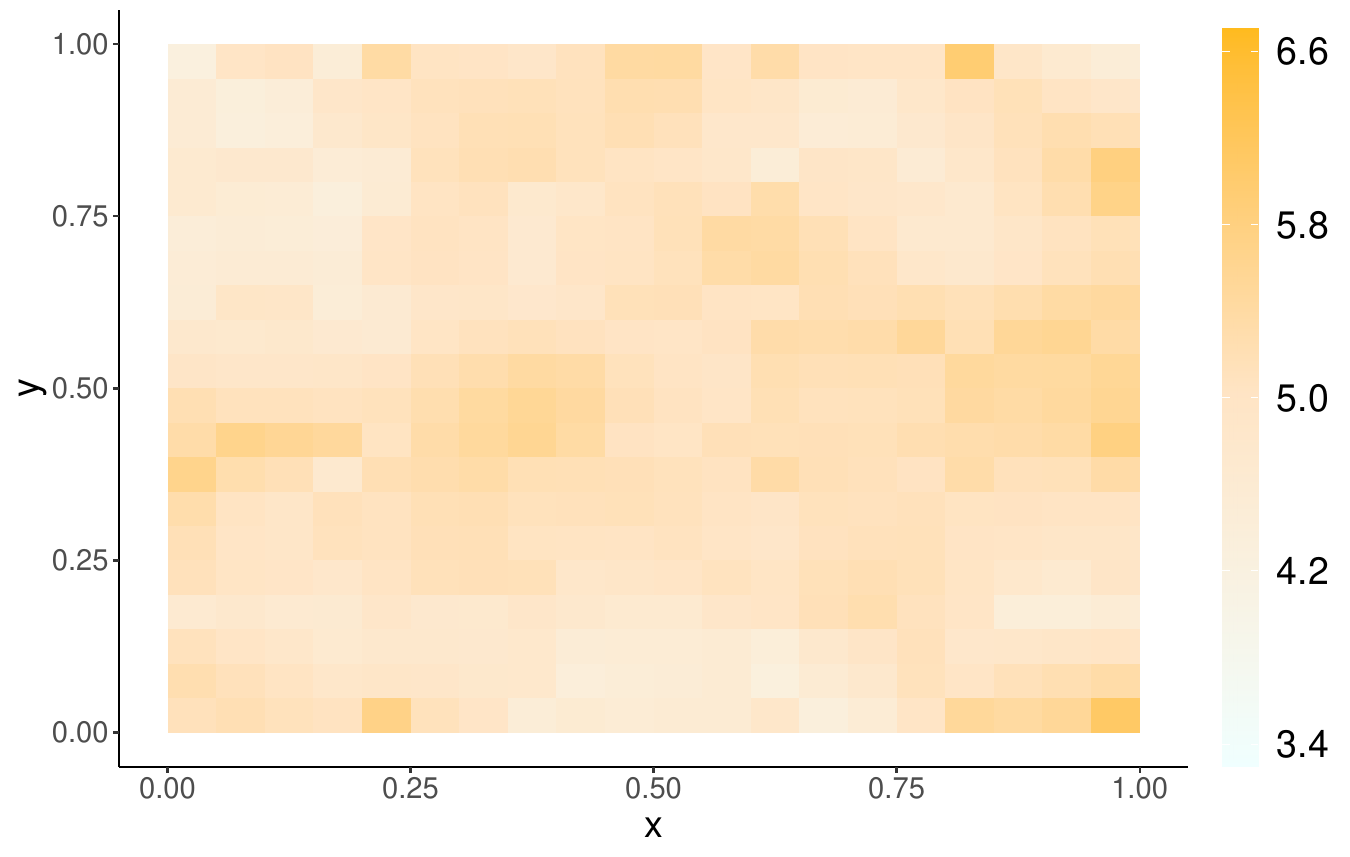} 
			\caption{MS-MCAR:Fit} \label{fig:m2_m2_y2_partition_pred}
		\end{subfigure}
		\hspace{0.1cm}
		\begin{subfigure}[t]{0.22\textwidth}
			\centering
			\includegraphics[width=\linewidth]{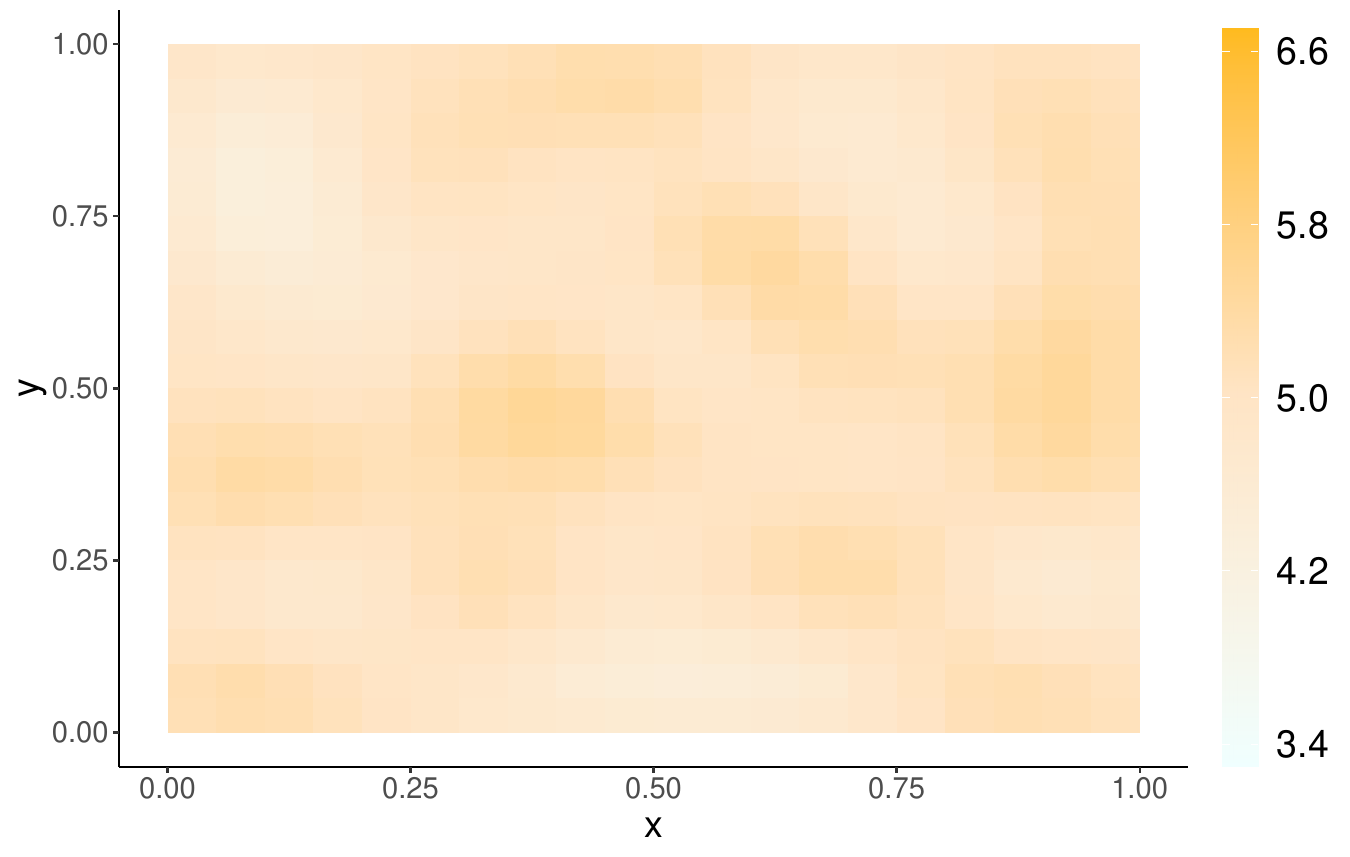} 
			\caption{MS-OH:Fit} \label{fig:m2_m3_y2_partition_pred}
		\end{subfigure}
		
		\vspace{0.1cm}
		\begin{subfigure}[t]{0.22\textwidth}
			\centering
			\includegraphics[width=\linewidth]{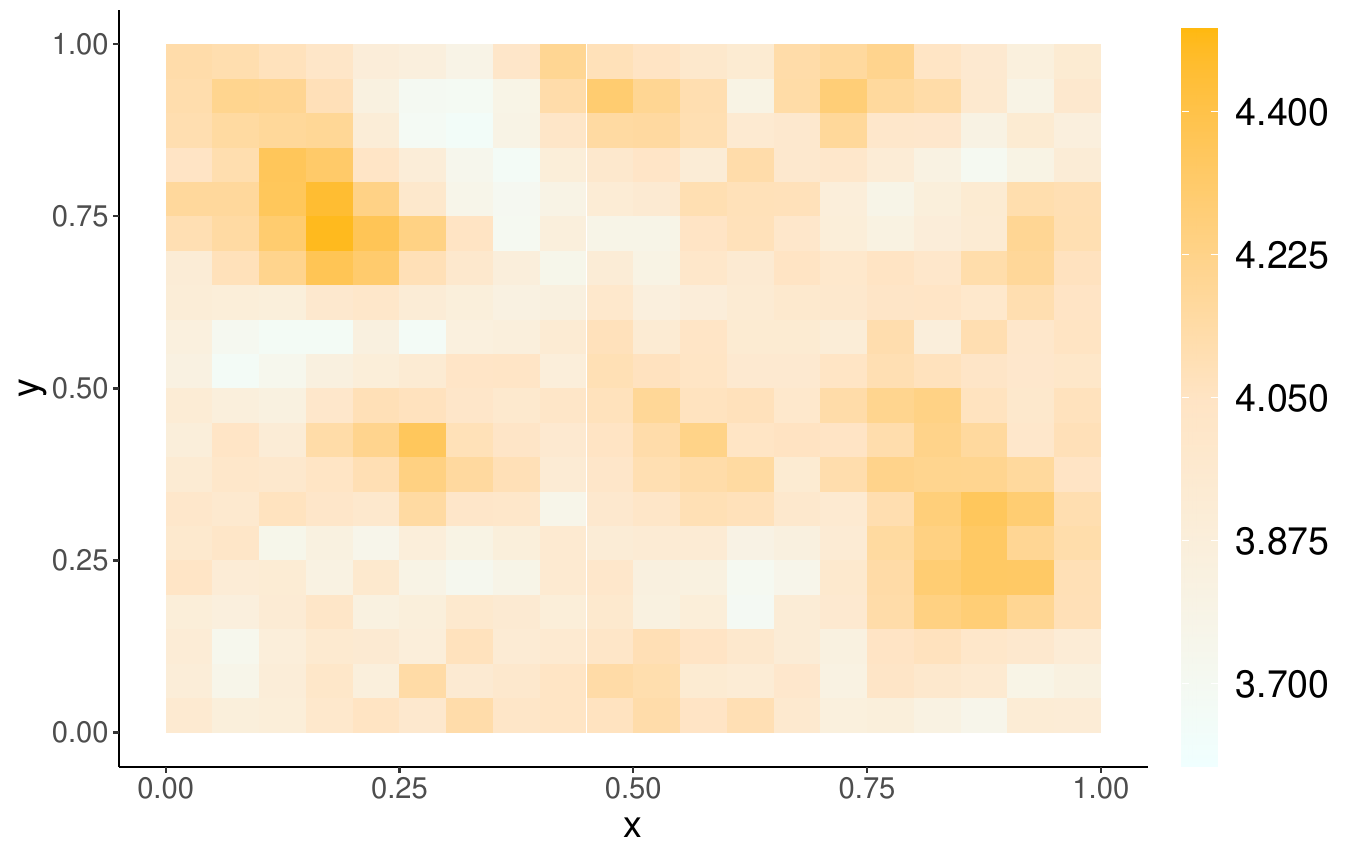} 
			\caption{MS-OH:Truth} \label{fig:m3_y2_partition_obs}
		\end{subfigure}
		\begin{subfigure}[t]{0.22\textwidth}
			\centering
			\includegraphics[width=\linewidth]{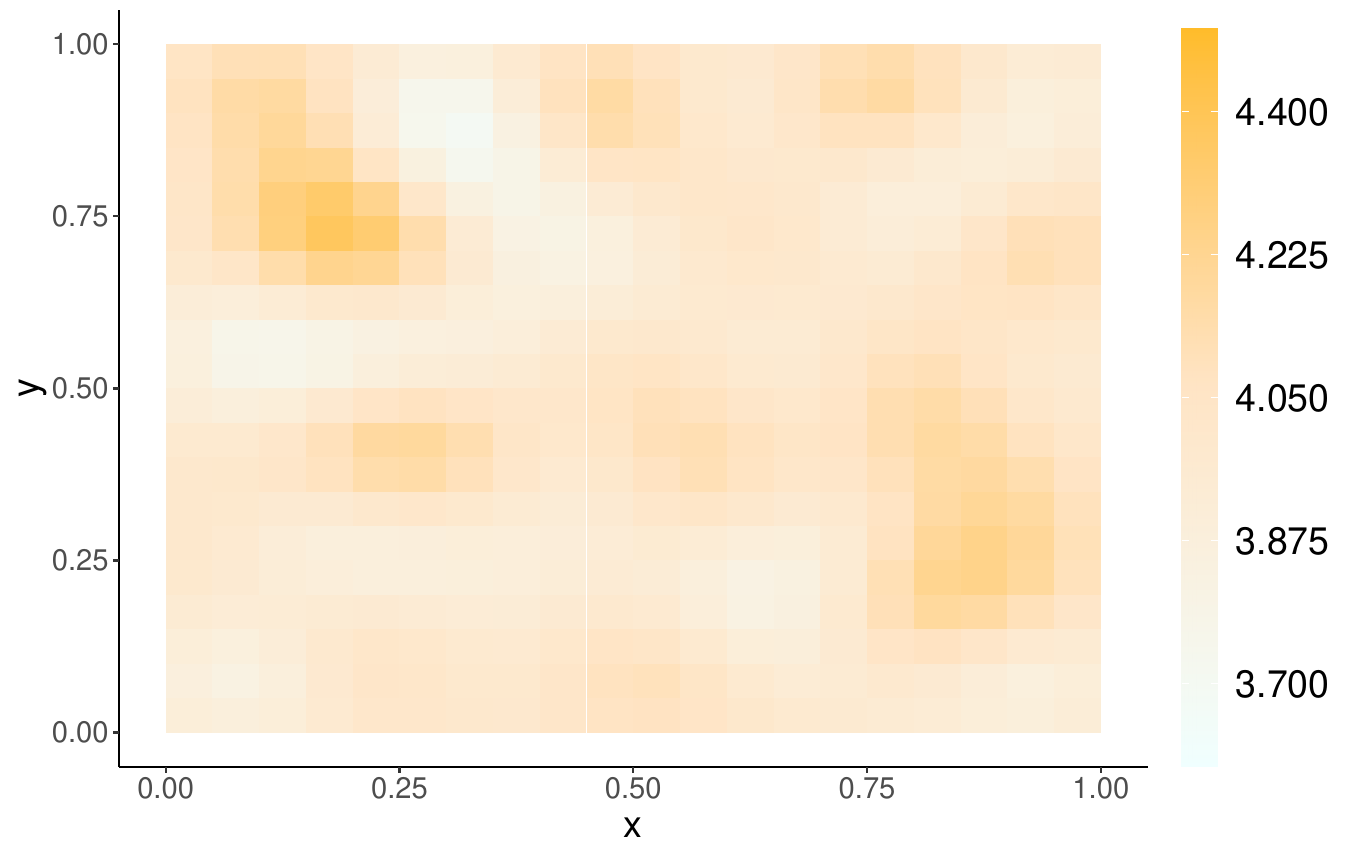} 
			\caption{MS-SRE:Fit} \label{fig:m3_m1_y2_partition_pred}
		\end{subfigure}
		\hspace{0.1cm}
		\begin{subfigure}[t]{0.22\textwidth}
			\centering
			\includegraphics[width=\linewidth]{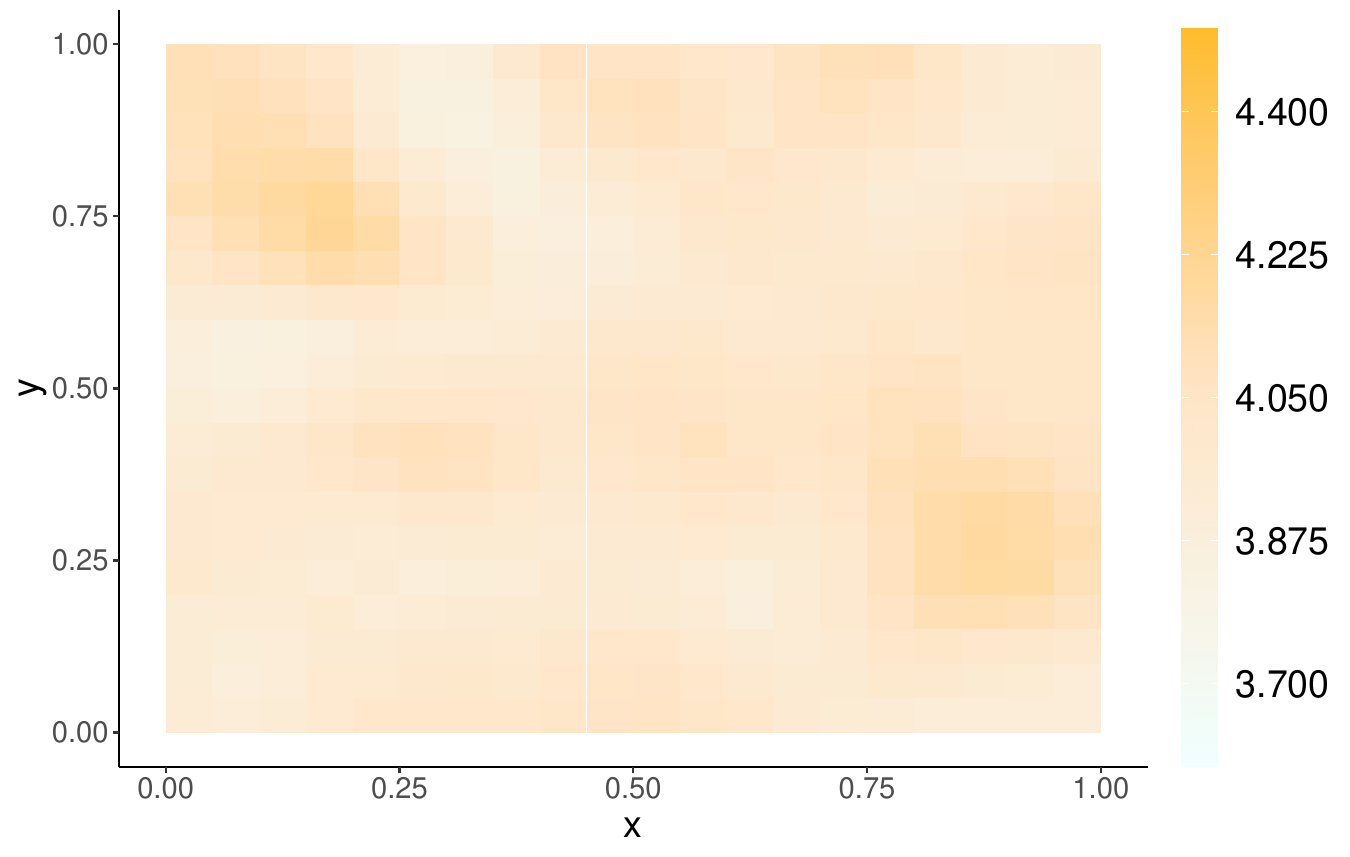} 
			\caption{MS-MCAR:Fit} \label{fig:m3_m2_y2_partition_pred}
		\end{subfigure}
		\hspace{0.1cm}
		\begin{subfigure}[t]{0.22\textwidth}
			\centering
			\includegraphics[width=\linewidth]{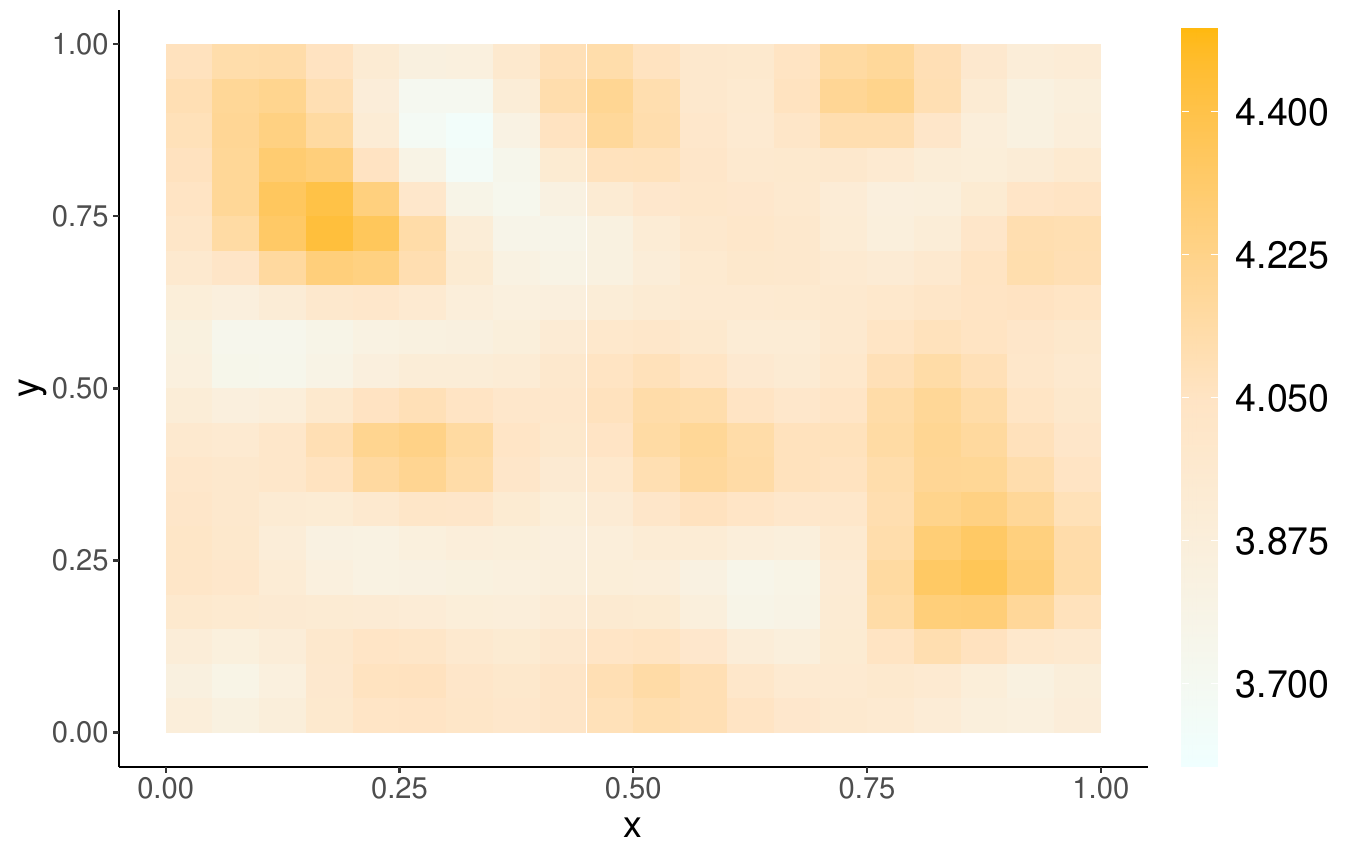} 
			\caption{MS-OH:Fit} \label{fig:m3_m3_y2_partition_pred}
		\end{subfigure}
		\caption{Data vs. prediction for $Y_2$ on $D_A$. The first column presents the simulated data on the partition scale $D_A$ where each subcaption indicates the truth (i.e., from which model the data is simulated). The subsequent three columns give the predictions obtained by fitting the model indicated in the subcaption to the data with truth specified in the first plot of the corresponding row.}
		\label{fig:simulationpred1-y2}
	\end{figure}

 	\begin{table}
		\centering
		\begin{tabular}{c@{\hskip 0.2in} c@{\hskip 0.2in} c@{\hskip 0.2in} c}
			\hline
			\hline
			Variable & Scale & Truth & Univariate RMSE \\
			\hline
			\multirow{6}{*}{$Y_1$} & \multirow{3}{*}{$D_A$} & MS-SRE & 0.022 (0.019,0.025)\\
			& & MS-MCAR & 0.420 (0.416,0.425) \\
			& & MS-OH & 0.022 (0.020,0.025) \\\cline{2-4}
			& \multirow{3}{*}{$D_1$} & MS-SRE &  0.020 (0.018,0.023) \\
			&  & MS-MCAR & 0.192 (0.185, 0.200)\\
			& & MS-OH & 0.020 (0.018,0.022) \\
			\hline
			\multirow{6}{*}{$Y_2$} & \multirow{3}{*}{$D_A$} & MS-SRE & 0.018 (0.015,0.020) \\
			& & MS-MCAR & 0.421 (0.415,0.428) \\ 
			& & MS-OH & 0.029 (0.023,0.034)\\\cline{2-4}
			& \multirow{3}{*}{$D_2$}& MS-SRE &  0.017 (0.014,0.019) \\
			& & MS-MCAR & 0.236 (0.225,0.248) \\
			& & MS-OH & 0.027 (0.022,0.032) \\
			\hline
			\hline
		\end{tabular} 
    \caption{\label{tab:simresults2}The average and confidence interval (i.e., average plus or minus two standard deviations) of RMSE by variable for univariate multiscale models when the data is generated from the corresponding bivariate multiscale models (MS-SRE, MS-MCAR, MS-OH).}
	\end{table}
	We plot the prediction results for one simulated dataset from each model. The first column in Figures \ref{fig:simulationpred1-y1} and \ref{fig:simulationpred1-y2} shows the simulated $Y_1$ and $Y_2$, respectively. For each of the two variables, the data are presented on the partition scale for comparison with the prediction plots. Recall we do not use the partition scale to fit our model, since in practice it is not observed, and the goal is to predict the process on the partition scale (i.e., COS). The prediction plots are shown in the remaining three columns of Figure \ref{fig:simulationpred1-y1} and Figure \ref{fig:simulationpred1-y2} for the first and second variable, respectively. Prediction plots
	\ref{fig:m1_m2_y1_partition_pred} and \ref{fig:m1_m3_y1_partition_pred} appear similar to each other. Spatial patterns shown in \ref{fig:m1_m1_y1_partition_pred} can also be detected in these two prediction plots, suggesting that MS-MCAR and MS-OH perform similarly (although not as good) to MS-SRE when data is generated from MS-SRE. For data generated from MS-MCAR, all prediction plots \ref{fig:m2_m1_y1_partition_pred}, \ref{fig:m2_m2_y1_partition_pred}, and \ref{fig:m2_m3_y1_partition_pred} seem to be relatively smooth; however,  \ref{fig:m2_m1_y1_partition_pred} and \ref{fig:m2_m3_y1_partition_pred} are still less favorable compared to \ref{fig:m2_m2_y1_partition_pred}. When comparing prediction plots \ref{fig:m3_m1_y1_partition_pred}, \ref{fig:m3_m2_y1_partition_pred}, and \ref{fig:m3_m3_y1_partition_pred} to \ref{fig:m3_y1_partition_obs}, we see that MS-SRE overestimates large values when MS-OH is the truth, although this does not seem to hold for the second variable. Other conclusions made remain valid for $Y_2$ in Figure \ref{fig:simulationpred1-y2}. Computationally, both MS-SRE and MS-OH outperform MS-MCAR in terms of efficiency. For fitting one simulated dataset with 5,000 iterations, the MS-SRE model needed 219.21 seconds to run, MS-OH model required 235.53 seconds, and MS-MCAR model required a much longer run-time of 9884.72 seconds.

	\section{Application}
	
	The Dartmouth Atlas Project reports health care data collected over United States annually. Specifically, they provide data on primary care access to determine the quality of hospitals within certain regions. Several quality measures are included and annual data are available on different areal scales. In our study, we are mainly interested in the blood test monitoring data used as hospital quality measures for Texas for the year 2015, and aim to jointly model the average annual percent of diabetic Medicare enrollees aged 65-75 having hemoglobin A1c test and average annual percent of diabetic Medicare enrollees aged 65-75 having blood lipids (LDL-C) test. These two tests are particularly important for senior diabetic patients, since they can be used to assess high blood sugar and high LDL-C (bad cholesterol) levels, which are known to be prevalent amongst diabetic patients \mbox{\citep{HA03}}. We focus on age group 65-75, as this group is at higher risk for complications \citep{Kirkman12}. Our bivariate multiscale models combine the first variable observed over Texas hospital service areas (HSA) with the second variable observed over Texas counties to produce higher resolution estimates of both variables. 
	
	We plot the data in Figure \ref{fig:analysisobs}. Figure \ref{A1c_hsa_obs} shows the average annual percentage of diabetic Medicare enrollees aged 65-75 having hemoglobin A1c test by the 208 HSA. Notice that some of the HSA can go beyond the boundary of Texas. The eastern and southern regions appear to have higher response values (in purple), suggesting more diabetic patients taking the hemoglobin A1c test and thus better hospital quality there. Figure \ref{lipid_county_obs} illustrates the average annual percentage of diabetic Medicare enrollees aged 65-75 having blood lipids test over counties. The gray regions indicate the data is missing for that county. There are 254 counties in total in the state of Texas, out of which 246 counties have data observed. The data in Figure \ref{lipid_county_obs} appears smoother in the plot than that in Figure \ref{A1c_hsa_obs}. Nevertheless, the purple regions are mostly seen in the eastern and southern Texas as well. To apply our models we assume the averages presented in Figure \ref{fig:analysisobs} are roughly normal. We plot histograms of the standardized residuals from our models (see Appendix E) and observe that they are unimodal and symmetric, which provides basic empirical evidence for our assumption that the data is normally distributed.

	Our inferential goal is to predict each variable on the partitioning formed by HSA and counties in Texas, which consists of 1,229 areal units in total. We fit the MS-SRE, MS-MCAR, and MS-OH models to the dataset and compare their prediction performance both through visually assessing the prediction plots and examining the widely applicable information criterion (WAIC, \citealp{Watanabe10}). Again we choose $r=150$ equally spaced knots for the models MS-SRE and MS-OH. The specifications of the hyperparameters for all three models are identical to those in the simulation study (see Section 4). The first 2,000 of 10,000 samples are discarded as burn-in, and the remaining 8,000 samples are used for analysis. Trace plots and Gelman-Rubin dignostics \citep{GR92} suggest that there is no issue with convergence.

 The scales $\{B_i\}$ (i.e., HSA ) and $C_j$ (i.e., counties) are often administratively defined, however, they are more susceptible to ecological fallacies and the modifiable areal unit problem (MAUP), creating difficulties for inference on these scales. The MAUP is a well-known issue in multiscale spatial statistics. In general, the MAUP is a type of Simpson's paradox and refers to the problem that one can obtain differing conclusions when summarizing data on different spatial scales. For example, one might conclude a particular HSA has a high average percentage of diabetic Medicare enrollees aged 65-75 that take the hemoglobin test, when in fact there is only one partition within that HSA with a high average percentage. This motivates the need for predictions on a finer scale would allow policy decision makers to have more thorough information on each particular HSA. In general, one should perform inference on the finest scale available (i.e., arising from the scales in which the data is observed) to avoid this issue \citep{bradley2015spatio,bradley16,bradley17}.
	
\begin{figure}
		\centering
		\begin{subfigure}[t]{0.45\textwidth}
			\centering
			\includegraphics[width=\linewidth]{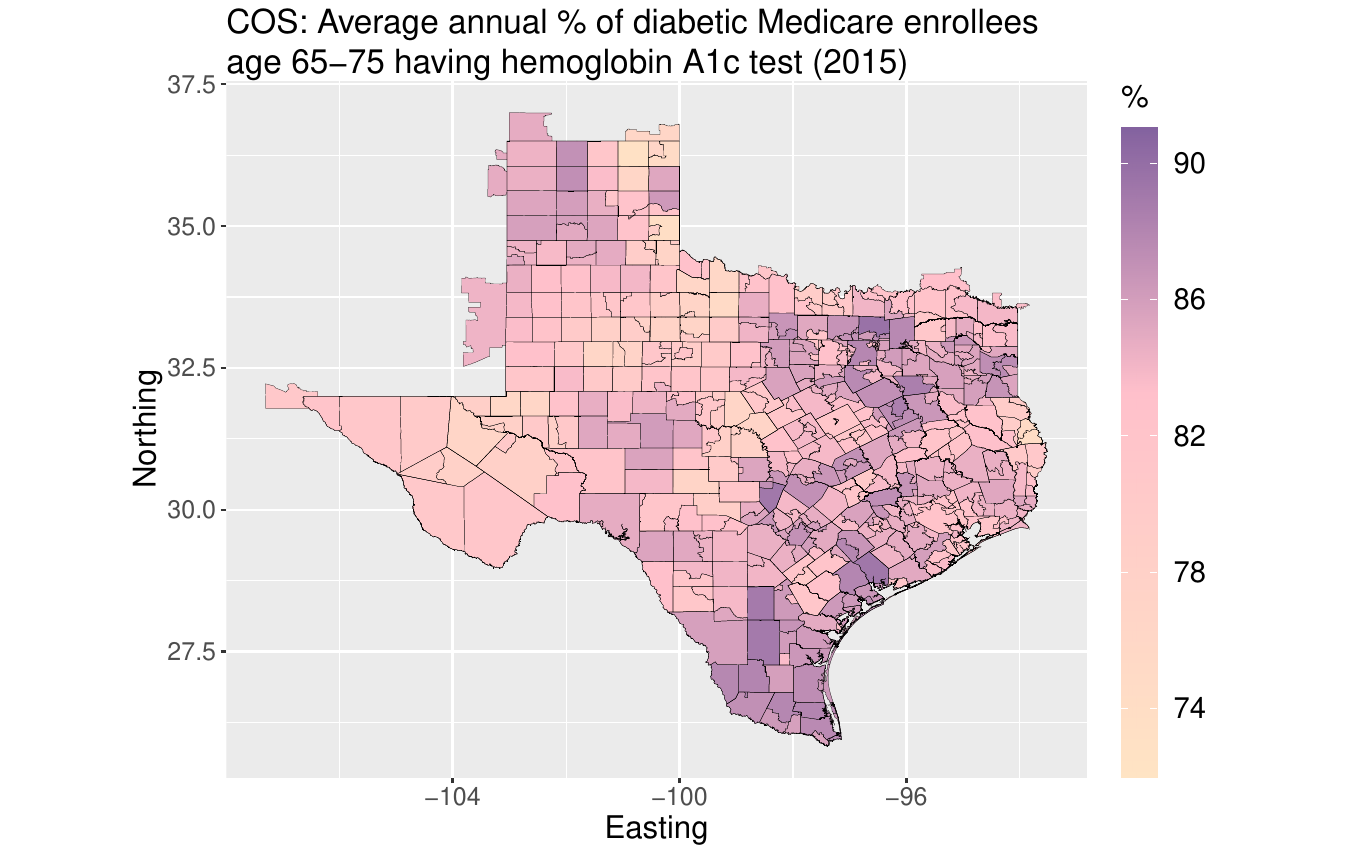} 
			\caption{MS-SRE} \label{fig:m1_A1c_partition_cos}
		\end{subfigure}
		\hspace{0.001cm}
		\begin{subfigure}[t]{0.45\textwidth}
			\centering
			\includegraphics[width=\linewidth]{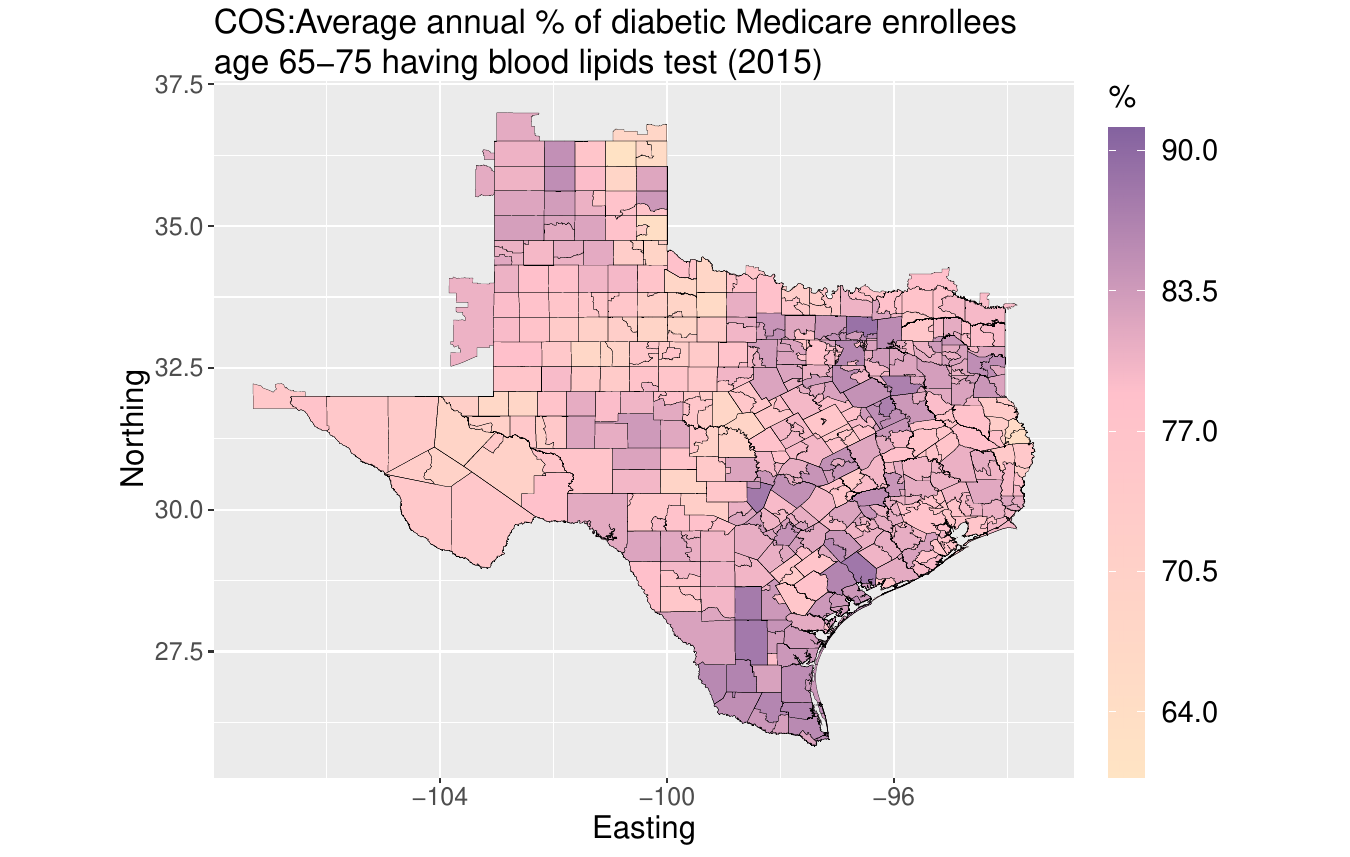} 
			\caption{MS-SRE} \label{fig:m1_lipid_partition_cos}
		\end{subfigure}
		
		\vspace{1cm}
		\begin{subfigure}[t]{0.45\textwidth}
			\centering
			\includegraphics[width=\linewidth]{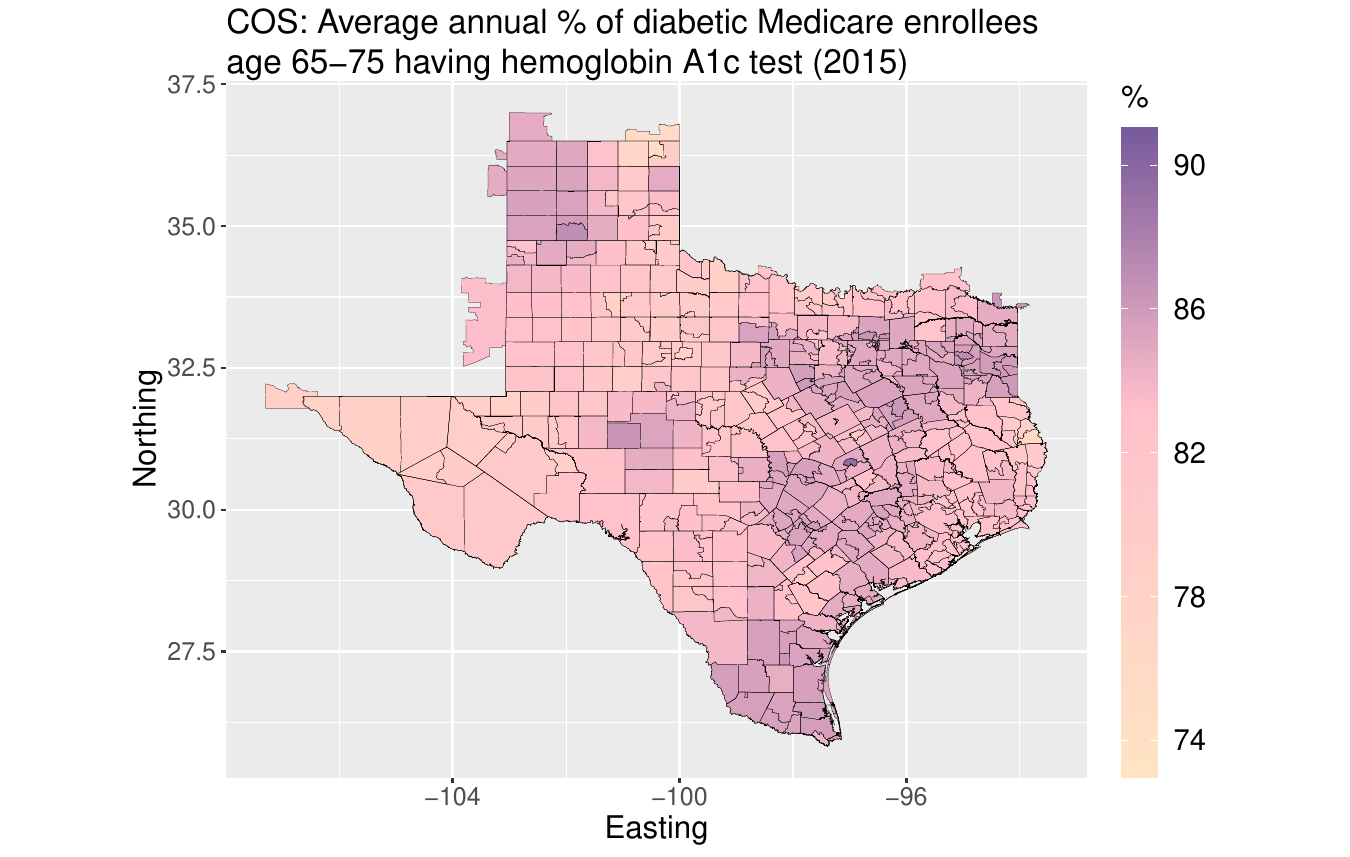} 
			\caption{MS-MCAR} \label{fig:m2_A1c_partition_cos}
		\end{subfigure}
		\hspace{0.001cm}
		\begin{subfigure}[t]{0.45\textwidth}
			\centering
			\includegraphics[width=\linewidth]{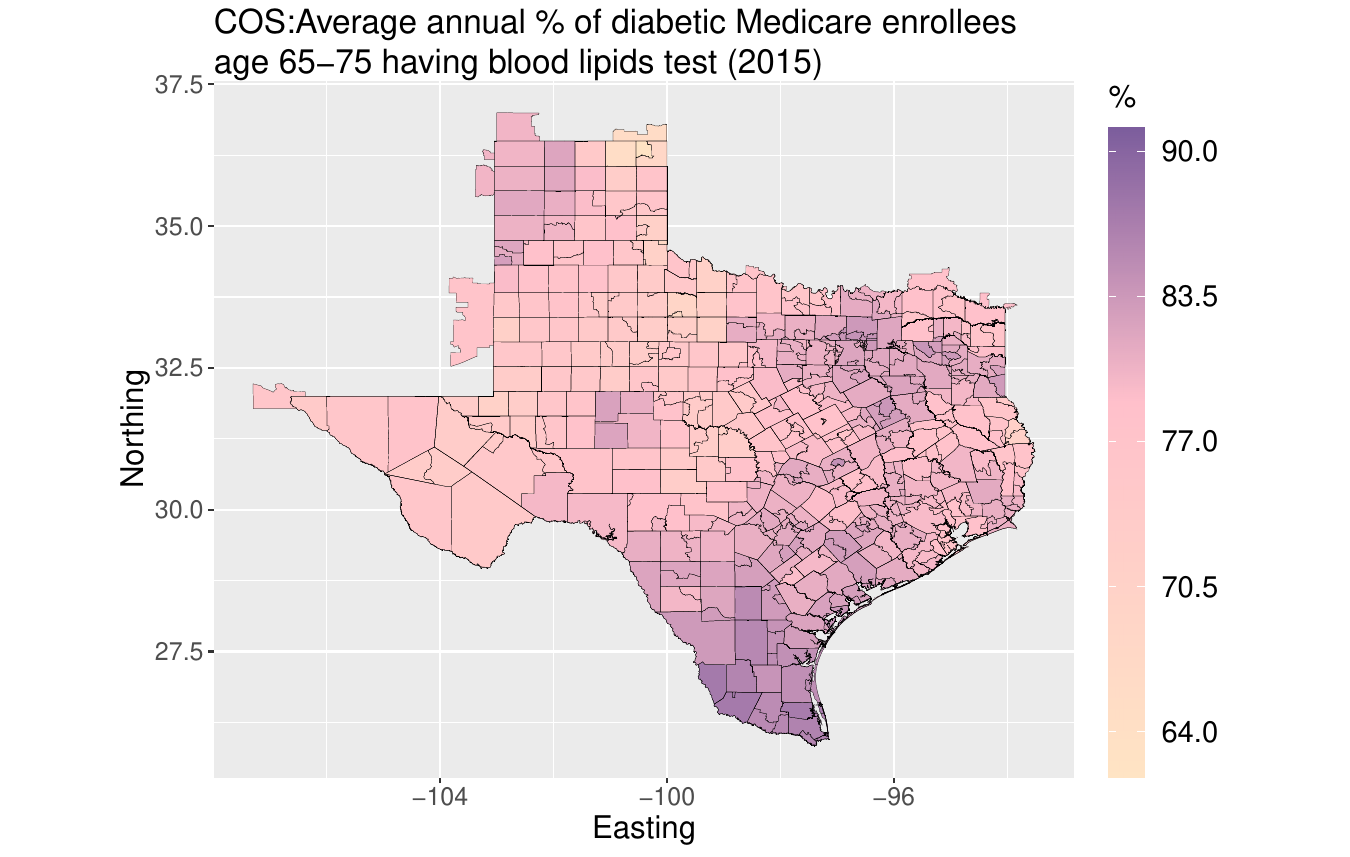} 
			\caption{MS-MCAR} \label{fig:m2_lipid_partition_cos}
		\end{subfigure}
		
		\vspace{1cm}
		\begin{subfigure}[t]{0.45\textwidth}
			\centering
			\includegraphics[width=\linewidth]{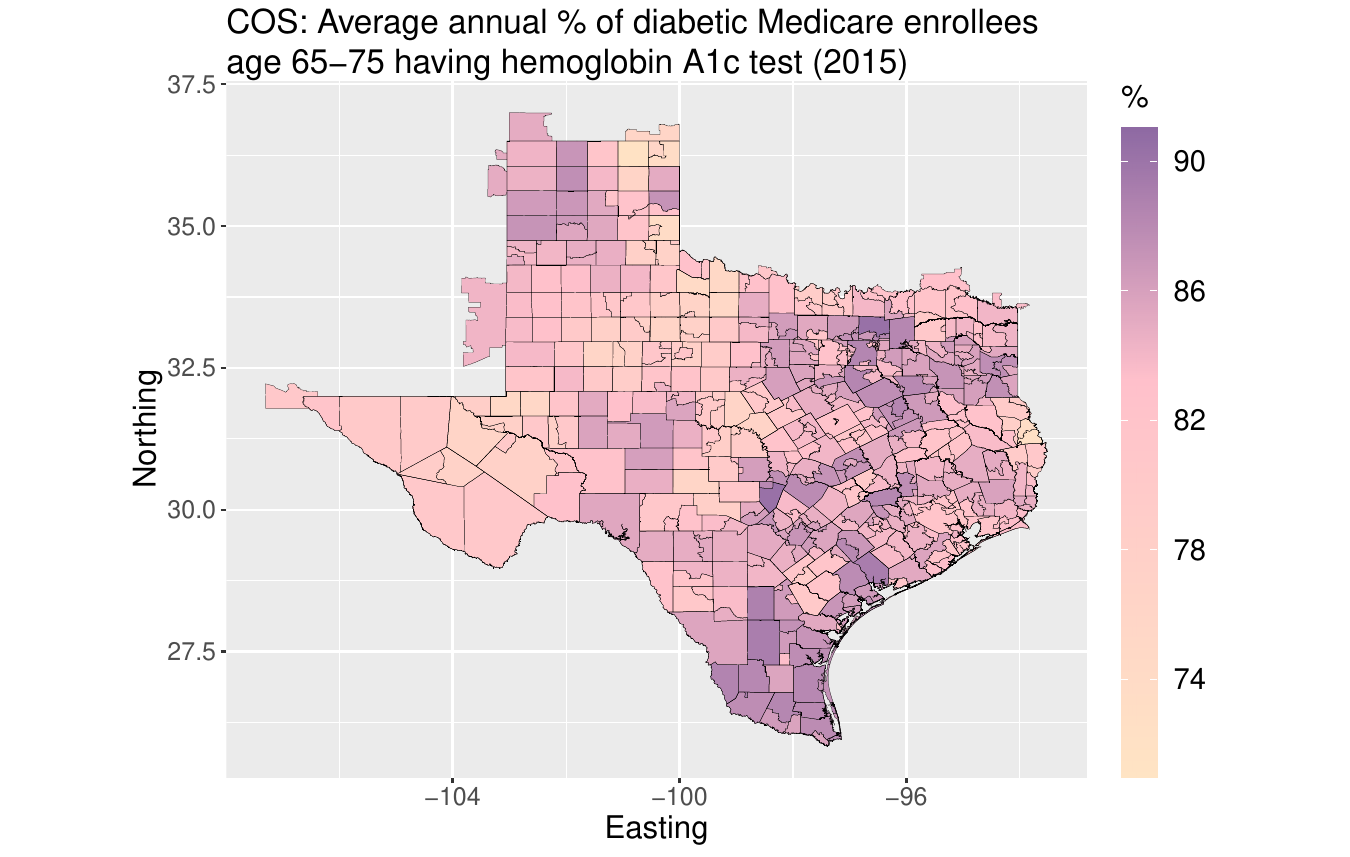} 
			\caption{MS-OH} \label{fig:m3_A1c_partition_cos}
		\end{subfigure}
		\hspace{0.001cm}
		\begin{subfigure}[t]{0.45\textwidth}
			\centering
			\includegraphics[width=\linewidth]{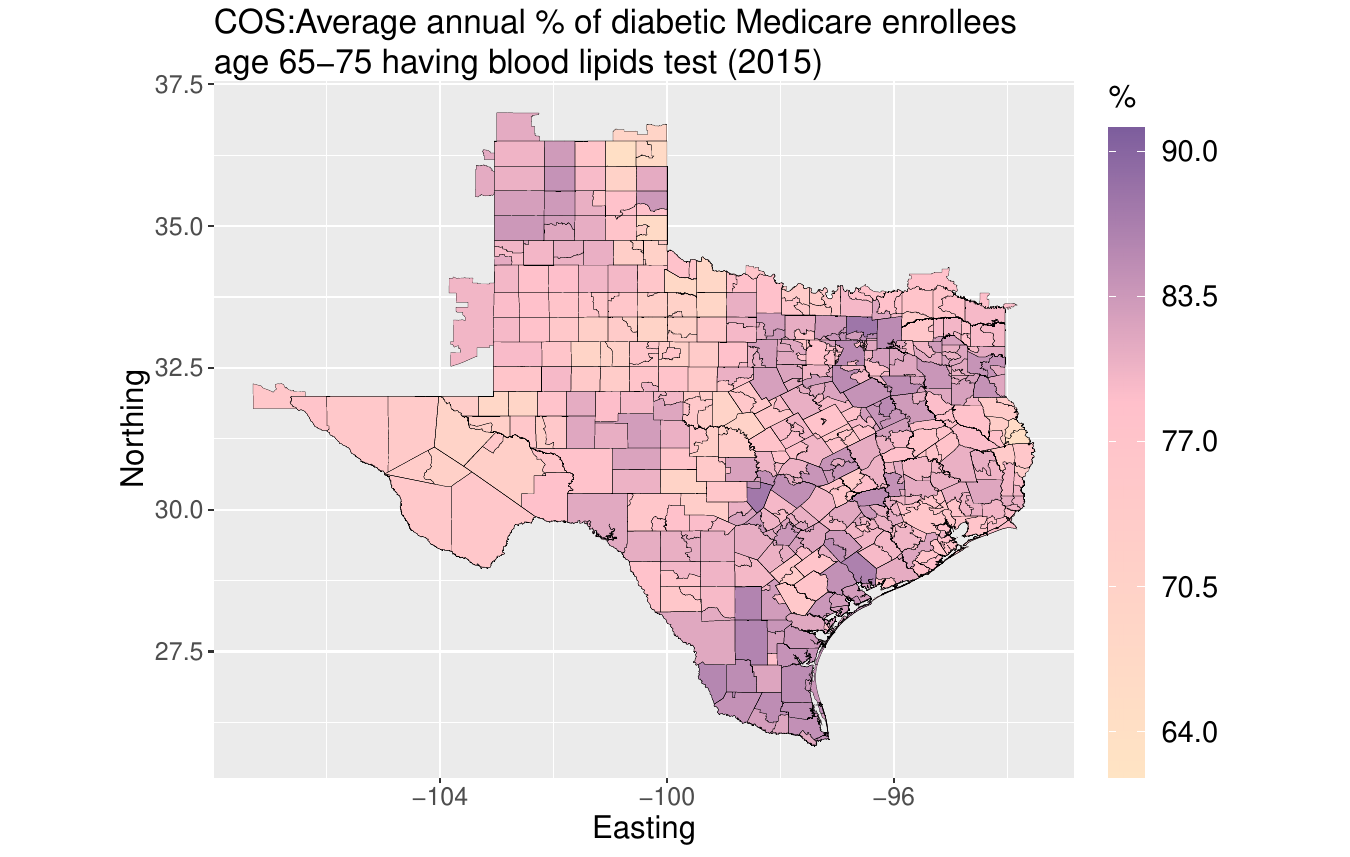} 
			\caption{MS-OH} \label{fig:m3_lipid_partition_cos}
		\end{subfigure}
		\caption{Prediction plots for Texas blood test monitoring data with captions indicating the model being fit. The first column gives predictions for average annual percentage of diabetic Medicare enrollees age 65-75 having hemoglobin A1c test. The second column gives predictions for average annual percentage of diabetic Medicare enrollees age 65-75 having blood lipids test. For both variables, the predictions from all three models are on the partition of HSA and counties of Texas.}
		\label{fig:analysispred}
	\end{figure}

\begin{figure}
		\centering
		\begin{subfigure}[t]{0.45\textwidth}
			\centering
			\includegraphics[width=\linewidth]{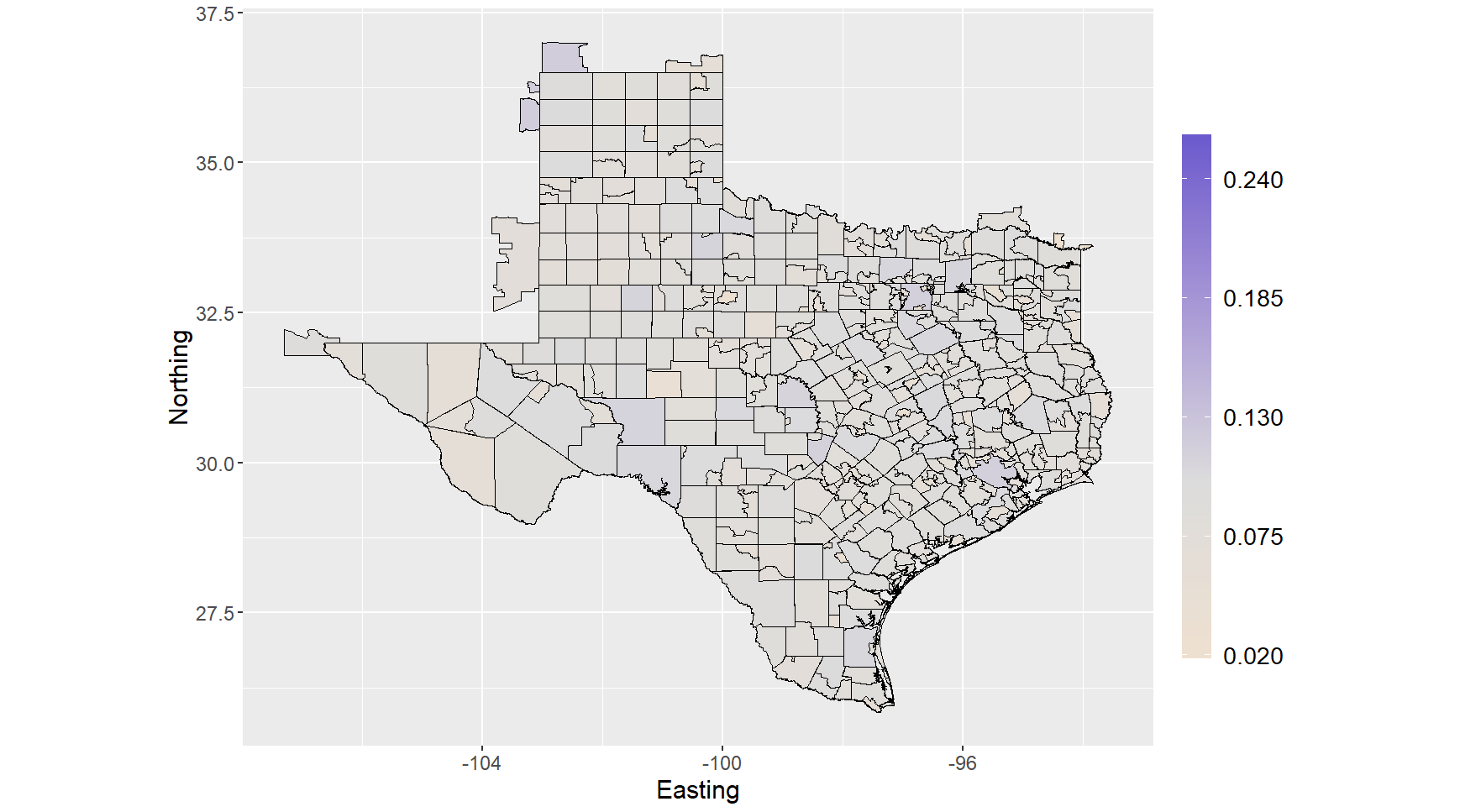} 
			\caption{MS-SRE} \label{fig:m1_y1_postsd}
		\end{subfigure}
		\hspace{0.001cm}
		\begin{subfigure}[t]{0.45\textwidth}
			\centering
			\includegraphics[width=\linewidth]{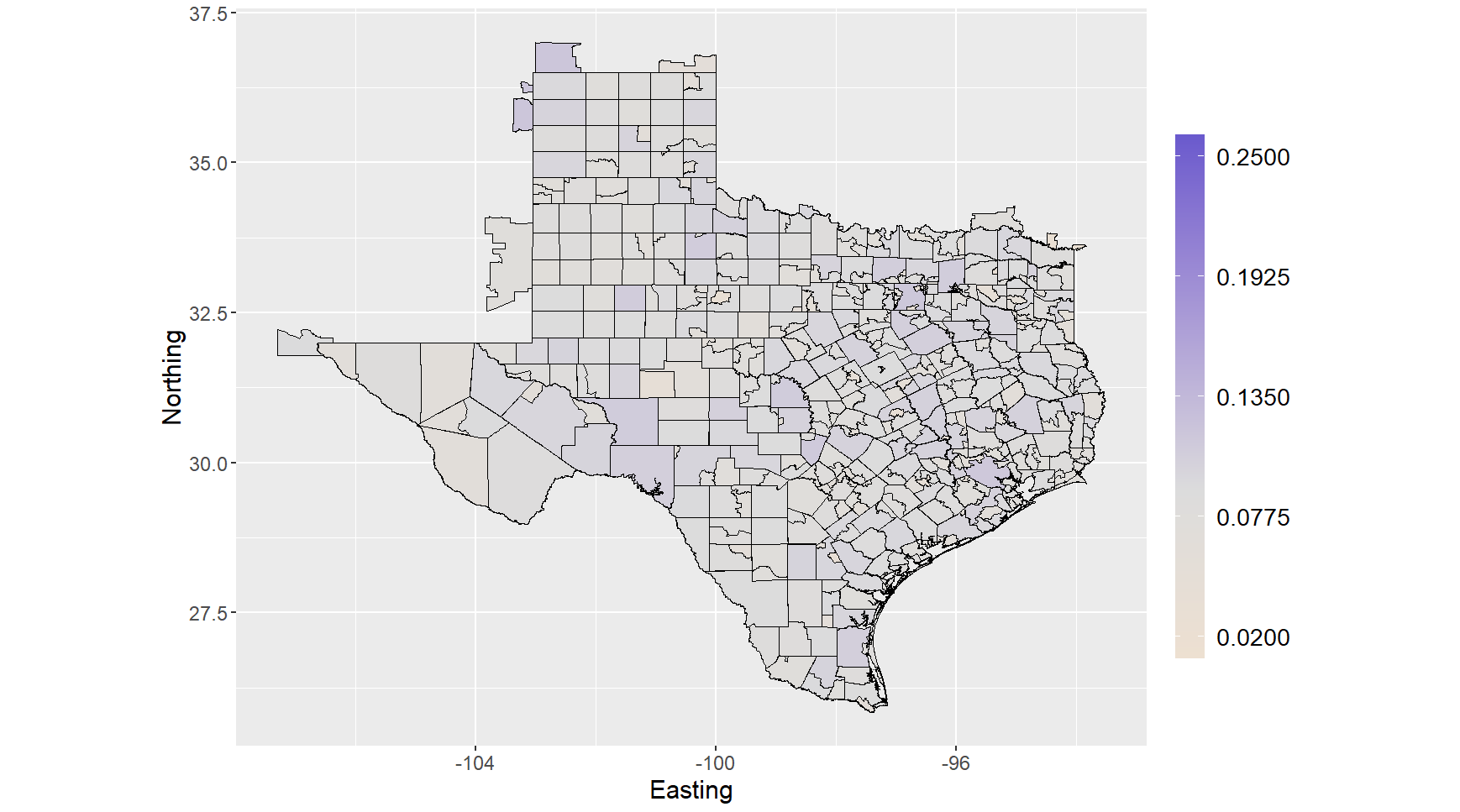} 
			\caption{MS-SRE} \label{fig:m1_y2_postsd}
		\end{subfigure}
		
		\vspace{1cm}
		\begin{subfigure}[t]{0.45\textwidth}
			\centering
			\includegraphics[width=\linewidth]{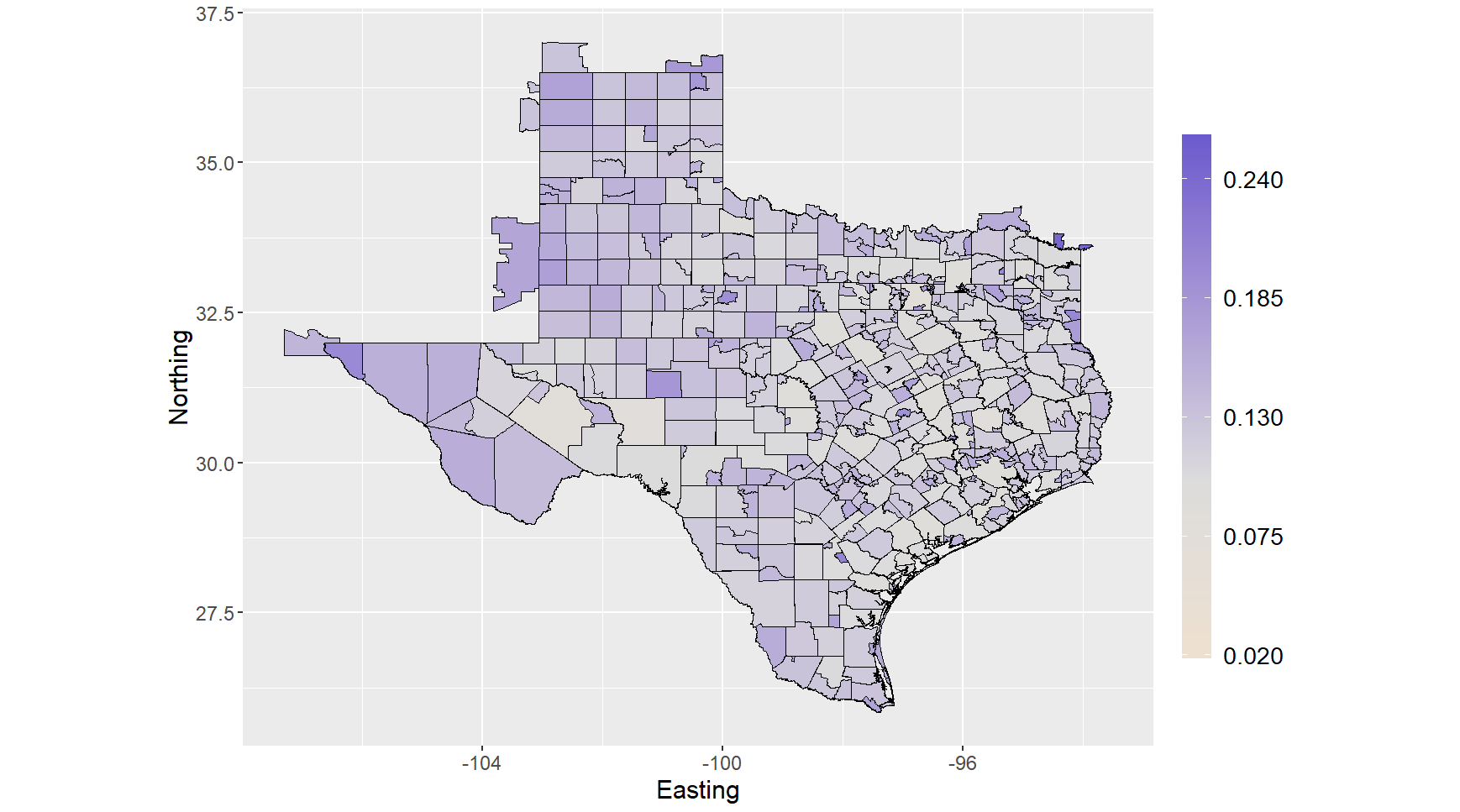} 
			\caption{MS-MCAR} \label{fig:m2_y1_postsd}
		\end{subfigure}
		\hspace{0.001cm}
		\begin{subfigure}[t]{0.45\textwidth}
			\centering
			\includegraphics[width=\linewidth]{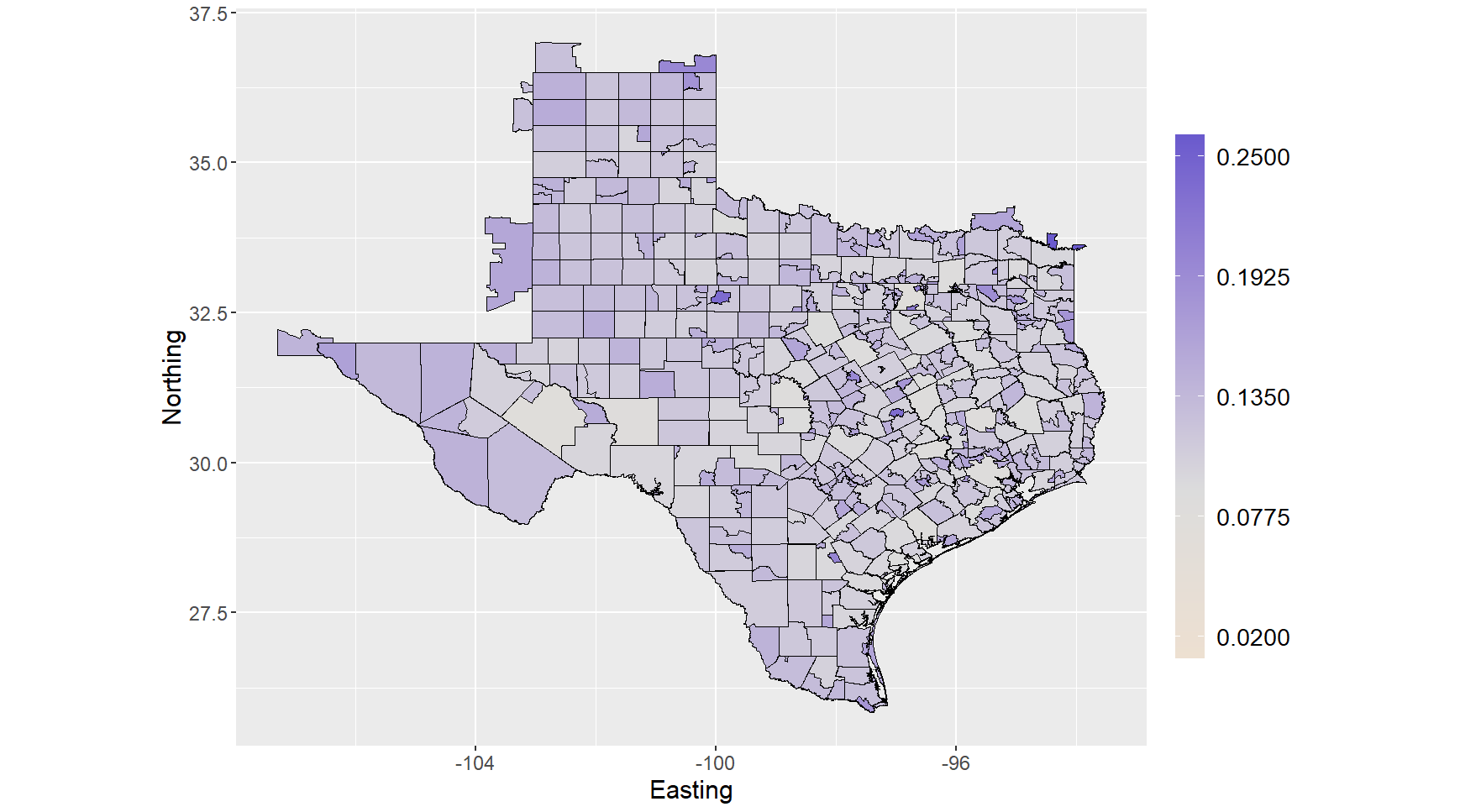} 
			\caption{MS-MCAR} \label{fig:m2_y2_postsd}
		\end{subfigure}
		
		\vspace{1cm}
		\begin{subfigure}[t]{0.45\textwidth}
			\centering
			\includegraphics[width=\linewidth]{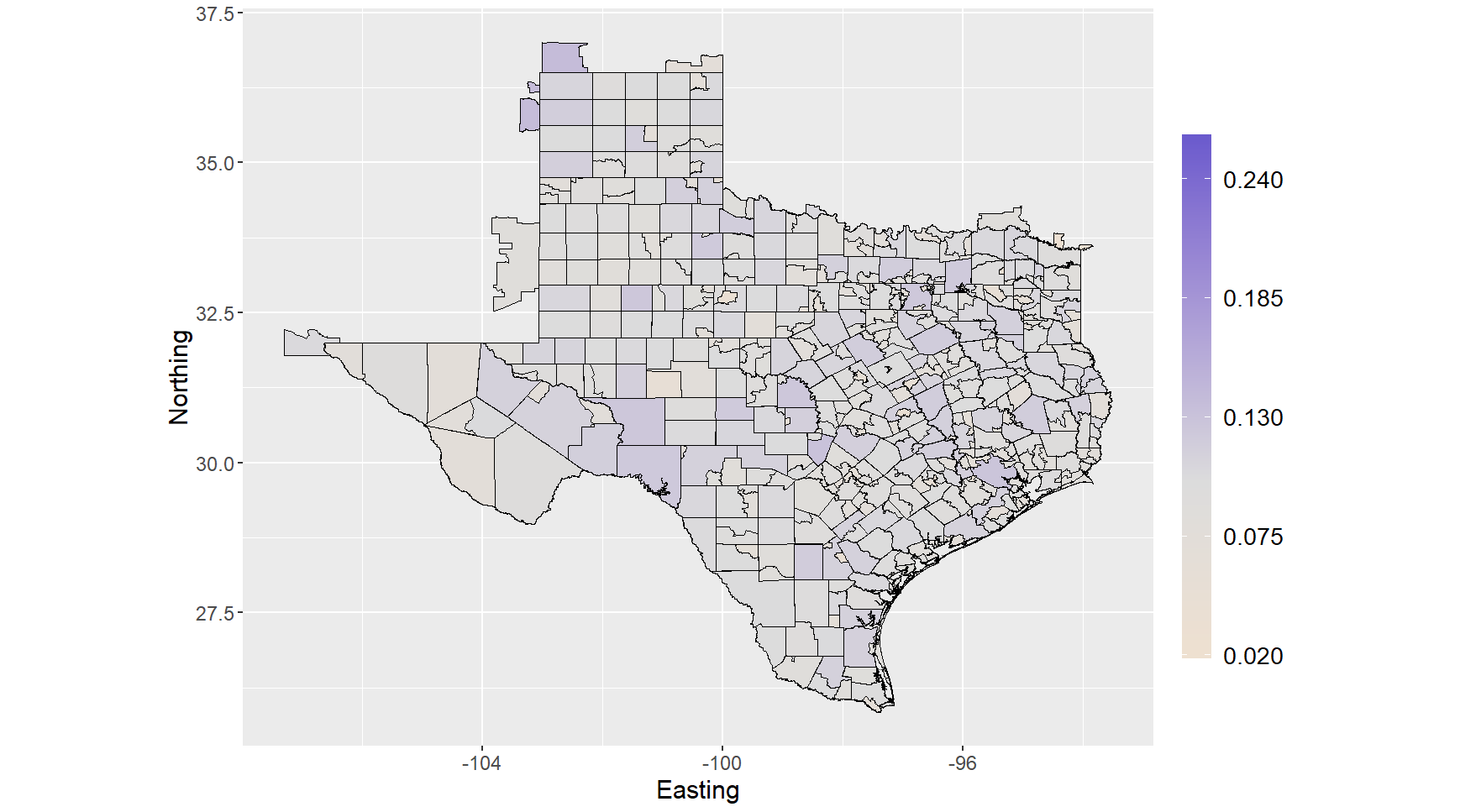} 
			\caption{MS-OH} \label{fig:m3_y1_postsd}
		\end{subfigure}
		\hspace{0.001cm}
		\begin{subfigure}[t]{0.45\textwidth}
			\centering
			\includegraphics[width=\linewidth]{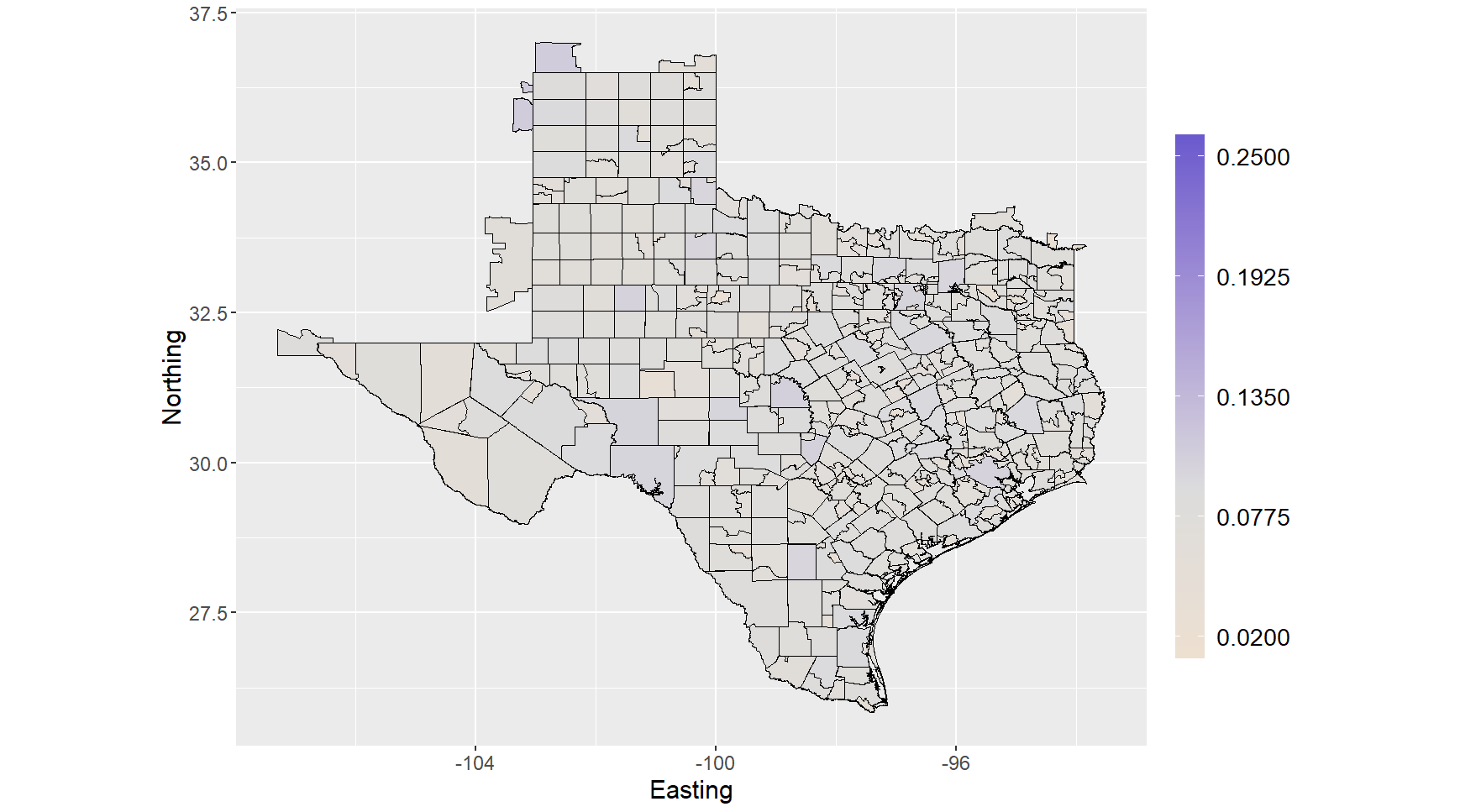} 
			\caption{MS-OH} \label{fig:m3_y2_postsd}
		\end{subfigure}
		\caption{Posterior standard deviations plots for Texas blood test monitoring data with captions indicating the model being fit. The first column gives posterior standard deviations for average annual percentage of diabetic Medicare enrollees age 65-75 having hemoglobin A1c test. The second column gives posterior standard deviations for average annual percentage of diabetic Medicare enrollees age 65-75 having blood lipids test. For both variables, the posterior standard deviations from all three models are on the partition of HSA and counties of Texas.}
		\label{fig:analysispostsd}
	\end{figure}

	We may obtain predictions of both variables at their observed scales, respectively. In such cases, the prediction of the second variable at the counties with missing data could be obtained through the posterior predictive replicates. Using MS-SRE as an example, we would sample from $Y_2(C_m)|\beta_2, \boldsymbol{\eta},\sigma^2_2\sim N(\beta_2+\boldsymbol{g}_2(C_m)'\boldsymbol{\eta},\sigma^2_2(\boldsymbol{P}_2\boldsymbol{P}_2')_{mm})$, where $C_m$ denotes the county with missing data and $(\boldsymbol{P}_2\boldsymbol{P}_2')_{mm}$ is the $m$-th diagonal element of the diagonal matrix $\boldsymbol{P}_2\boldsymbol{P}_2'$. We are particularly interested in performing COS and obtaining predictions at finer resolutions to avoid issues with the MAUP. We therefore give prediction plots with posterior standard deviations on the partition scale in Figure \ref{fig:analysispred} and \ref{fig:analysispostsd}, respectively. The predictions in all six plots are given on the 1,229 partitioned units, with the first column showing the prediction for the average annual percentage of diabetic Medicare enrollees aged 65-75 having hemoglobin A1c test and the second column showing the prediction for the average annual percentage of diabetic Medicare enrollees aged 65-75 having blood lipids test. The captions of the plots provide the name of the model from which the prediction is obtained. From the plot, all three models appear to perform similarly with similar spatial patterns in predictions. The posterior standard deviations are all relatively small (as compared to their predictions) while MS-MCAR has the largest posterior standard deviations.
 
 In general, the WAIC can only assess predictive performance on the observed scales. Consequently, the WAIC assesses the model fit on the regions $\{B_i\}$ for $Y_1$ and $\{C_j\}$ for $Y_2$. In general, the WAIC is equivalent to a leave-one-out cross-validation analysis in high dimensions \citep{VGG2017}, and hence, provides an out-of-sample assessment of the predictions on the observed scale. The WAIC reported in Table \ref{tab:waic}a shows that MS-SRE has the smallest WAIC for the A1c test and MS-MCAR has the smallest WAIC for the lipid test. Overall, MS-MCAR has the best prediction results considering the average of WAICs for both variables. We also compare the WAICs of bivariate COS with that of univariate COS. Note that the WAICs of univariate MS-OH are omitted in the table since MS-SRE and MS-OH are equivalent in the univariate case. For each of the three models, the bivariate WAIC for each variable is smaller than the corresponding univariate WAIC, as including the information of another variable would improve the prediction accuracy when there is a correlation between the two responses. The data suggests such correlation exists as MS-MCAR model's correlation parameter ($\tau$) has a (non-zero) posterior mean of 0.251 with 95\% credible interval $(0.156,0.327)$ that does not contain zero, which suggest a significant positive correlation.

 We also compare the continuous rank probability score \citep[CRPS, ][]{GR2007}, which assesses both prediction error and prediction uncertainty with smaller values to be preferable. Based on the results in Table \ref{tab:waic}b, MS-SRE and MS-OH appear to perform similarly well for both variables. MS-MCAR has the smallest CRPS for the A1c test, but becomes the least favorable for the lipid test. Similar to the WAIC, univariate CRPS are consistently larger than the corresponding bivariate CRPS for both variables and all models, suggesting the bivariate analysis is preferable to univariate analysis.
	
	\begin{table}
        \begin{subtable}[h]{1.0\textwidth}
		\centering
		\begin{tabular}{c@{\hskip 0.2in} c@{\hskip 0.2in} c@{\hskip 0.2in} c c@{\hskip 0.2in} c@{\hskip 0.2in}}
			\hline
			\hline
			& \multicolumn{2}{c}{Bivariate} &  & \multicolumn{2}{c}{Univariate}\\\cline{2-3} \cline{5-6}
			& $Y_1$ (A1c test) &  $Y_2$ (lipid test) &  & $Y_1$ (A1c test) &  $Y_2$ (lipid test)\\
			\hline
			MS-SRE & -225.343 & -269.240 &  & -166.164 & -215.343\\
			MS-MCAR & -217.195 & -299.990 &  & -201.288 & -281.975\\
			MS-OH & -224.192 & -266.550 &  &  & \\
			\hline
			\hline
		\end{tabular}
        \caption{WAIC}
        \end{subtable}

        \begin{subtable}[h]{1.0\textwidth}
		\centering
		\begin{tabular}{c@{\hskip 0.2in} c@{\hskip 0.2in} c@{\hskip 0.2in} c c@{\hskip 0.2in} c@{\hskip 0.2in}}
			\hline
			\hline
			& \multicolumn{2}{c}{Bivariate} &  & \multicolumn{2}{c}{Univariate}\\\cline{2-3} \cline{5-6}
			& $Y_1$ (A1c test) &  $Y_2$ (lipid test) &  & $Y_1$ (A1c test) &  $Y_2$ (lipid test)\\
			\hline
			MS-SRE & 0.066 & 0.062 &  & 0.100 & 0.093\\
			MS-MCAR & 0.057 & 0.087 &  & 0.087 & 0.094\\
			MS-OH & 0.063 & 0.064 &  &  & \\
			\hline
			\hline
		\end{tabular}
       \caption{CRPS}
        \end{subtable} 
		\caption{\label{tab:waic}Widely applicable information criterion (WAIC) and continuous rank probability score (CRPS) for MS-SRE, MS-MCAR, and MS-OH. In each subtable, the column ``bivariate" contains the WAIC/CRPS when fitting our bivariate multiscale model, and the column ``univariate" contains the WAIC/CRPS when fitting the univariate multiscale model.}
	\end{table}
	
	\begin{figure}
		\centering
		\begin{subfigure}[t]{0.30\textwidth}
			\centering
			\includegraphics[width=\linewidth]{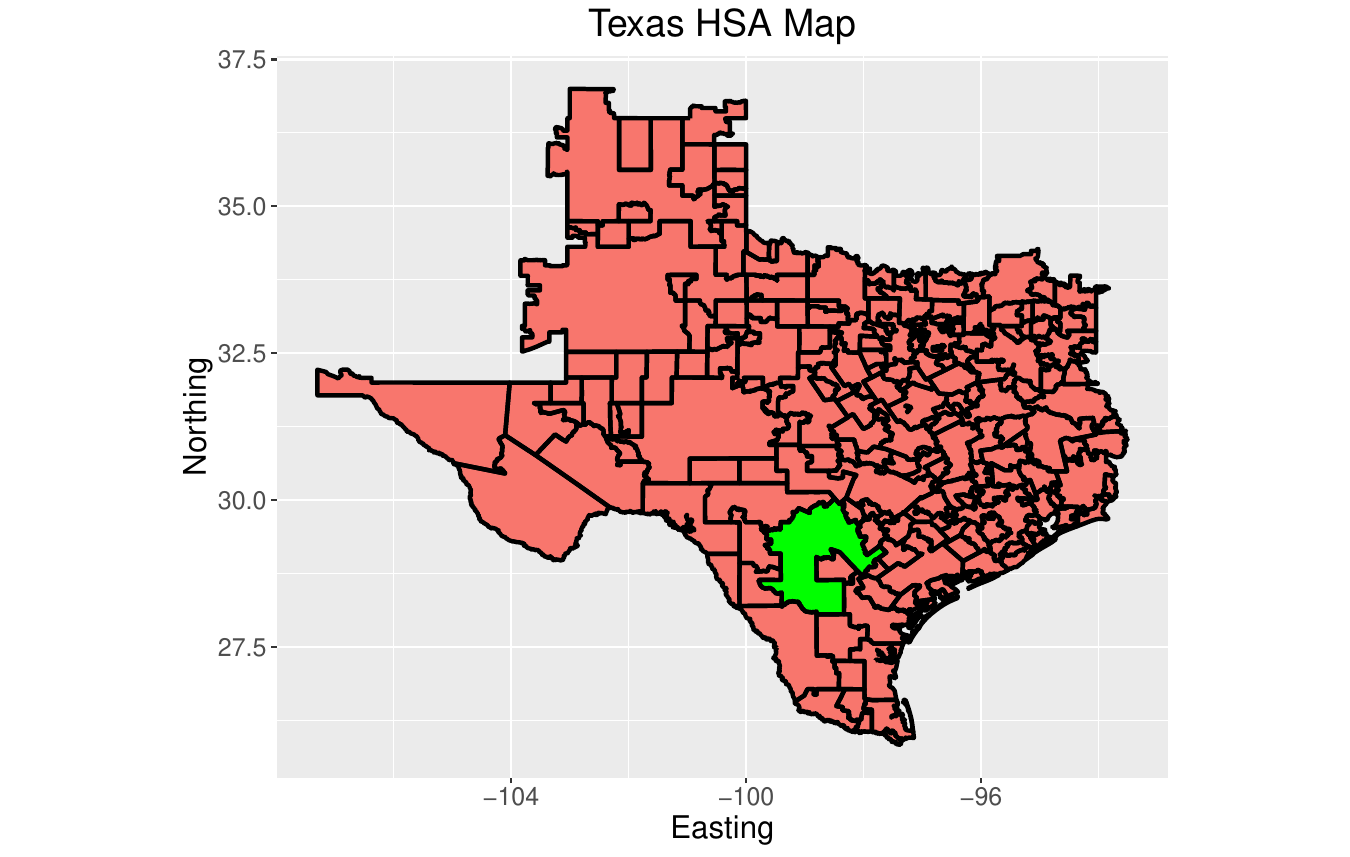} 
			\caption{}
			\label{hsamap}
		\end{subfigure}
		\hspace{0.001cm}
		\begin{subfigure}[t]{0.30\textwidth}
			\centering
			\includegraphics[width=\linewidth]{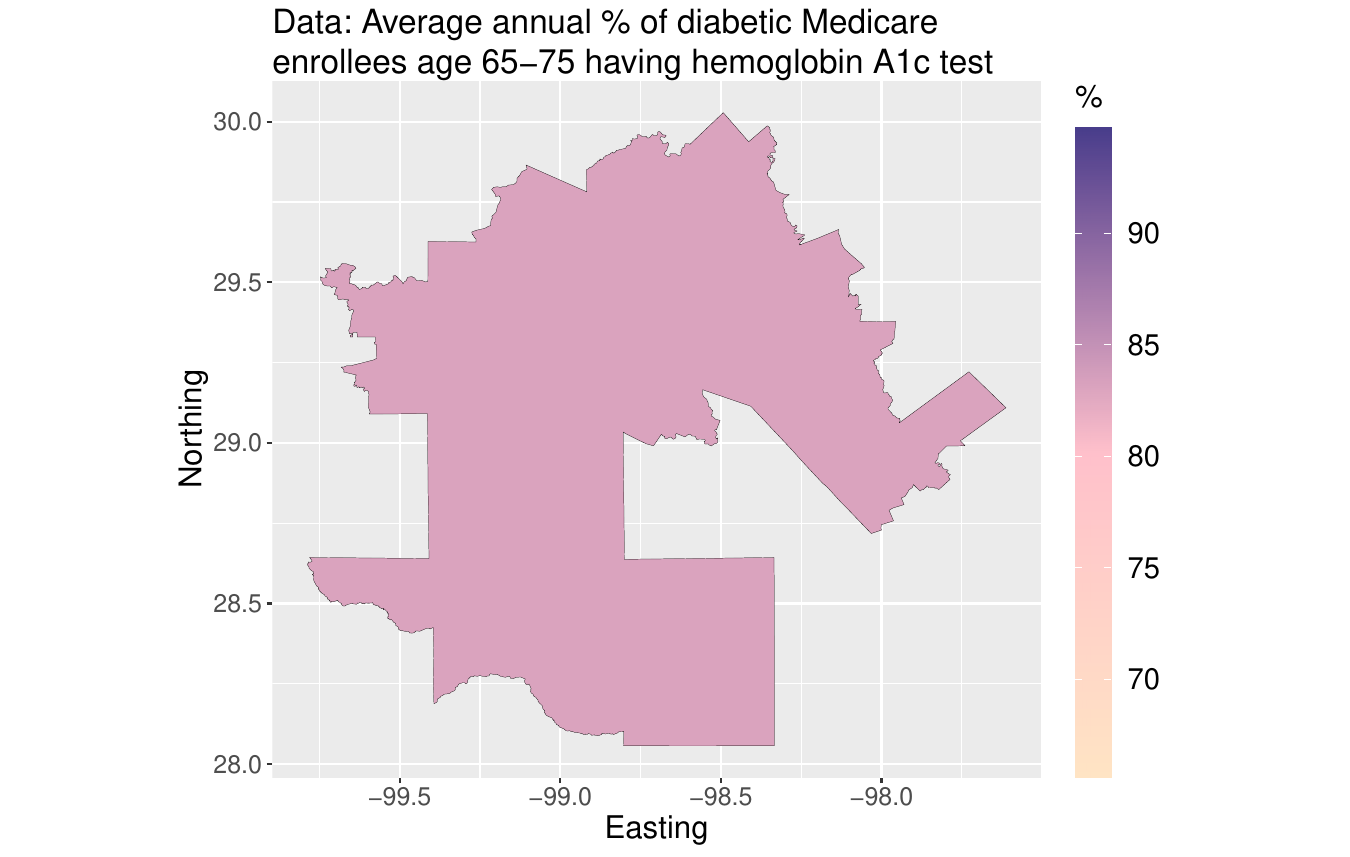} 
			\caption{}
			\label{Data_san antonio}
		\end{subfigure}
		\hspace{0.001cm}
		\begin{subfigure}[t]{0.30\textwidth}
			\centering
			\includegraphics[width=\linewidth]{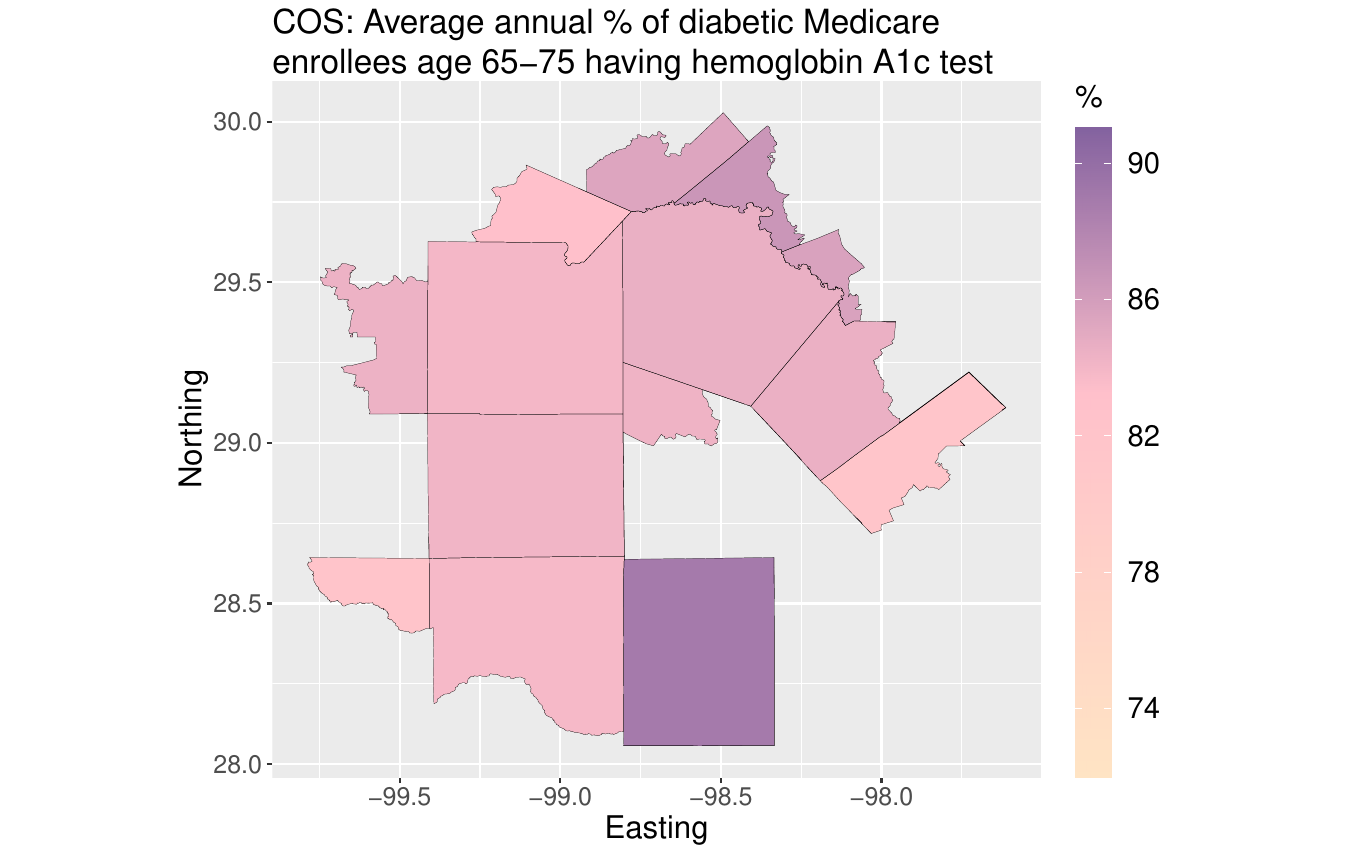} 
			\caption{}
			\label{COS_san antonio}
		\end{subfigure}
		\caption{Observed data and COS prediction from MS-SRE for one HSA (i.e., San Antonio) (a) A map that shows all Texas HSA. The green region is the HSA named San Antonio. (b) Average annual percentage of diabetic Medicare enrollees age 65-75 having hemoglobin A1c test --- observed on the HSA San Antonio. (c) Average annual percentage of diabetic Medicare enrollees age 65-75 having hemoglobin A1c test --- predicted on the partitioning units of San Antonio.}
		\label{fig:sanantonio}
	\end{figure}

	Our work is particularly important for those who are interested in making regulatory decisions (e.g., allocating funds for improvement in healthcare) based on hospital quality data such as blood tests. Our models allow us to obtain predictions at finer resolution, which can provide important insights that cannot be seen in the data observed on the original scale. For example, Figure \ref{fig:sanantonio} shows the difference between the data observed and the COS predictions obtained from MS-SRE for one HSA unit for the average annual percentage of diabetic Medicare enrollees aged 65-75 having A1c test. For illustration, we choose the HSA named San Antonio, whose location is highlighted in green in Figure \ref{hsamap}. Note that San Antonio is the name for this particular HSA, and hence, it has a border that is different from the border of the City of San Antonio, Texas (although two regions overlap). Figures \ref{Data_san antonio} and \ref{COS_san antonio} are segments of Figures \ref{A1c_hsa_obs} and \ref{fig:m1_A1c_partition_cos}, respectively. The light purple color in Figure \ref{Data_san antonio} suggests that San Antonio has a relatively high average annual percentage. However, from the finer resolution predictions in Figure \ref{COS_san antonio}, we see that only one partitioning unit has a high predicted value.

	\section{Discussion}
	
	In this paper, we have developed three novel bivariate multiscale spatial models through combining several existing bivariate spatial models with a multiscale approach. Specifically, we present MS-SRE, MS-MCAR, and MS-OH for the multiscale data observed from two misaligned areal supports. We run a comprehensive simulation study to compare the three new models where we simulate data from two different areal supports and aim to predict on the partition of the two areal supports. We compare the RMSE, coverage, and prediction plots, and find that MS-MCAR appears to have higher prediction errors and are robust under model misspecification while is computationally at a disadvantage compared to MS-SRE and MS-OH. 
	
	We include a motivating data analysis which shows the benefits of our bivariate multiscale approaches. Specifically, we look at the blood test monitoring data in Texas from the Dartmouth Atlas Project. We fit the MS-SRE, MS-MCAR, and MS-OH to jointly model two different blood tests that can be used as a hospital quality measure. Each blood test considered has a different scale, as one of them is observed on Texas hospital service areas and the other is observed on Texas counties. Through visually examining the prediction plots on the partition scale and comparing the WAICs, we reach the conclusion that MS-MCAR performs the best in terms of prediction accuracy. Reasonable predictions can also be made from MS-SRE and MS-OH, which are computationally much more efficient than MS-MCAR as shown in the simulation.
	
	Notice that in our illustrations the COS is performed by choosing a target support that is different from the two source supports. Based on that choice, our models allow for the COS for both variables. It is important to note that our models do not restrict the choice of target support to be the one we use. For example, we could choose to predict on one of the source supports, in which case COS is done for only one of the variables by the models. Our models allow for COS to any scale we are interested, which can be achieved through specifying a different partition matrix $\boldsymbol{P}$ based on the partition of the source supports and target support chosen. The flexibility of our models allows them to be widely applied in other bivariate multiscale spatial settings.
	
	There are still many avenues for future research. One possible extension of our work is to consider non-Gaussian data. Our work is based on the assumption that both responses follow a Gaussian distribution. However, this may not always be true as, for example, survey data that sometimes requires multiscale analysis tends to be Poisson distributed (\citealp{bradley16}; \citealp{MC98}; \citealp{MC99}). Moreover, our models are presented in spatial-only and bivariate settings. Extension of our models to consider spatio-temporal or higher dimensional multivariate settings may be needed. The implementation of our models depends heavily on the MCMC. The Metropolis-Hastings steps within the Gibbs sampler for updating certain parameters create difficulty as the tuning requires substantial amount of time and effort. Considering these computational difficulties when the sample size is large, more efficient methods should be considered.

	\section*{Acknowledgments}
	
	This research was partially supported by the U.S. National Science Foundation (NSF) under NSF grant 2310756.

	\bigskip
	\bigskip

	\clearpage

	\begin{appendices}

        \section{Unit-Level Interpretation of Spatial COS}
Let $\textbf{s}_{k}$ be the latitude/longitude of the $k$-th  healthcare agency in the state of Texas, and let the average percentage of diabetic Medicare enrollees age 65-75 at this $k$-th healthcare agency who has taken the $m$-blood test be denoted with $Y_{m}(\textbf{s}_{k})$. This implies that the average annual percentage of diabetic Medicare enrollees aged 65 $\--$ 75 that have taken the hemoglobin A1c blood test in the $i$-th HSA is given by
\begin{equation*}
	Y_{1}(B_{i}) = \frac{1}{N(B_{i})}\underset{\{k:\textbf{s}_{k}\in B_{i}\}}{\sum}Y_{1}(\textbf{s}_{k}),
\end{equation*}
where $N(B_{i})$ is the total population of health-care agencies within HSA $B_{i}$ and $Y_{1}(\textbf{s}_{k})$ is the average annual percentage of diabetic Medicare enrollees aged 65 $\--$ 75 that have taken the hemoglobin A1c blood test for healthcare agency $k$. Similarly, the average annual percentage of diabetic Medicare enrollees aged 65 $\--$ 75 that have taken the lipid blood test in the $j$-th county $C_{j}$ is given by
\begin{equation*}
	Y_{2}(C_{j}) = \frac{1}{N(C_{j})}\underset{\{k:\textbf{s}_{k}\in C_{j}\}}{\sum}Y_{2}(\textbf{s}_{k}),
\end{equation*}
where $N(C_{j})$ is the total population of health-care agencies within HSA $B_{i}$ and $Y_{2}(\textbf{s}_{k})$ is the average annual percentage of diabetic Medicare enrollees aged 65 $\--$ 75 that have taken the lipid blood test for healthcare agency $k$. Recall, we only observe the data on the HSA scale (i.e., $\{Y_{1}(B_{i})\}$) and the county scale (i.e., $\{Y_{2}(C_{j})\}$)

The healthcare locations $\{\textbf{s}_{k}\}$ are not made available by Dartmouth Atlas Study, and the unit-level data $Y(\textbf{s}_{k})$ is not publicly available. We make the simplifying assumption that $\{\textbf{s}_{k}\}$ are roughly uniformly distributed throughout the HSAs and the counties. This assumption is strong, however, without any additional information, it is unclear how one could assume a systematic non-uniform spread. We also assume that the number of health care agencies within HSAs and counties are fairly large. Under these two assumptions we have from the central limit theorem (applied to $\textbf{s}_{k}$) gives 
\begin{align}\nonumber
	Y_{1}(B_{i}) &= \frac{1}{N(B_{i})}\underset{\{k:\textbf{s}_{k}\in B_{i}\}}{\sum}Y_{1}(\textbf{s}_{k})\approx \frac{1}{|B_{i}|}\int_{B_{i}} Y_{1}(\textbf{s})d\textbf{s},\\
	\label{eq:cos}
	Y_{2}(C_{j}) &= \frac{1}{N(C_{j})}\underset{\{k:\textbf{s}_{k}\in C_{j}\}}{\sum}Y_{2}(\textbf{s}_{k})\approx \frac{1}{|C_{j}|}\int_{C_{j}} Y_{2}(\textbf{s})d\textbf{s},
\end{align}
\noindent
which is the traditional spatial COS formula \citep{WG04}. We make the simplifying assumption that $Y(\textbf{s})$ is piece-wise constant on the partitioning $\{A_{1},\ldots, A_{n_{3}}\}$. This assumption is strong, however, we only observe data on $\{Y(B_{i})\}$ and $\{Y(C_{j})\}$ and we cannot reliably extrapolate to lower resolutions beyond the partitioning of the observed scales $\{B_{i}\}$ and $\{C_{j}\}$. The piece-wise constant assumption implies that
\begin{equation}\label{eq:constant}
	Y_{m}(\textbf{s}) = \sum_{l=1}^{n_{3}}Y_{m}(A_{l})I(\textbf{s}\in A_{l}); \hspace{2pt} l = 1,\ldots, n_{3}, m = 1,2,
\end{equation}
\noindent
where recall $I(\cdot)$ is the indicator function so that for $\textbf{s}\in A_{l}$ we have $Y_{m}(\textbf{s}) = Y_{m}(A_{l})$. Substituting (\ref{eq:constant}) into (\ref{eq:cos}) gives
\begin{align}\nonumber
	Y_{1}(B_{i}) &\approx \frac{1}{|B_{i}|}\int_{B_{i}} \sum_{l = 1}^{n_{3}}Y_1(A_{l})I(\textbf{s}\in A_{l})d\textbf{s} \\
 \label{eqn:geoweight}
	&=  \sum_{l = 1}^{n_{3}}\frac{\int_{B_{i}\cap A_{l}}1 d\textbf{s}}{|B_{i}|} Y_{1}(A_{l}) = \sum_{l = 1}^{n_{3}}\frac{|A_{l}\cap B_{i}|}{|B_{i}|} Y_{1}(A_{l}) = \textbf{p}_{iB}^{\prime}\textbf{Y}_{A1},
\end{align}
\noindent
where $\textbf{p}_{iB} = \left(\frac{|A_{1}\cap B_{i}|}{|B_{i}|},\ldots, \frac{|A_{n_{3}}\cap B_{i}|}{|B_{i}|}\right)^{\prime}$ is the $i$-th row of $\textbf{P}_{1}$. Stacking (\ref{eqn:geoweight}) across $i$ implies that
\begin{equation}\label{eq:fine:to:course}
	\textbf{Y}_{1} = \textbf{P}_{1}\textbf{Y}_{A1},
\end{equation}
\noindent
Equation (\ref{eqn:singlescalematrix}) and (\ref{eq:fine:to:course}) produces the model for $\textbf{Y}_{1}$ in Equation (\ref{eqn:multiscalematrix}). In a similar manner we have that our assumptions on $Y_{m}(\textbf{s})$, $\{\textbf{s}_{k}\}$, and $N(\cdot)$ imply
\begin{equation*}
	\textbf{Y}_{2} = \textbf{P}_{2}\textbf{Y}_{A2},
\end{equation*}
\noindent
yielding the statistical model in (\ref{eqn:multiscalematrix}) for $\textbf{Y}_{2}$.

		\section{Full Conditional Distributions}
		In this appendix, we provide the derivation of full conditional distributions for all three models, namely, MS-SRE, MS-MCAR, and MS-OH.
		
		\subsection{Derivation of full conditional distributions for MS-SRE}
		Consider the MS-SRE model specified in Table \ref{tab:topdownmodels}, the full conditional distribution for $\boldsymbol{\eta}$ can be written as,
		\begin{align*}
			f(\boldsymbol{\eta}|\cdot) &\propto \prod_{i=1}^{n_1} f(Y_1(B_i)|\beta_1,\boldsymbol{\eta},\sigma_1^2)\cdot \prod_{j=1}^{n_2} f(Y_2(C_j)|\beta_2,\boldsymbol{\eta},\sigma_2^2)\cdot f(\boldsymbol{\eta}|\sigma_{\eta}^2,\phi) \\
			&\propto \exp\left(\sum_{i=1}^{n_1}-\frac{1}{2\sigma_1^2(\boldsymbol{P}_1\boldsymbol{P}_1')_{ii}}(Y_1(B_i)-\beta_1-\boldsymbol{g}_1(B_i)'\boldsymbol{\eta})^2\right.\\
			&\quad\quad\left.-\sum_{j=1}^{n_2}\frac{1}{2\sigma_2^2(\boldsymbol{P}_2\boldsymbol{P}_2')_{jj}}(Y_2(C_j)-\beta_2-\boldsymbol{g}_2(C_j)'\boldsymbol{\eta})^2-\frac{1}{2}\boldsymbol{\eta}'\boldsymbol{K}^{-1}\boldsymbol{\eta}\right)\\
			&\propto \exp\left(\sum_{i=1}^{n_1}-\frac{1}{2\sigma_1^2(\boldsymbol{P}_1\boldsymbol{P}_1')_{ii}}\boldsymbol{\eta}'\boldsymbol{g}_1(B_i)\boldsymbol{g}_1(B_i)'\boldsymbol{\eta}+\sum_{i=1}^{n_1}\frac{1}{\sigma_1^2(\boldsymbol{P}_1\boldsymbol{P}_1')_{ii}}\boldsymbol{g}_1(B_i)'\boldsymbol{\eta}(Y_1(B_i)-\beta_1)\right.\\
			&\quad \left. - \sum_{j=1}^{n_2}\frac{1}{2\sigma_2^2(\boldsymbol{P}_2\boldsymbol{P}_2')_{jj}}\boldsymbol{\eta}'\boldsymbol{g}_2(C_j)\boldsymbol{g}_2(C_j)'\boldsymbol{\eta}+\sum_{j=1}^{n_2}\frac{1}{\sigma_2^2(\boldsymbol{P}_2\boldsymbol{P}_2')_{jj}}\boldsymbol{g}_2(C_j)'\boldsymbol{\eta}(Y_2(C_j)-\beta_2)-\frac{1}{2}\boldsymbol{\eta}'\boldsymbol{K}^{-1}\boldsymbol{\eta}\right)\\
			&=\exp\left\{-\frac{1}{2}\left[\boldsymbol{\eta}'\left(\sum_{i=1}^{n_1}\frac{\boldsymbol{g}_1(B_i)\boldsymbol{g}_1(B_i)'}{\sigma_1^2(\boldsymbol{P}_1\boldsymbol{P}_1')_{ii}}+\sum_{j=1}^{n_2}\frac{\boldsymbol{g}_2(C_j)\boldsymbol{g}_2(C_j)'}{\sigma_2^2(\boldsymbol{P}_2\boldsymbol{P}_2')_{jj}}+\boldsymbol{K}^{-1}\right)\boldsymbol{\eta}\right.\right. \\
			& \quad \left.\left.-2\left(\sum_{i=1}^{n_1}\frac{\boldsymbol{g}_1(B_i)'(Y_1(B_{i})-\beta_1)}{\sigma_1^2(\boldsymbol{P}_1\boldsymbol{P}_1')_{ii}}+\sum_{j=1}^{n_2}\frac{\boldsymbol{g}_2(C_j)'(Y_2(C_{j})-\beta_2)}{\sigma_2^2(\boldsymbol{P}_2\boldsymbol{P}_2')_{jj}}\right)\boldsymbol{\eta}\right]\right\} \\
			&= \exp\left(-\frac{1}{2}(\boldsymbol{\eta}'\boldsymbol{B}^{-1}\boldsymbol{\eta}-2\boldsymbol{b}'\boldsymbol{\eta})\right)\\
			&\propto \exp\left(-\frac{1}{2}(\boldsymbol{\eta}'\boldsymbol{B}^{-1}\boldsymbol{\eta}-2\boldsymbol{b}'\boldsymbol{B}\boldsymbol{B}^{-1}\boldsymbol{\eta}+\boldsymbol{b}'\boldsymbol{B}\boldsymbol{b})\right)\\
			&= \exp\left(-\frac{1}{2}(\boldsymbol{\eta}-\boldsymbol{Bb})'\boldsymbol{B}^{-1}(\boldsymbol{\eta}-\boldsymbol{Bb})\right)\\
			&\propto MVN(\boldsymbol{Bb},\boldsymbol{B})
		\end{align*}
		where $\boldsymbol{B}^{-1}=\sum_{i=1}^{n_1}\frac{\boldsymbol{g}_1(B_i)\boldsymbol{g}_1(B_i)'}{\sigma_1^2(\boldsymbol{P}_1\boldsymbol{P}_1')_{ii}}+\sum_{j=1}^{n_2}\frac{\boldsymbol{g}_2(C_j)\boldsymbol{g}_2(C_j)'}{\sigma_2^2(\boldsymbol{P}_2\boldsymbol{P}_2')_{jj}}+\boldsymbol{K}^{-1}$ and $\boldsymbol{b}'=\sum_{i=1}^{n_1}\frac{\boldsymbol{g}_1(B_i)'(Y_1(B_{i})-\beta_1)}{\sigma_1^2(\boldsymbol{P}_1\boldsymbol{P}_1')_{ii}}+\sum_{j=1}^{n_2}\frac{\boldsymbol{g}_2(C_j)'(Y_2(C_{j})-\beta_2)}{\sigma_2^2(\boldsymbol{P}_2\boldsymbol{P}_2')_{jj}}$. Similarly, the full conditional distributions for $\beta_1,\beta_2,\sigma^2_{\eta},\sigma^2_1,\sigma^2_2$, and $\phi$ can be derived as follows,
		\begin{align*}
			f(\beta_1|\cdot) &\propto \prod_{i=1}^{n_1}f(Y_1(B_i)|\beta_1,\boldsymbol{\eta},\sigma_1^2)\cdot f(\beta_1)\\
			&\propto \exp\left(\sum_{i=1}^{n_1}-\frac{1}{2\sigma_1^2(\boldsymbol{P}_1\boldsymbol{P}_1')_{ii}}(Y_1(B_i)-\beta_1-\boldsymbol{g}_1(B_i)'\boldsymbol{\eta})^2-\frac{1}{2\sigma^2_{\beta}}\beta_1^2\right)\\
			&\propto \exp\left[\beta_1^2\left(-\sum_{i=1}^{n_1}\frac{1}{2\sigma^2_1(\boldsymbol{P}_1\boldsymbol{P}_1')_{ii}}-\frac{1}{2\sigma^2_{\beta}}\right)-\beta_1\left(-\sum_{i=1}^{n_1}\frac{Y_1(B_i)-\boldsymbol{g}_1(B_i)'\boldsymbol{\eta}}{\sigma^2_1(\boldsymbol{P}_1\boldsymbol{P}_1')_{ii}}\right)\right]\\
			&= \exp(a\beta_1^2-b\beta_1) \\
			&= \exp\left(\frac{\beta_1^2-\frac{b}{a}\beta_1}{1/a}\right)\\
			&= \exp\left(\frac{\beta_1^2-2\frac{b}{2a}\beta_1+(\frac{b}{2a})^2-(\frac{b}{2a})^2}{1/a}\right)\\
			&\propto \exp\left(\frac{(\beta_1-\frac{b}{2a})^2}{1/a}\right) \\
			&= \exp\left(\frac{(\beta_1-\frac{b}{2a})^2}{-2(-\frac{1}{2a})}\right)\\
			&\propto N\left(\frac{b}{2a},-\frac{1}{2a}\right) \quad \textrm{where} \,\,a=-\sum_{i=1}^{n_1}\frac{1}{2\sigma^2_1(\boldsymbol{P}_1\boldsymbol{P}_1')_{ii}}-\frac{1}{2\sigma^2_{\beta}}\,\,\textrm{and}\,\, b=-\sum_{i=1}^{n_1}\frac{Y_1(B_i)-\boldsymbol{g}_1(B_i)'\boldsymbol{\eta}}{\sigma^2_1(\boldsymbol{P}_1\boldsymbol{P}_1')_{ii}}\\
			f(\beta_2|\cdot) &\propto \prod_{j=1}^{n_2}f(Y_2(C_j)|\beta_2,\boldsymbol{\eta},\sigma_2^2)\cdot f(\beta_2)\\
			&\propto \exp\left(\sum_{j=1}^{n_2}-\frac{1}{2\sigma_2^2(\boldsymbol{P}_2\boldsymbol{P}_2')_{jj}}(Y_2(C_j)-\beta_2-\boldsymbol{g}_2(C_j)'\boldsymbol{\eta})^2-\frac{1}{2\sigma^2_{\beta}}\beta_2^2\right)\\
			&\propto \exp\left[\beta_2^2\left(-\sum_{j=1}^{n_2}\frac{1}{2\sigma^2_2(\boldsymbol{P}_2\boldsymbol{P}_2')_{jj}}-\frac{1}{2\sigma^2_{\beta}}\right)-\beta_2\left(-\sum_{j=1}^{n_2}\frac{Y_2(C_j)-\boldsymbol{g}_2(C_j)'\boldsymbol{\eta}}{\sigma^2_2(\boldsymbol{P}_2\boldsymbol{P}_2')_{jj}}\right)\right]\\
			&= \exp(a\beta_2^2-b\beta_2) \\
			&= \exp\left(\frac{\beta_2^2-\frac{b}{a}\beta_2}{1/a}\right)\\
			&= \exp\left(\frac{\beta_2^2-2\frac{b}{2a}\beta_2+(\frac{b}{2a})^2-(\frac{b}{2a})^2}{1/a}\right)\\
			&\propto \exp\left(\frac{(\beta_2-\frac{b}{2a})^2}{1/a}\right) \\
			&= \exp\left(\frac{(\beta_2-\frac{b}{2a})^2}{-2(-\frac{1}{2a})}\right)\\
			&\propto N\left(\frac{b}{2a},-\frac{1}{2a}\right) \quad \textrm{where} \,\,a=-\sum_{j=1}^{n_2}\frac{1}{2\sigma^2_2(\boldsymbol{P}_2\boldsymbol{P}_2')_{jj}}-\frac{1}{2\sigma^2_{\beta}}\,\,\textrm{and}\,\, b=-\sum_{j=1}^{n_2}\frac{Y_2(C_j)-\boldsymbol{g}_2(C_j)'\boldsymbol{\eta}}{\sigma^2_2(\boldsymbol{P}_2\boldsymbol{P}_2')_{jj}}\\
			f(\sigma^2_{\eta}|\cdot) &\propto f(\boldsymbol{\eta}|\sigma^2_{\eta},\phi)\cdot f(\sigma^2_{\eta})\\
			&\propto (\sigma^2_{\eta})^{-a_{\eta}-1}\exp\left(-\frac{b_{\eta}}{\sigma^2_{\eta}}\right)\cdot|\{\sigma^2_{\eta}\exp(-\phi\|\boldsymbol{c}_i-\boldsymbol{c}_j\|)\}|^{-1/2}\exp\left(-\frac{1}{2}\boldsymbol{\eta}'\{\sigma^2_{\eta}\exp(-\phi\|\boldsymbol{c}_i-\boldsymbol{c}_j\|)\}^{-1}\boldsymbol{\eta}\right)\\
			&\propto (\sigma^2_{\eta})^{-a_{\eta}-1}\exp\left(-\frac{b_{\eta}}{\sigma^2_{\eta}}\right)(\sigma^2_{\eta})^{-r/2}\exp\left(-\frac{1}{2\sigma^2_{\eta}}\boldsymbol{\eta}'\{\exp(-\phi\|\boldsymbol{c}_i-\boldsymbol{c}_j\|)\}^{-1}\boldsymbol{\eta}\right)\\
			&= (\sigma^2_{\eta})^{-(a_{\eta}+\frac{r}{2})-1}\exp\left(-\frac{b_{\eta}+\frac{1}{2}\boldsymbol{\eta}'\{\exp(-\phi\|\boldsymbol{c}_i-\boldsymbol{c}_j\|)\}^{-1}\boldsymbol{\eta}}{\sigma^2_{\eta}}\right)\\
			&\propto IG\left(a_{\eta}+\frac{r}{2},b_{\eta}+\frac{1}{2}\boldsymbol{\eta}'\{\exp(-\phi\|\boldsymbol{c}_i-\boldsymbol{c}_j\|)\}^{-1}\boldsymbol{\eta}\right)\\
			f(\sigma^2_1|\cdot) &\propto \prod_{i=1}^{n_1}f(Y_1(B_i)|\beta_1,\boldsymbol{\eta},\sigma_1^2)\cdot f(\sigma^2_1)\\
			&\propto \prod_{i=1}^{n_1} \left(\frac{1}{\sqrt{2\pi\sigma^2_1(\boldsymbol{P}_1\boldsymbol{P}_1')_{ii}}}\exp\left(-\frac{1}{2\sigma^2_1(\boldsymbol{P}_1\boldsymbol{P}_1')_{ii}}(Y_1(B_i)-\beta_1-\boldsymbol{g}_1(B_i)'\boldsymbol{\eta})^2\right)\right)\\
			&\quad\quad\cdot (\sigma^2_1)^{-a_{\sigma}-1}\exp\left(-\frac{b_{\sigma}}{\sigma^2_1}\right)\\
			&\propto (\sigma^2_1)^{-(a_{\sigma}+\frac{n_1}{2})-1}\exp\left(-\frac{b_{\sigma}+\sum_{i=1}^{n_1}\frac{1}{2(\boldsymbol{P}_1\boldsymbol{P}_1')_{ii}}(Y_1(B_i)-\beta_1-\boldsymbol{g}_1(B_i)'\boldsymbol{\eta})^2}{\sigma^2_1}\right)\\
			&\propto IG\left(a_{\sigma}+\frac{n_1}{2},b_{\sigma}+\sum_{i=1}^{n_1}\frac{1}{2(\boldsymbol{P}_1\boldsymbol{P}_1')_{ii}}(Y_1(B_i)-\beta_1-\boldsymbol{g}_1(B_i)'\boldsymbol{\eta})^2\right)\\
			f(\sigma^2_2|\cdot) &\propto \prod_{j=1}^{n_2}f(Y_2(C_j)|\beta_2,\boldsymbol{\eta},\sigma_2^2)\cdot f(\sigma^2_2)\\
			&\propto \prod_{j=1}^{n_2} \left(\frac{1}{\sqrt{2\pi\sigma^2_2(\boldsymbol{P}_2\boldsymbol{P}_2')_{jj}}}\exp\left(-\frac{1}{2\sigma^2_2(\boldsymbol{P}_2\boldsymbol{P}_2')_{jj}}(Y_2(C_j)-\beta_2-\boldsymbol{g}_2(C_j)'\boldsymbol{\eta})^2\right)\right) \\
			&\quad\quad \cdot(\sigma^2_2)^{-a_{\sigma}-1}\exp\left(-\frac{b_{\sigma}}{\sigma^2_2}\right)\\
			&\propto (\sigma^2_2)^{-(a_{\sigma}+\frac{n_2}{2})-1}\exp\left(-\frac{b_{\sigma}+\sum_{j=1}^{n_2}\frac{1}{2(\boldsymbol{P}_2\boldsymbol{P}_2')_{jj}}(Y_2(C_j)-\beta_2-\boldsymbol{g}_2(C_j)'\boldsymbol{\eta})^2}{\sigma^2_2}\right)\\
			&\propto IG\left(a_{\sigma}+\frac{n_2}{2},b_{\sigma}+\sum_{j=1}^{n_2}\frac{1}{2(\boldsymbol{P}_2\boldsymbol{P}_2')_{jj}}(Y_2(C_j)-\beta_2-\boldsymbol{g}_2(C_j)'\boldsymbol{\eta})^2\right)\\
			f(\phi|\cdot) &\propto f(\boldsymbol{\eta}|\sigma^2_{\eta},\phi)\cdot f(\phi)\\
			&\propto |\{\sigma^2_{\eta}\exp(-\phi\|\boldsymbol{c}_i-\boldsymbol{c}_j\|)\}|^{-1/2}\exp\left(-\frac{1}{2}\boldsymbol{\eta}'\{\sigma^2_{\eta}\exp(-\phi\|\boldsymbol{c}_i-\boldsymbol{c}_j\|)\}^{-1}\boldsymbol{\eta}\right)\cdot \frac{1}{b_{\phi}-a_{\phi}}
		\end{align*}
		Note that the full conditional distribution for $\phi$ does not have a closed form. Therefore, the update of $\phi$ requires a Metropolis-Hastings step within the Gibbs sampler.

		\subsection{Derivation of full conditional distributions for MS-MCAR}
		We first derive the full conditional distribution for $\boldsymbol{\psi}$ in the MS-MCAR model specified in Table \ref{tab:topdownmodels},
		\begin{align*}
			f(\boldsymbol{\psi}|\cdot) &\propto \prod_{i=1}^{n_1}f(Y_1(B_i)|\beta_1,\boldsymbol{\psi},\sigma^2_1)\cdot\prod_{j=1}^{n_2}f(Y_2(C_j)|\beta_2,\boldsymbol{\psi},\sigma^2_2)\cdot f(\boldsymbol{\psi}|\rho,\tau,\nu^2) \\
			&\propto \exp\left(\sum_{i=1}^{n_1}-\frac{1}{2\sigma^2_1(\boldsymbol{P}_1\boldsymbol{P}_1')_{ii}}(Y_1(B_i)-\beta_1-\boldsymbol{p}(B_i)'\boldsymbol{\psi})^2\right.\\
			&\quad\quad\quad\left.-\sum_{j=1}^{n_2}\frac{1}{2\sigma^2_2(\boldsymbol{P}_2\boldsymbol{P}_2')_{jj}}(Y_2(C_j)-\beta_2-\boldsymbol{p}(C_j)'\boldsymbol{\psi})^2-\frac{1}{2}\boldsymbol{\psi}'(\boldsymbol{\Sigma}\otimes (\boldsymbol{D}-\rho\boldsymbol{W})^{-1})^{-1}\boldsymbol{\psi}\right) \\
			&\propto\exp\left(-\frac{1}{2}\left[\boldsymbol{\psi}'\left(\sum_{i=1}^{n_1}\frac{\boldsymbol{p}(B_i)\boldsymbol{p}(B_i)'}{\sigma^2_1(\boldsymbol{P}_1\boldsymbol{P}_1')_{ii}}+\sum_{j=1}^{n_2}\frac{\boldsymbol{p}(C_j)\boldsymbol{p}(C_j)'}{\sigma^2_2(\boldsymbol{P}_2\boldsymbol{P}_2')_{jj}}+(\boldsymbol{\Sigma}\otimes (\boldsymbol{D}-\rho\boldsymbol{W})^{-1})^{-1}\right)\boldsymbol{\psi}\right.\right.\\
			&\quad\quad\quad\left.\left.-2\left(\sum_{i=1}^{n_1}\frac{\boldsymbol{p}(B_i)'(Y_1(B_i)-\beta_1)}{\sigma^2_1(\boldsymbol{P}_1\boldsymbol{P}_1')_{ii}}+\sum_{j=1}^{n_2}\frac{\boldsymbol{p}(C_j)'(Y_2(C_j)-\beta_2)}{\sigma^2_2(\boldsymbol{P}_2\boldsymbol{P}_2')_{jj}}\right)\boldsymbol{\psi}\right]\right)\\
			&=\exp\left(-\frac{1}{2}(\boldsymbol{\psi}'\boldsymbol{B}^{-1}\boldsymbol{\psi}-2\boldsymbol{b}'\boldsymbol{\psi})\right)\\
			&\propto \exp\left(-\frac{1}{2}(\boldsymbol{\psi}'\boldsymbol{B}^{-1}\boldsymbol{\psi}-2\boldsymbol{b}'\boldsymbol{B}\boldsymbol{B}^{-1}\boldsymbol{\psi}+\boldsymbol{b}'\boldsymbol{B}\boldsymbol{b})\right)\\
			&= \exp\left(-\frac{1}{2}(\boldsymbol{\psi}-\boldsymbol{Bb})'\boldsymbol{B}^{-1}(\boldsymbol{\psi}-\boldsymbol{Bb})\right)\\
			&\propto MVN(\boldsymbol{Bb},\boldsymbol{B})
		\end{align*}
		where $\boldsymbol{B}^{-1}=\sum_{i=1}^{n_1}\frac{\boldsymbol{p}(B_i)\boldsymbol{p}(B_i)'}{\sigma^2_1(\boldsymbol{P}_1\boldsymbol{P}_1')_{ii}}+\sum_{j=1}^{n_2}\frac{\boldsymbol{p}(C_j)\boldsymbol{p}(C_j)'}{\sigma^2_2(\boldsymbol{P}_2\boldsymbol{P}_2')_{jj}}+(\boldsymbol{\Sigma}\otimes (\boldsymbol{D}-\rho\boldsymbol{W})^{-1})^{-1}$ and $\boldsymbol{b}'=\sum_{i=1}^{n_1}\frac{\boldsymbol{p}(B_i)'(Y_1(B_i)-\beta_1)}{\sigma^2_1(\boldsymbol{P}_1\boldsymbol{P}_1')_{ii}}+\sum_{j=1}^{n_2}\frac{\boldsymbol{p}(C_j)'(Y_2(C_j)-\beta_2)}{\sigma^2_2(\boldsymbol{P}_2\boldsymbol{P}_2')_{jj}}$. The full conditional distributions for $\beta_1,\beta_2,\rho,\tau,\nu^2,\sigma^2_1$ and $\sigma^2_2$ can be derived similarly as follows,
		\begin{align*}
			f(\beta_1|\cdot)&\propto \prod_{i=1}^{n_1}f(Y_1(B_i)|\beta_1,\boldsymbol{\psi},\sigma^2_1)\cdot f(\beta_1)\\
			&\propto \exp\left(-\sum_{i=1}^{n_1}\frac{1}{2\sigma^2_1(\boldsymbol{P}_1\boldsymbol{P}_1')_{ii}}(Y_1(B_i)-\beta_1-\boldsymbol{p}(B_i)'\boldsymbol{\psi})^2-\frac{1}{2\sigma^2_{\beta}}\beta_1^2\right)\\
			&\propto \exp\left(\beta_1^2\left(-\sum_{i=1}^{n_1}\frac{1}{2\sigma^2_1(\boldsymbol{P}_1\boldsymbol{P}_1')_{ii}}-\frac{1}{2\sigma^2_{\beta}}\right)-\beta_1\left(-\sum_{i=1}^{n_1}\frac{Y_1(B_i)-\boldsymbol{p}(B_i)'\boldsymbol{\psi}}{\sigma^2_1(\boldsymbol{P}_1\boldsymbol{P}_1')_{ii}}\right)\right)\\
			&= \exp(a\beta_1^2-b\beta_1) \\
			&= \exp\left(\frac{\beta_1^2-\frac{b}{a}\beta_1}{1/a}\right)\\
			&= \exp\left(\frac{\beta_1^2-2\frac{b}{2a}\beta_1+(\frac{b}{2a})^2-(\frac{b}{2a})^2}{1/a}\right)\\
			&\propto \exp\left(\frac{(\beta_1-\frac{b}{2a})^2}{1/a}\right) \\
			&= \exp\left(\frac{(\beta_1-\frac{b}{2a})^2}{-2(-\frac{1}{2a})}\right)\\
			&\propto N\left(\frac{b}{2a},-\frac{1}{2a}\right) \quad \textrm{where} \,\,a=-\sum_{i=1}^{n_1}\frac{1}{2\sigma^2_1(\boldsymbol{P}_1\boldsymbol{P}_1')_{ii}}-\frac{1}{2\sigma^2_{\beta}}\,\,\textrm{and}\,\, b=-\sum_{i=1}^{n_1}\frac{Y_1(B_i)-\boldsymbol{p}(B_i)'\boldsymbol{\psi}}{\sigma^2_1(\boldsymbol{P}_1\boldsymbol{P}_1')_{ii}}\\
			f(\beta_2|\cdot)&\propto \prod_{j=1}^{n_2}f(Y_2(C_j)|\beta_2,\boldsymbol{\psi},\sigma^2_2)\cdot f(\beta_2)\\
			&\propto \exp\left(-\sum_{j=1}^{n_2}\frac{1}{2\sigma^2_2(\boldsymbol{P}_2\boldsymbol{P}_2')_{jj}}(Y_2(C_j)-\beta_2-\boldsymbol{p}(C_j)'\boldsymbol{\psi})^2-\frac{1}{2\sigma^2_{\beta}}\beta_2^2\right)\\
			&\propto \exp\left(\beta_2^2\left(-\sum_{j=1}^{n_2}\frac{1}{2\sigma^2_2(\boldsymbol{P}_2\boldsymbol{P}_2')_{jj}}-\frac{1}{2\sigma^2_{\beta}}\right)-\beta_2\left(-\sum_{j=1}^{n_2}\frac{Y_2(C_j)-\boldsymbol{p}(C_j)'\boldsymbol{\psi}}{\sigma^2_2(\boldsymbol{P}_2\boldsymbol{P}_2')_{jj}}\right)\right)\\
			&= \exp(a\beta_2^2-b\beta_2) \\
			&= \exp\left(\frac{\beta_2^2-\frac{b}{a}\beta_2}{1/a}\right)\\
			&= \exp\left(\frac{\beta_2^2-2\frac{b}{2a}\beta_2+(\frac{b}{2a})^2-(\frac{b}{2a})^2}{1/a}\right)\\
			&\propto \exp\left(\frac{(\beta_2-\frac{b}{2a})^2}{1/a}\right) \\
			&= \exp\left(\frac{(\beta_2-\frac{b}{2a})^2}{-2(-\frac{1}{2a})}\right)\\
			&\propto N\left(\frac{b}{2a},-\frac{1}{2a}\right) \quad \textrm{where} \,\,a=-\sum_{j=1}^{n_2}\frac{1}{2\sigma^2_2(\boldsymbol{P}_2\boldsymbol{P}_2')_{jj}}-\frac{1}{2\sigma^2_{\beta}}\,\,\textrm{and}\,\, b=-\sum_{j=1}^{n_2}\frac{Y_2(C_j)-\boldsymbol{p}(C_j)'\boldsymbol{\psi}}{\sigma^2_2(\boldsymbol{P}_2\boldsymbol{P}_2')_{jj}}\\
			f(\rho|\cdot) & \propto f(\boldsymbol{\psi}|\rho,\tau,\nu^2) \cdot f(\rho)\\
			&\propto |\boldsymbol{\Sigma}\otimes (\boldsymbol{D}-\rho\boldsymbol{W})^{-1}|^{-1/2}\exp\left(-\frac{1}{2}\boldsymbol{\psi}'[\boldsymbol{\Sigma}\otimes (\boldsymbol{D}-\rho\boldsymbol{W})^{-1}]^{-1}\boldsymbol{\psi}\right)\cdot \frac{1}{b_{\rho}-a_{\rho}}\\
			f(\tau|\cdot) & \propto f(\boldsymbol{\psi}|\rho,\tau,\nu^2) \cdot f(\tau)\\
			&\propto |\boldsymbol{\Sigma}\otimes (\boldsymbol{D}-\rho\boldsymbol{W})^{-1}|^{-1/2}\exp\left(-\frac{1}{2}\boldsymbol{\psi}'[\boldsymbol{\Sigma}\otimes (\boldsymbol{D}-\rho\boldsymbol{W})^{-1}]^{-1}\boldsymbol{\psi}\right)\cdot \frac{1}{b_{\tau}-a_{\tau}}\\
			f(\nu^2|\cdot) & \propto f(\boldsymbol{\psi}|\rho,\tau,\nu^2) \cdot f(\nu^2)\\
			&\propto |\boldsymbol{\Sigma}\otimes (\boldsymbol{D}-\rho\boldsymbol{W})^{-1}|^{-1/2}\exp\left(-\frac{1}{2}\boldsymbol{\psi}'[\boldsymbol{\Sigma}\otimes (\boldsymbol{D}-\rho\boldsymbol{W})^{-1}]^{-1}\boldsymbol{\psi}\right)\cdot (\nu^2)^{-a_{\nu}-1}\exp\left(-\frac{b_{\nu}}{\nu^2}\right)\\
			&\propto (\nu^2)^{-2n_3/2}\exp\left(-\frac{1}{2\nu^2}\boldsymbol{\psi}'[ \boldsymbol{T}(\tau)^{-1}\otimes (\boldsymbol{D}-\rho\boldsymbol{W})]\boldsymbol{\psi}\right)\cdot (\nu^2)^{-a_{\nu}-1}\exp\left(-\frac{b_{\nu}}{\nu^2}\right)\\
			&\propto (\nu^2)^{-n_3-a_{\nu}-1}\exp\left(-\frac{b_{\nu}+\frac{1}{2}\boldsymbol{\psi}'[\boldsymbol{T}(\tau)^{-1}\otimes (\boldsymbol{D}-\rho\boldsymbol{W})]\boldsymbol{\psi}}{\nu^2}\right)\\
			&\propto IG\left(a_{\nu}+n_3,b_{\nu}+\frac{1}{2}\boldsymbol{\psi}'[ \boldsymbol{T}(\tau)^{-1}\otimes (\boldsymbol{D}-\rho\boldsymbol{W})]\boldsymbol{\psi}\right)\\
			f(\sigma^2_1|\cdot) &\propto \prod_{i=1}^{n_1}f(Y_1(B_i)|\beta_1,\boldsymbol{\psi},\sigma_1^2)\cdot f(\sigma^2_1)\\
			&\propto \prod_{i=1}^{n_1} \left(\frac{1}{\sqrt{2\pi\sigma^2_1(\boldsymbol{P}_1\boldsymbol{P}_1')_{ii}}}\exp\left(-\frac{1}{2\sigma^2_1(\boldsymbol{P}_1\boldsymbol{P}_1')_{ii}}(Y_1(B_i)-\beta_1-\boldsymbol{p}(B_i)'\boldsymbol{\psi})^2\right)\right)\\
			&\quad\quad\cdot (\sigma^2_1)^{-a_{\sigma}-1}\exp\left(-\frac{b_{\sigma}}{\sigma^2_1}\right)\\
			&\propto (\sigma^2_1)^{-(a_{\sigma}+\frac{n_1}{2})-1}\exp\left(-\frac{b_{\sigma}+\sum_{i=1}^{n_1}\frac{1}{2(\boldsymbol{P}_1\boldsymbol{P}_1')_{ii}}(Y_1(B_i)-\beta_1-\boldsymbol{p}(B_i)'\boldsymbol{\psi})^2}{\sigma^2_1}\right)\\
			&\propto IG\left(a_{\sigma}+\frac{n_1}{2},b_{\sigma}+\sum_{i=1}^{n_1}\frac{1}{2(\boldsymbol{P}_1\boldsymbol{P}_1')_{ii}}(Y_1(B_i)-\beta_1-\boldsymbol{p}(B_i)'\boldsymbol{\eta})^2\right)\\
			f(\sigma^2_2|\cdot) &\propto \prod_{j=1}^{n_2}f(Y_2(C_j)|\beta_2,\boldsymbol{\psi},\sigma_2^2)\cdot f(\sigma^2_2)\\
			&\propto \prod_{j=1}^{n_2} \left(\frac{1}{\sqrt{2\pi\sigma^2_2(\boldsymbol{P}_2\boldsymbol{P}_2')_{jj}}}\exp\left(-\frac{1}{2\sigma^2_2(\boldsymbol{P}_2\boldsymbol{P}_2')_{jj}}(Y_2(C_j)-\beta_2-\boldsymbol{p}(C_j)'\boldsymbol{\psi})^2\right)\right) \\
			&\quad\quad \cdot(\sigma^2_2)^{-a_{\sigma}-1}\exp\left(-\frac{b_{\sigma}}{\sigma^2_2}\right)\\
			&\propto (\sigma^2_2)^{-(a_{\sigma}+\frac{n_2}{2})-1}\exp\left(-\frac{b_{\sigma}+\sum_{j=1}^{n_2}\frac{1}{2(\boldsymbol{P}_2\boldsymbol{P}_2')_{jj}}(Y_2(C_j)-\beta_2-\boldsymbol{p}(C_j)'\boldsymbol{\psi})^2}{\sigma^2_2}\right)\\
			&\propto IG\left(a_{\sigma}+\frac{n_2}{2},b_{\sigma}+\sum_{j=1}^{n_2}\frac{1}{2(\boldsymbol{P}_2\boldsymbol{P}_2')_{jj}}(Y_2(C_j)-\beta_2-\boldsymbol{p}(C_j)'\boldsymbol{\psi})^2\right)
		\end{align*}
		Note that the full conditional distributions for $\rho$ and $\tau$ do not have closed forms. Therefore, Metropolis-Hastings steps are required for updating $\rho$ and $\tau$.
		
		\subsection{Derivation of full conditional distributions for MS-OH}
		Given the MS-OH model specified in Table \ref{tab:topdownmodels}, we can derive the full conditional distributions for $\boldsymbol{\eta},\beta_1,\beta_2,\beta_0,\sigma^2_{\eta},\sigma^2_1,\sigma^2_2$, and $\phi$ as follows,
		\begin{align*}
			f(\boldsymbol{\eta}|\cdot) &\propto \prod_{i=1}^{n_1} f(Y_1(B_i)|\beta_1,\boldsymbol{\eta},\sigma_1^2)\cdot \prod_{j=1}^{n_2} f(Y_2(C_j)|\beta_0,\beta_1,\beta_2,\boldsymbol{\eta},\sigma_2^2)\cdot f(\boldsymbol{\eta}|\sigma_{\eta}^2,\phi) \\
			&\propto \exp\left(\sum_{i=1}^{n_1}-\frac{1}{2\sigma_1^2(\boldsymbol{P}_1\boldsymbol{P}_1')_{ii}}(Y_1(B_i)-\beta_1-\boldsymbol{g}_1(B_i)'\boldsymbol{\eta})^2\right.\\
			&\quad\quad\quad\left.-\sum_{j=1}^{n_2}\frac{1}{2\sigma_2^2(\boldsymbol{P}_2\boldsymbol{P}_2')_{jj}}(Y_2(C_j)-\beta_0-\beta_1\beta_2-\beta_2\boldsymbol{g}_1(C_j)'\boldsymbol{\eta})^2 -\frac{1}{2}\boldsymbol{\eta}'\boldsymbol{K}^{-1}\boldsymbol{\eta}\right)\\
			&\propto \exp\left(\sum_{i=1}^{n_1}-\frac{1}{2\sigma_1^2(\boldsymbol{P}_1\boldsymbol{P}_1')_{ii}}\left[\boldsymbol{\eta}'\boldsymbol{g}_1(B_i)\boldsymbol{g}_1(B_i)'\boldsymbol{\eta}-2\boldsymbol{g}_1(B_i)'\boldsymbol{\eta}(Y_1(B_i)-\beta_1)\right]\right.\\
			&\quad\quad\quad\left. -\sum_{j=1}^{n_2}\frac{1}{2\sigma_2^2(\boldsymbol{P}_2\boldsymbol{P}_2')_{jj}}\left[\beta_2^2\boldsymbol{\eta}'\boldsymbol{g}_1(C_j)\boldsymbol{g}_1(C_j)'\boldsymbol{\eta}-2\beta_2\boldsymbol{g}_1(C_j)'\boldsymbol{\eta}(Y_2(C_j)-\beta_0-\beta_1\beta_2)\right] -\frac{1}{2}\boldsymbol{\eta}'\boldsymbol{K}^{-1}\boldsymbol{\eta}\right)\\
			&=\exp\left\{-\frac{1}{2}\left[\boldsymbol{\eta}'\left(\sum_{i=1}^{n_1}\frac{\boldsymbol{g}_1(B_i)\boldsymbol{g}_1(B_i)'}{\sigma_1^2(\boldsymbol{P}_1\boldsymbol{P}_1')_{ii}}+\sum_{j=1}^{n_2}\frac{\beta_2^2\boldsymbol{g}_1(C_j)\boldsymbol{g}_1(C_j)'}{\sigma_2^2(\boldsymbol{P}_2\boldsymbol{P}_2')_{jj}}+\boldsymbol{K}^{-1}\right)\boldsymbol{\eta}\right.\right.\\
			&\quad\quad\quad \left.\left.-2\left(\sum_{i=1}^{n_1}\frac{\boldsymbol{g}_1(B_i)'(Y_1(B_{i})-\beta_1)}{\sigma_1^2(\boldsymbol{P}_1\boldsymbol{P}_1')_{ii}}+\sum_{j=1}^{n_2}\frac{\beta_2\boldsymbol{g}_1(C_j)'(Y_2(C_{j})-\beta_0-\beta_1\beta_2)}{\sigma_2^2(\boldsymbol{P}_2\boldsymbol{P}_2')_{jj}}\right)\boldsymbol{\eta}\right]\right\} \\
			&= \exp\left(-\frac{1}{2}(\boldsymbol{\eta}'\boldsymbol{B}^{-1}\boldsymbol{\eta}-2\boldsymbol{b}'\boldsymbol{\eta})\right)\\
			&\propto \exp\left(-\frac{1}{2}(\boldsymbol{\eta}'\boldsymbol{B}^{-1}\boldsymbol{\eta}-2\boldsymbol{b}'\boldsymbol{B}\boldsymbol{B}^{-1}\boldsymbol{\eta}+\boldsymbol{b}'\boldsymbol{B}\boldsymbol{b})\right)\\
			&= \exp\left(-\frac{1}{2}(\boldsymbol{\eta}-\boldsymbol{Bb})'\boldsymbol{B}^{-1}(\boldsymbol{\eta}-\boldsymbol{Bb})\right)\\
			&\propto MVN(\boldsymbol{Bb},\boldsymbol{B})\quad \textrm{where} \,\,\boldsymbol{B}^{-1}=\sum_{i=1}^{n_1}\frac{\boldsymbol{g}_1(B_i)\boldsymbol{g}_1(B_i)'}{\sigma_1^2(\boldsymbol{P}_1\boldsymbol{P}_1')_{ii}}+\sum_{j=1}^{n_2}\frac{\beta_2^2\boldsymbol{g}_1(C_j)\boldsymbol{g}_1(C_j)'}{\sigma_2^2(\boldsymbol{P}_2\boldsymbol{P}_2')_{jj}}+\boldsymbol{K}^{-1}\\
			& \quad \quad \quad \quad\quad \quad\quad \quad\quad  \textrm{and} \,\,\boldsymbol{b}'=\sum_{i=1}^{n_1}\frac{\boldsymbol{g}_1(B_i)'(Y_1(B_{i})-\beta_1)}{\sigma_1^2(\boldsymbol{P}_1\boldsymbol{P}_1')_{ii}}+\sum_{j=1}^{n_2}\frac{\beta_2\boldsymbol{g}_1(C_j)'(Y_2(C_{j})-\beta_0-\beta_1\beta_2)}{\sigma_2^2(\boldsymbol{P}_2\boldsymbol{P}_2')_{jj}}\\
			f(\beta_1|\cdot) &\propto \prod_{i=1}^{n_1}f(Y_1(B_i)|\beta_1,\boldsymbol{\eta},\sigma_1^2)\cdot\prod_{j=1}^{n_2} f(Y_2(C_j)|\beta_0,\beta_1,\beta_2,\boldsymbol{\eta},\sigma_2^2)\cdot f(\beta_1)\\
			&\propto \exp\left(\sum_{i=1}^{n_1}-\frac{1}{2\sigma_1^2(\boldsymbol{P}_1\boldsymbol{P}_1')_{ii}}(Y_1(B_i)-\beta_1-\boldsymbol{g}_1(B_i)'\boldsymbol{\eta})^2\right.\\
			&\quad\quad\quad\left.-\sum_{j=1}^{n_2}\frac{1}{2\sigma_2^2(\boldsymbol{P}_2\boldsymbol{P}_2')_{jj}}(Y_2(C_j)-\beta_0-\beta_1\beta_2-\beta_2\boldsymbol{g}_1(C_j)'\boldsymbol{\eta})^2-\frac{1}{2\sigma^2_{\beta}}\beta_1^2\right)\\
			&\propto \exp\left[\beta_1^2\left(-\sum_{i=1}^{n_1}\frac{1}{2\sigma^2_1(\boldsymbol{P}_1\boldsymbol{P}_1')_{ii}}-\sum_{j=1}^{n_2}\frac{\beta_2^2}{2\sigma^2_2(\boldsymbol{P}_2\boldsymbol{P}_2')_{jj}}-\frac{1}{2\sigma^2_{\beta}}\right)\right.\\
			&\quad\quad\quad\left.-\beta_1\left(-\sum_{i=1}^{n_1}\frac{Y_1(B_i)-\boldsymbol{g}_1(B_i)'\boldsymbol{\eta}}{\sigma^2_1(\boldsymbol{P}_1\boldsymbol{P}_1')_{ii}}-\sum_{j=1}^{n_2}\frac{\beta_2(Y_2(C_j)-\beta_0-\beta_2\boldsymbol{g}_1(C_j)'\boldsymbol{\eta}}{\sigma^2_2(\boldsymbol{P}_2\boldsymbol{P}_2')_{jj}}\right)\right]\\
			&= \exp(a\beta_1^2-b\beta_1) \\
			&= \exp\left(\frac{\beta_1^2-\frac{b}{a}\beta_1}{1/a}\right)\\
			&= \exp\left(\frac{\beta_1^2-2\frac{b}{2a}\beta_1+(\frac{b}{2a})^2-(\frac{b}{2a})^2}{1/a}\right)\\
			&\propto \exp\left(\frac{(\beta_1-\frac{b}{2a})^2}{1/a}\right) \\
			&= \exp\left(\frac{(\beta_1-\frac{b}{2a})^2}{-2(-\frac{1}{2a})}\right)\\
			&\propto N\left(\frac{b}{2a},-\frac{1}{2a}\right) \quad \textrm{where} \,\,a=-\sum_{i=1}^{n_1}\frac{1}{2\sigma^2_1(\boldsymbol{P}_1\boldsymbol{P}_1')_{ii}}-\sum_{j=1}^{n_2}\frac{\beta_2^2}{2\sigma^2_2(\boldsymbol{P}_2\boldsymbol{P}_2')_{jj}}-\frac{1}{2\sigma^2_{\beta}}\\
			&\quad \quad \quad \quad\quad \quad\quad \quad\quad \textrm{and}\,\, b=-\sum_{i=1}^{n_1}\frac{Y_1(B_i)-\boldsymbol{g}_1(B_i)'\boldsymbol{\eta}}{\sigma^2_1(\boldsymbol{P}_1\boldsymbol{P}_1')_{ii}}-\sum_{j=1}^{n_2}\frac{\beta_2(Y_2(C_j)-\beta_0-\beta_2\boldsymbol{g}_1(C_j)'\boldsymbol{\eta}}{\sigma^2_2(\boldsymbol{P}_2\boldsymbol{P}_2')_{jj}}\\
			f(\beta_2|\cdot) &\propto \prod_{j=1}^{n_2}f(Y_2(C_j)|\beta_0,\beta_1,\beta_2,\boldsymbol{\eta},\sigma_2^2)\cdot f(\beta_2)\\
			&\propto \exp\left(\sum_{j=1}^{n_2}-\frac{1}{2\sigma_2^2(\boldsymbol{P}_2\boldsymbol{P}_2')_{jj}}(Y_2(C_j)-\beta_0-\beta_1\beta_2-\beta_2\boldsymbol{g}_1(C_j)'\boldsymbol{\eta})^2-\frac{1}{2\sigma^2_{\beta}}\beta_2^2\right)\\
			&\propto \exp\left[\beta_2^2\left(\sum_{j=1}^{n_2}-\frac{(\beta_1+\boldsymbol{g}_1(C_j)'\boldsymbol{\eta})^2}{2\sigma^2_2(\boldsymbol{P}_2\boldsymbol{P}_2')_{jj}}-\frac{1}{2\sigma^2_{\beta}}\right)-\beta_2\left(-\sum_{j=1}^{n_2}\frac{(\beta_1+\boldsymbol{g}_1(C_j)'\boldsymbol{\eta})(Y_2(C_j)-\beta_0)}{\sigma^2_2(\boldsymbol{P}_2\boldsymbol{P}_2')_{jj}}\right)\right]\\
			&= \exp(a\beta_2^2-b\beta_2) \\
			&= \exp\left(\frac{\beta_2^2-\frac{b}{a}\beta_2}{1/a}\right)\\
			&= \exp\left(\frac{\beta_2^2-2\frac{b}{2a}\beta_2+(\frac{b}{2a})^2-(\frac{b}{2a})^2}{1/a}\right)\\
			&\propto \exp\left(\frac{(\beta_2-\frac{b}{2a})^2}{1/a}\right) \\
			&= \exp\left(\frac{(\beta_2-\frac{b}{2a})^2}{-2(-\frac{1}{2a})}\right)\\
			&\propto N\left(\frac{b}{2a},-\frac{1}{2a}\right) \quad \textrm{where} \,\,a=\sum_{j=1}^{n_2}-\frac{(\beta_1+\boldsymbol{g}_1(C_j)'\boldsymbol{\eta})^2}{2\sigma^2_2(\boldsymbol{P}_2\boldsymbol{P}_2')_{jj}}-\frac{1}{2\sigma^2_{\beta}}\\
			&\quad \quad \quad \quad\quad \quad\quad \quad\quad\textrm{and}\,\, b=-\sum_{j=1}^{n_2}\frac{(\beta_1+\boldsymbol{g}_1(C_j)'\boldsymbol{\eta})(Y_2(C_j)-\beta_0)}{\sigma^2_2(\boldsymbol{P}_2\boldsymbol{P}_2')_{jj}}\\
			f(\beta_0|\cdot) &\propto \prod_{j=1}^{n_2}f(Y_2(C_j)|\beta_0,\beta_1,\beta_2,\boldsymbol{\eta},\sigma_2^2)\cdot f(\beta_0)\\
			&\propto \exp\left(\sum_{j=1}^{n_2}-\frac{1}{2\sigma_2^2(\boldsymbol{P}_2\boldsymbol{P}_2')_{jj}}(Y_2(C_j)-\beta_0-\beta_1\beta_2-\beta_2\boldsymbol{g}_1(C_j)'\boldsymbol{\eta})^2-\frac{1}{2\sigma^2_{\beta}}\beta_0^2\right)\\
			&\propto \exp\left[\beta_0^2\left(\sum_{j=1}^{n_2}-\frac{1}{2\sigma^2_2(\boldsymbol{P}_2\boldsymbol{P}_2')_{jj}}-\frac{1}{2\sigma^2_{\beta}}\right)-\beta_0\left(-\sum_{j=1}^{n_2}\frac{Y_2(C_j)-\beta_1\beta_2-\beta_2\boldsymbol{g}_1(C_j)'\boldsymbol{\eta}}{\sigma^2_2(\boldsymbol{P}_2\boldsymbol{P}_2')_{jj}}\right)\right]\\
			&= \exp(a\beta_0^2-b\beta_0) \\
			&= \exp\left(\frac{\beta_0^2-\frac{b}{a}\beta_0}{1/a}\right)\\
			&= \exp\left(\frac{\beta_0^2-2\frac{b}{2a}\beta_0+(\frac{b}{2a})^2-(\frac{b}{2a})^2}{1/a}\right)\\
			&\propto \exp\left(\frac{(\beta_0-\frac{b}{2a})^2}{1/a}\right) \\
			&= \exp\left(\frac{(\beta_0-\frac{b}{2a})^2}{-2(-\frac{1}{2a})}\right)\\
			&\propto N\left(\frac{b}{2a},-\frac{1}{2a}\right) \quad \textrm{where} \,\,a=\sum_{j=1}^{n_2}-\frac{1}{2\sigma^2_2(\boldsymbol{P}_2\boldsymbol{P}_2')_{jj}}-\frac{1}{2\sigma^2_{\beta}}\\
			&\quad \quad \quad \quad\quad \quad\quad \quad\quad\textrm{and}\,\, b=-\sum_{j=1}^{n_2}\frac{Y_2(C_j)-\beta_1\beta_2-\beta_2\boldsymbol{g}_1(C_j)'\boldsymbol{\eta}}{\sigma^2_2(\boldsymbol{P}_2\boldsymbol{P}_2')_{jj}}\\
			f(\sigma^2_{\eta}|\cdot) &\propto f(\boldsymbol{\eta}|\sigma^2_{\eta},\phi)\cdot f(\sigma^2_{\eta})\\
			&\propto (\sigma^2_{\eta})^{-a_{\eta}-1}\exp\left(-\frac{b_{\eta}}{\sigma^2_{\eta}}\right)\cdot|\{\sigma^2_{\eta}\exp(-\phi\|\boldsymbol{c}_i-\boldsymbol{c}_j\|)\}|^{-1/2}\exp\left(-\frac{1}{2}\boldsymbol{\eta}'\{\sigma^2_{\eta}\exp(-\phi\|\boldsymbol{c}_i-\boldsymbol{c}_j\|)\}^{-1}\boldsymbol{\eta}\right)\\
			&\propto (\sigma^2_{\eta})^{-a_{\eta}-1}\exp\left(-\frac{b_{\eta}}{\sigma^2_{\eta}}\right)(\sigma^2_{\eta})^{-r/2}\exp\left(-\frac{1}{2\sigma^2_{\eta}}\boldsymbol{\eta}'\{\exp(-\phi\|\boldsymbol{c}_i-\boldsymbol{c}_j\|)\}^{-1}\boldsymbol{\eta}\right)\\
			&= (\sigma^2_{\eta})^{-(a_{\eta}+\frac{r}{2})-1}\exp\left(-\frac{b_{\eta}+\frac{1}{2}\boldsymbol{\eta}'\{\exp(-\phi\|\boldsymbol{c}_i-\boldsymbol{c}_j\|)\}^{-1}\boldsymbol{\eta}}{\sigma^2_{\eta}}\right)\\
			&\propto IG\left(a_{\eta}+\frac{r}{2},b_{\eta}+\frac{1}{2}\boldsymbol{\eta}'\{\exp(-\phi\|\boldsymbol{c}_i-\boldsymbol{c}_j\|)\}^{-1}\boldsymbol{\eta}\right)\\
			f(\sigma^2_1|\cdot) &\propto \prod_{i=1}^{n_1}f(Y_1(B_i)|\beta_1,\boldsymbol{\eta},\sigma_1^2)\cdot f(\sigma^2_1)\\
			&\propto \prod_{i=1}^{n_1} \left(\frac{1}{\sqrt{2\pi\sigma^2_1(\boldsymbol{P}_1\boldsymbol{P}_1')_{ii}}}\exp\left(-\frac{(Y_1(B_i)-\beta_1-\boldsymbol{g}_1(B_i)'\boldsymbol{\eta})^2}{2\sigma^2_1(\boldsymbol{P}_1\boldsymbol{P}_1')_{ii}}\right)\right)\cdot (\sigma^2_1)^{-a_{\sigma}-1}\exp\left(-\frac{b_{\sigma}}{\sigma^2_1}\right)\\
			&\propto (\sigma^2_1)^{-(a_{\sigma}+\frac{n_1}{2})-1}\exp\left(-\frac{b_{\sigma}+\sum_{i=1}^{n_1}\frac{1}{2(\boldsymbol{P}_1\boldsymbol{P}_1')_{ii}}(Y_1(B_i)-\beta_1-\boldsymbol{g}_1(B_i)'\boldsymbol{\eta})^2}{\sigma^2_1}\right)\\
			&\propto IG\left(a_{\sigma}+\frac{n_1}{2},b_{\sigma}+\sum_{i=1}^{n_1}\frac{1}{2(\boldsymbol{P}_1\boldsymbol{P}_1')_{ii}}(Y_1(B_i)-\beta_1-\boldsymbol{g}_1(B_i)'\boldsymbol{\eta})^2\right)\\
			f(\sigma^2_2|\cdot) &\propto \prod_{j=1}^{n_2}f(Y_2(C_j)|\beta_0,\beta_1,\beta_2,\boldsymbol{\eta},\sigma_2^2)\cdot f(\sigma^2_2)\\
			&\propto \prod_{j=1}^{n_2} \left(\frac{1}{\sqrt{2\pi\sigma^2_2(\boldsymbol{P}_2\boldsymbol{P}_2')_{jj}}}\exp\left(-\frac{(Y_2(C_j)-\beta_0-\beta_1\beta_2-\beta_2\boldsymbol{g}_1(C_j)'\boldsymbol{\eta})^2}{2\sigma^2_2(\boldsymbol{P}_2\boldsymbol{P}_2')_{jj}}\right)\right)\cdot (\sigma^2_2)^{-a_{\sigma}-1}\exp\left(-\frac{b_{\sigma}}{\sigma^2_2}\right)\\
			&\propto (\sigma^2_2)^{-(a_{\sigma}+\frac{n_2}{2})-1}\exp\left(-\frac{b_{\sigma}+\sum_{j=1}^{n_2}\frac{1}{2(\boldsymbol{P}_2\boldsymbol{P}_2')_{jj}}(Y_2(C_j)-\beta_0-\beta_1\beta_2-\beta_2\boldsymbol{g}_1(C_j)'\boldsymbol{\eta})^2}{\sigma^2_2}\right)\\
			&\propto IG\left(a_{\sigma}+\frac{n_2}{2},b_{\sigma}+\sum_{j=1}^{n_2}\frac{1}{2(\boldsymbol{P}_2\boldsymbol{P}_2')_{jj}}(Y_2(C_j)-\beta_0-\beta_1\beta_2-\beta_2\boldsymbol{g}_1(C_j)'\boldsymbol{\eta})^2\right)\\
			f(\phi|\cdot) &\propto f(\boldsymbol{\eta}|\sigma^2_{\eta},\phi)\cdot f(\phi)\\
			&\propto |\{\sigma^2_{\eta}\exp(-\phi\|\boldsymbol{c}_i-\boldsymbol{c}_j\|)\}|^{-1/2}\exp\left(-\frac{1}{2}\boldsymbol{\eta}'\{\sigma^2_{\eta}\exp(-\phi\|\boldsymbol{c}_i-\boldsymbol{c}_j\|)\}^{-1}\boldsymbol{\eta}\right)\cdot \frac{1}{b_{\phi}-a_{\phi}}
		\end{align*}
		Again, $\phi$ does not have a closed form full conditional distribution, and thus requires a Metropolis-Hasting step to update.

        \subsection{Details on Metropolis-Hastings Step}
        In our models, there are four parameters that require Metropolis-Hastings steps to update (i.e., $\phi$ in MS-SRE, $\rho$ and $\tau$ in MS-MCAR, and $\phi$ in MS-OH). That is, at each iteration, we would simulate a value from a proposal distribution of our choice and compute the acceptance ratio to determine whether we reject or accept the simulated value. In the simulation, we choose the proposal densities to be truncated inverse gamma for $\phi$ in the MS-SRE and MS-OH with shape=3 for finite variance. The scale parameter is tuned such that the acceptance rate is between 30-50\%. Similarly, we adopt a truncated normal for $\rho$ and $\tau$, and adjust the variance parameter to reach 30-50\% acceptance rate. In general, the strategy is to increase the variance of the proposal distribution if the acceptance rate is too high while decreasing it if the acceptance rate is too low. Choosing a proposal distribution and tuning is relatively straightforward in the simulation study when knowing the truth of parameters. In the real data analysis, we adopt an adaptive approach by using truncated normal as proposal densities for all parameters that require Metropolis-Hastings steps, with mean specified to be the previous value in the chain and variance constantly adjusted to reach an optimal acceptance rate. We fix the proposal distributions after we reach an acceptance rate between 30-50\%, and the replicates involved with tuning are included as part of the burn-in.

        \section{Convergence and Mixing}
        In this appendix, we include Gelman-Rubin statistics, trace plots, and effective sample sizes that are used to assess the convergence and mixing of MCMC samples. We show these information for one replicate of each model.

        \begin{table}[h]
		\centering
		\begin{tabular}{c@{\hskip 0.2in} c@{\hskip 0.2in} c}
			\hline
			\hline
			Model & Parameter & Gelman-Rubin Statistic \\
            \hline
			\multirow{5}{*}{MS-SRE} & $\eta_1$ &  1.00\\
			& $\beta_1$ & 1.00\\
                & $\sigma^2_{\eta}$ & 1.04 \\
                & $\sigma^2_1$ & 1.01 \\
                & $\phi$ & 1.03 \\
                \hline
                \multirow{6}{*}{MS-MCAR} & $\psi_{11}$ &  1.00\\
                & $\beta_1$ & 1.08 \\
                & $\rho$ & 1.00 \\
                & $\tau$ & 1.00 \\
                & $\nu^2$ & 1.00 \\
                & $\sigma^2_1$ & 1.01 \\
                \hline
                \multirow{5}{*}{MS-OH} & $\eta_1$ &  1.06\\
			& $\beta_1$ & 1.00\\
                & $\sigma^2_{\eta}$ & 1.01 \\
                & $\sigma^2_1$ & 1.05 \\
                & $\phi$ & 1.04 \\
			\hline
			\hline
		\end{tabular}
	\caption{\label{tab:gelmanrubin}Gelman-Rubin statistic for one replicate of each model computed using 2 chains and 2000 iterations each with first 1000 discarded for burn-in.}
	\end{table}

 \begin{table}[h]
		\centering
		\begin{tabular}{c@{\hskip 0.2in} c@{\hskip 0.2in} c@{\hskip 0.2in} c}
			\hline
			\hline
			Model & Statistic & Parameter & Prediction \\
            \hline
			\multirow{2}{*}{MS-SRE} & median &  2000 & 2000\\
			& mean & 1872 & 1949\\
                \hline
                \multirow{2}{*}{MS-MCAR} & median &  1437 & 2000\\
			& mean & 1390 & 1912\\
                \hline
                \multirow{2}{*}{MS-OH} & median &  1874 & 1649\\
			& mean & 1594 & 1580\\
			\hline
			\hline
		\end{tabular}
	\caption{\label{tab:ess}Effective sample size (ESS) for one replicate of each model computed using 2 chains and 2000 iterations each with first 1000 discarded for burn-in. We report the median and mean ESS over all parameters and predictions ($Y_1$ and $Y_2$ pooled).}
	\end{table}

 \begin{figure}[H]
		\centering
		\begin{subfigure}[t]{0.4\textwidth}
			\centering
			\includegraphics[width=\linewidth]{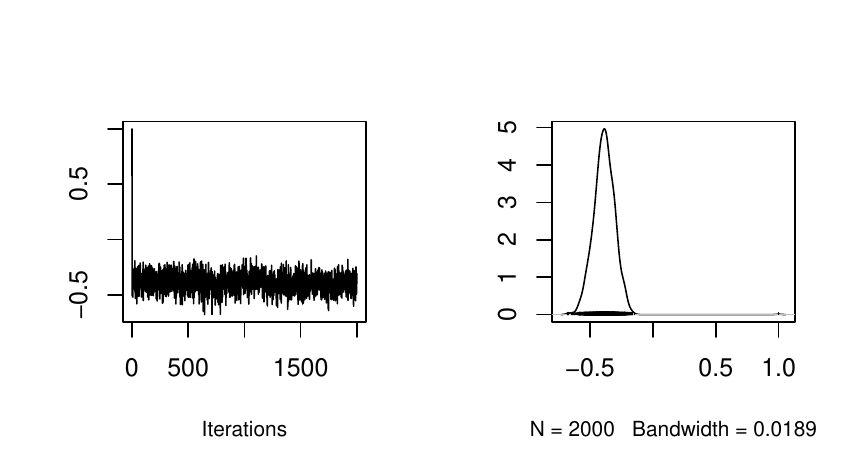} 
			\caption{$\eta_1$} \label{fig:m1_eta}
		\end{subfigure}
		\begin{subfigure}[t]{0.4\textwidth}
			\centering
			\includegraphics[width=\linewidth]{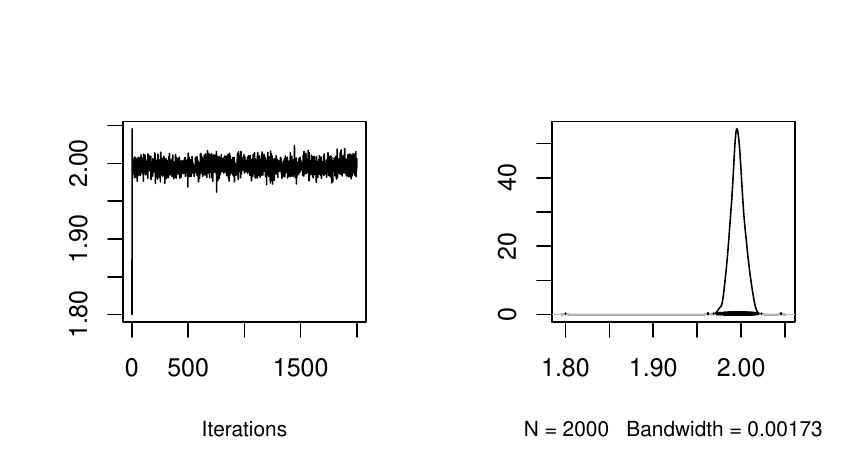} 
			\caption{$\beta_1$} \label{fig:m1_beta}
		\end{subfigure}
		
		\vspace{0.1cm}
		\begin{subfigure}[t]{0.4\textwidth}
			\centering
			\includegraphics[width=\linewidth]{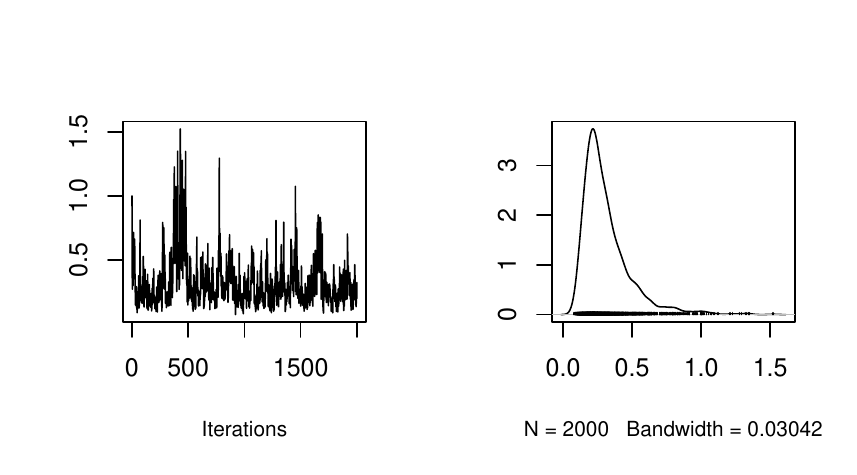} 
			\caption{$\sigma^2_{\eta}$} \label{fig:m1_sigma2eta}
		\end{subfigure}
		\begin{subfigure}[t]{0.4\textwidth}
			\centering
			\includegraphics[width=\linewidth]{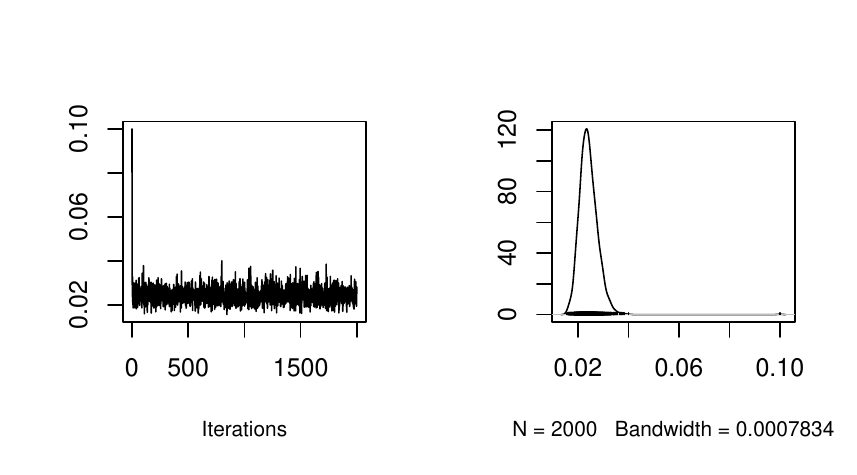} 
			\caption{$\sigma^2_1$} \label{fig:m1_sigma2}
		\end{subfigure}
		
		\vspace{0.1cm}
		\begin{subfigure}[t]{0.4\textwidth}
			\centering
			\includegraphics[width=\linewidth]{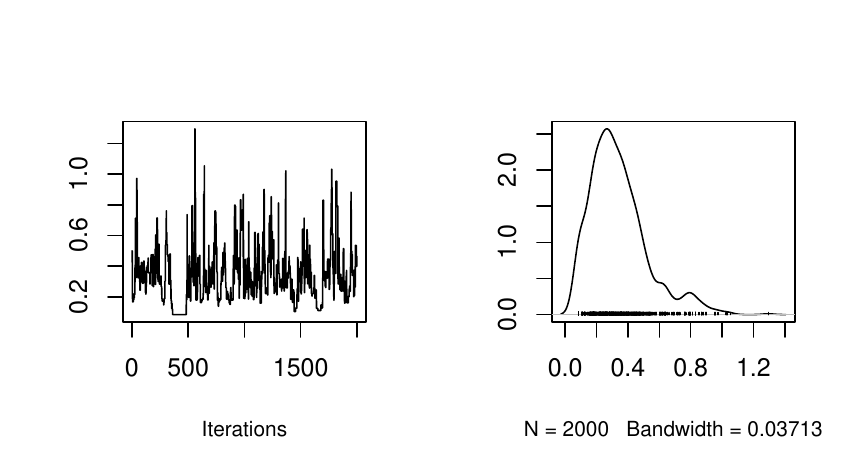} 
			\caption{$\phi$} \label{fig:m1_phi}
		\end{subfigure}
		
		\caption{Trace plots and density plots of 2000 sampled values (the first 1000 are discarded as burn-in for inference but are included in these trace plots) for parameters in MS-SRE.}
		\label{fig:trace_sre}
	\end{figure}

 \begin{figure}[H]
		\centering
		\begin{subfigure}[t]{0.4\textwidth}
			\centering
			\includegraphics[width=\linewidth]{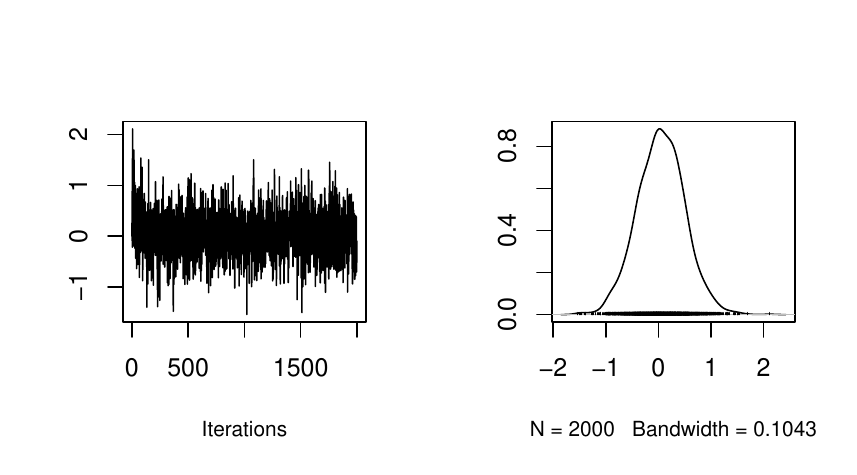} 
			\caption{$\psi_{11}$} \label{fig:m2_psi}
		\end{subfigure}
		\begin{subfigure}[t]{0.4\textwidth}
			\centering
			\includegraphics[width=\linewidth]{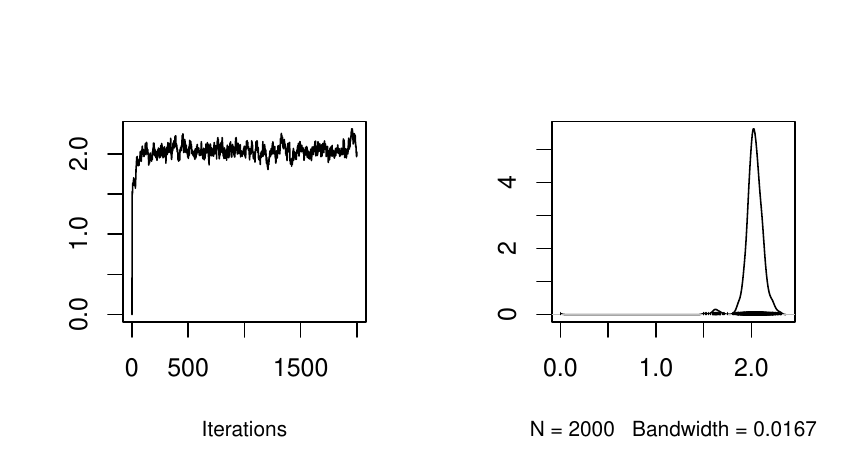} 
			\caption{$\beta_1$} \label{fig:m2_beta}
		\end{subfigure}
		
		\vspace{0.1cm}
		\begin{subfigure}[t]{0.4\textwidth}
			\centering
			\includegraphics[width=\linewidth]{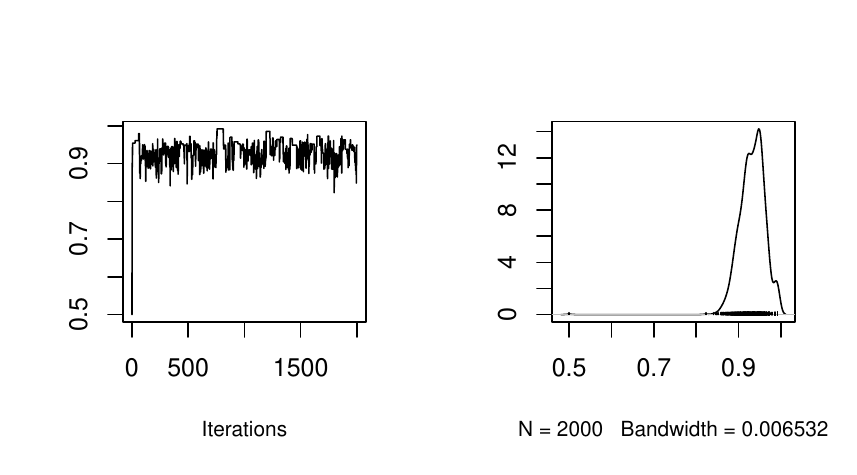} 
			\caption{$\rho$} \label{fig:m2_rho}
		\end{subfigure}
		\begin{subfigure}[t]{0.4\textwidth}
			\centering
			\includegraphics[width=\linewidth]{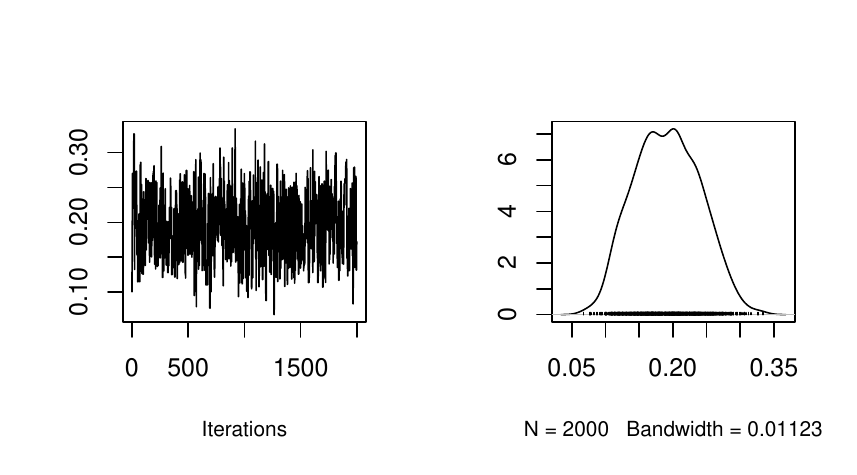} 
			\caption{$\tau$} \label{fig:m2_tau}
		\end{subfigure}
		
		\vspace{0.1cm}
		\begin{subfigure}[t]{0.4\textwidth}
			\centering
			\includegraphics[width=\linewidth]{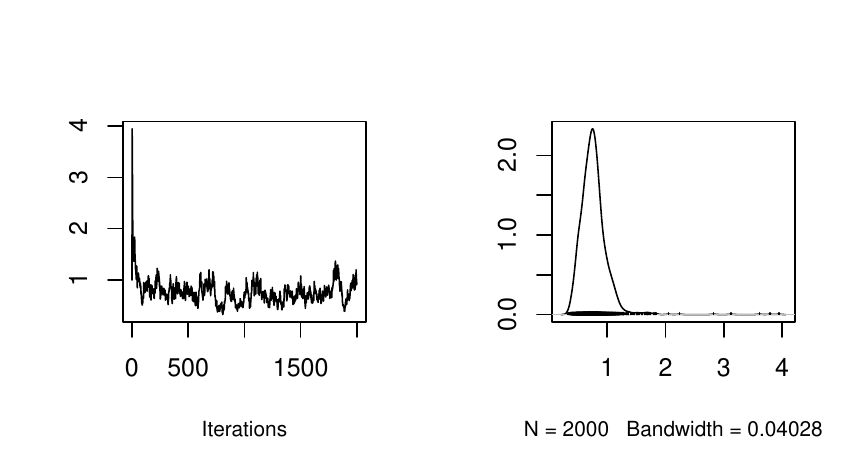} 
			\caption{$\nu^2$} \label{fig:m2_nu2}
		\end{subfigure}
  \begin{subfigure}[t]{0.4\textwidth}
			\centering
			\includegraphics[width=\linewidth]{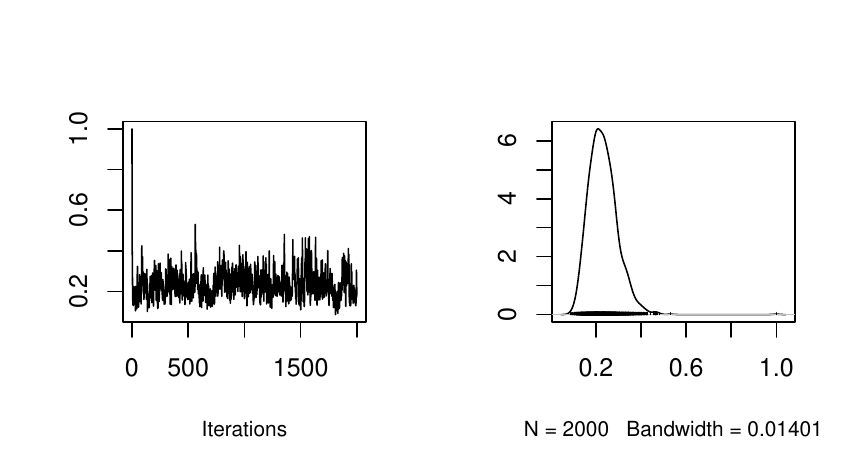} 
			\caption{$\sigma^2_1$} \label{fig:m2_sigma2}
		\end{subfigure}
		
		\caption{Trace plots and density plots of 2000 sampled values (the first 1000 are discarded as burn-in for inference but are included in these trace plots) for parameters in MS-MCAR.}
		\label{fig:trace_mcar}
	\end{figure}

 \begin{figure}[H]
		\centering
		\begin{subfigure}[t]{0.4\textwidth}
			\centering
			\includegraphics[width=\linewidth]{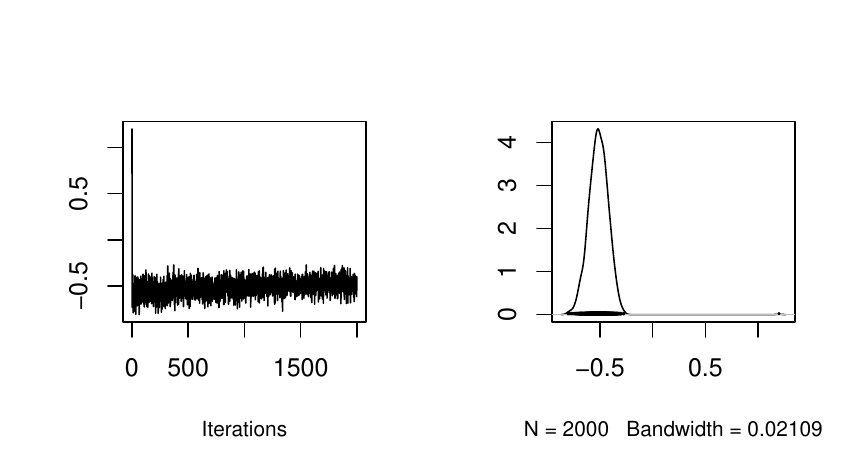} 
			\caption{$\eta_1$} \label{fig:m3_eta}
		\end{subfigure}
		\begin{subfigure}[t]{0.4\textwidth}
			\centering
			\includegraphics[width=\linewidth]{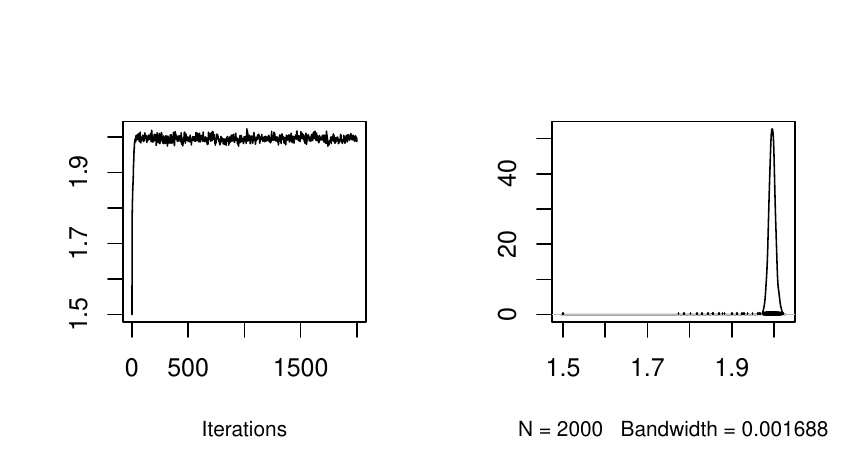} 
			\caption{$\beta_1$} \label{fig:m3_beta}
		\end{subfigure}
		
		\vspace{0.1cm}
		\begin{subfigure}[t]{0.4\textwidth}
			\centering
			\includegraphics[width=\linewidth]{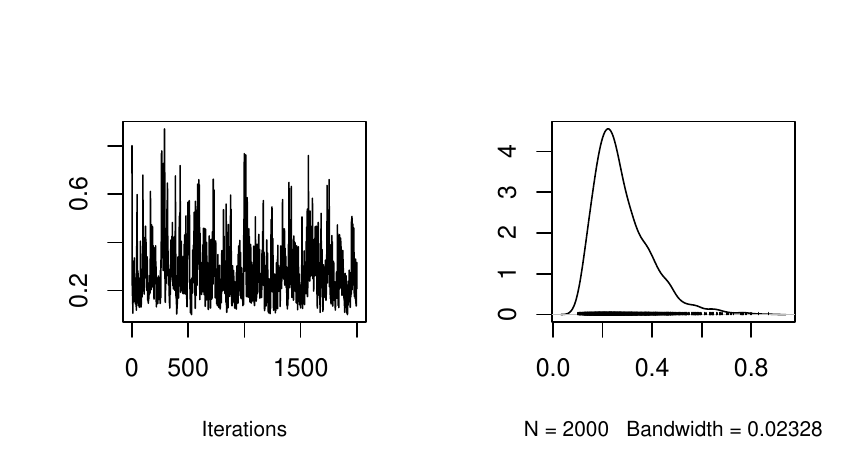} 
			\caption{$\sigma^2_{\eta}$} \label{fig:m3_sigma2eta}
		\end{subfigure}
		\begin{subfigure}[t]{0.4\textwidth}
			\centering
			\includegraphics[width=\linewidth]{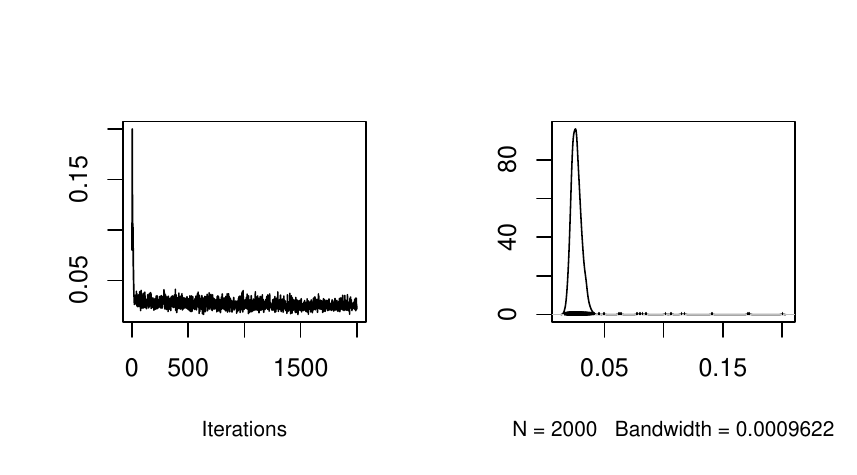} 
			\caption{$\sigma^2_1$} \label{fig:m3_sigma2}
		\end{subfigure}
		
		\vspace{0.1cm}
		\begin{subfigure}[t]{0.4\textwidth}
			\centering
			\includegraphics[width=\linewidth]{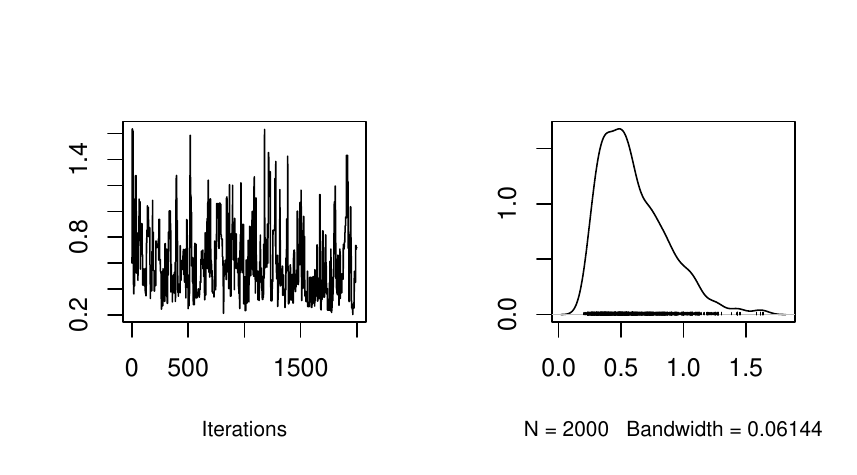} 
			\caption{$\phi$} \label{fig:m3_phi}
		\end{subfigure}
		
		\caption{Trace plots and density plots of 2000 sampled values (the first 1000 are discarded as burn-in for inference but are included in these trace plots) for parameters in MS-OH.}
		\label{fig:trace_oh}
	\end{figure}

 \section{Sensitivity Analysis: Comparisons when no Bivariate Correlation is Present}

 In practice, one should always check the assumption of bivariate correlations (e.g., through Bayesian hypothesis testing of the correlation) when making use of any bivariate statistical method that assumes such structure. In our empirical analysis we find a significant correlation with a credible interval that does not cover zero. In this appendix, we provide example consequences to the RMSE when one ignores this diagnostic check. In particular, we simulate $Y_1$ and $Y_2$ independently from a univariate MS-SRE model, and fit both univariate and bivariate MS-SRE, MS-MCAR, and MS-OH model to the data simulated. We report the RMSE by variable and scale in the following table. We see that the bivariate MS-MCAR is the most robust in terms of RMSE to miss-specifying bivariate dependence (the bivariate MS-MCAR produces similar RMSE to the univariate model), and the bivariate MS-SRE and bivariate MS-OH produce worse predictions for $Y_1$.

 \begin{table}
		\centering
		\begin{tabular}{c@{\hskip 0.2in} c@{\hskip 0.2in} c@{\hskip 0.2in} c@{\hskip 0.2in} c}
			\hline
			\hline
			Variable & Scale & Truth & Univariate & Bivariate \\
			\hline
			\multirow{6}{*}{$Y_1$} & \multirow{3}{*}{$D_A$} & MS-SRE & 0.030 & 0.176\\
			& & MS-MCAR & 0.037 & 0.068 \\
			& & MS-OH &  & 0.127 \\\cline{2-5}
			& \multirow{3}{*}{$D_1$} & MS-SRE &  0.027 & 0.159 \\
			&  & MS-MCAR & 0.023 & 0.058\\
			& & MS-OH & & 0.115 \\
			\hline
			\multirow{6}{*}{$Y_2$} & \multirow{3}{*}{$D_A$} & MS-SRE & 0.019 & 0.028 \\
			& & MS-MCAR & 0.043 & 0.071 \\ 
			& & MS-OH & & 0.020\\\cline{2-5}
			& \multirow{3}{*}{$D_2$}& MS-SRE & 0.018 & 0.026 \\
			& & MS-MCAR & 0.040 & 0.066 \\
			& & MS-OH & & 0.019 \\
			\hline
			\hline
		\end{tabular} 
    \caption{\label{tab:sensitivity}The RMSE obtained by fitting univariate and bivariate MS-SRE, MS-MCAR, and MS-OH to data with no bivariate dependence. Note that the results for univariate MS-OH are omitted as univariate MS-OH is identical to univariate MS-SRE.}
	\end{table}
        
		        \section{Residual Diagnostics}
        In this appendix, we include the histograms of standardized residuals from each model.

\begin{figure}[H]
		\centering
		\begin{subfigure}[t]{0.4\textwidth}
			\centering
			\includegraphics[width=\linewidth]{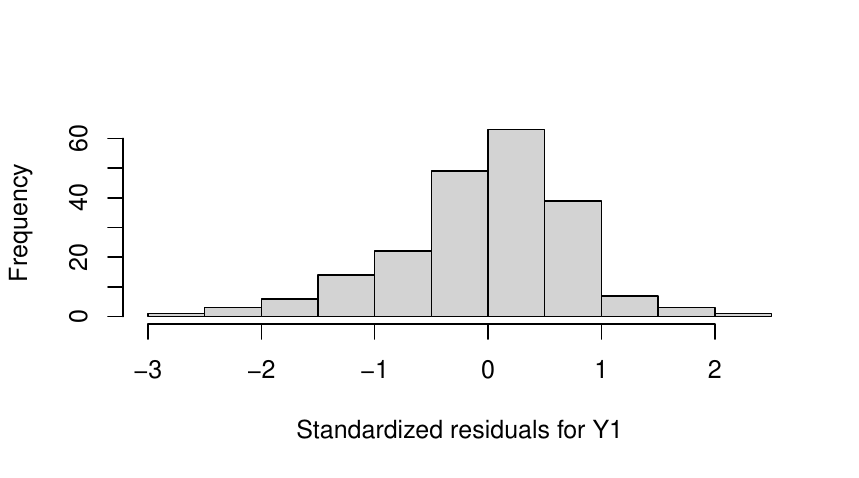} 
			\caption{MS-SRE} 
		\end{subfigure}
		\begin{subfigure}[t]{0.4\textwidth}
			\centering
			\includegraphics[width=\linewidth]{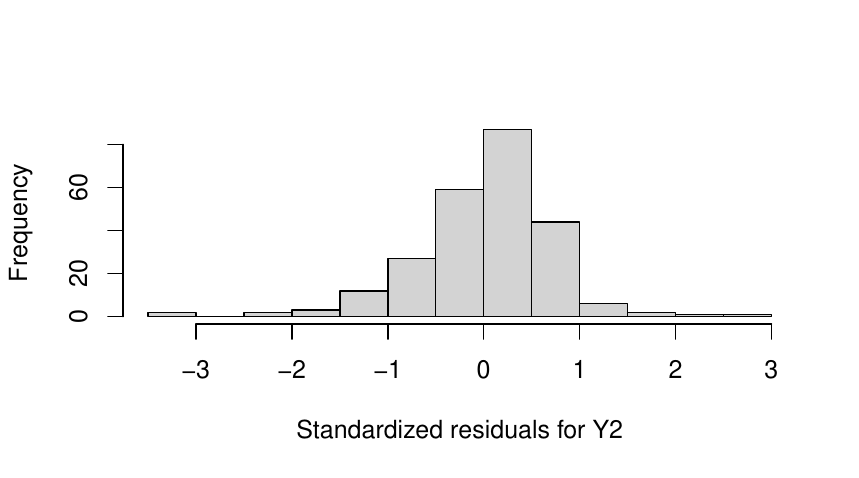} 
			\caption{MS-SRE} 
		\end{subfigure}
		
		\vspace{0.1cm}
		\begin{subfigure}[t]{0.4\textwidth}
			\centering
			\includegraphics[width=\linewidth]{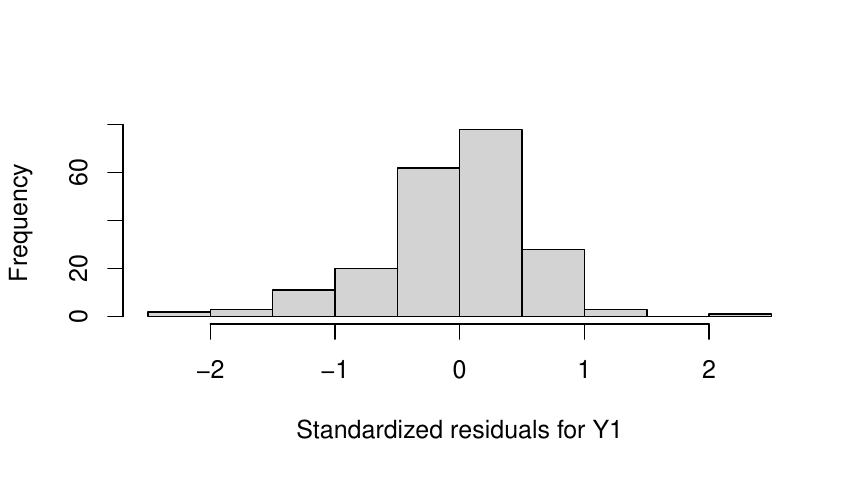} 
			\caption{MS-MCAR} 
		\end{subfigure}
		\begin{subfigure}[t]{0.4\textwidth}
			\centering
			\includegraphics[width=\linewidth]{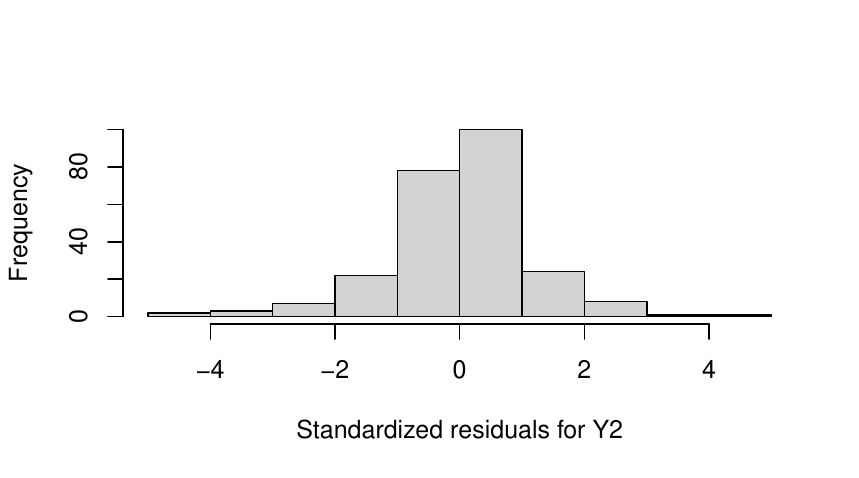} 
			\caption{MS-MCAR} 
		\end{subfigure}
		
		\vspace{0.1cm}
		\begin{subfigure}[t]{0.4\textwidth}
			\centering
			\includegraphics[width=\linewidth]{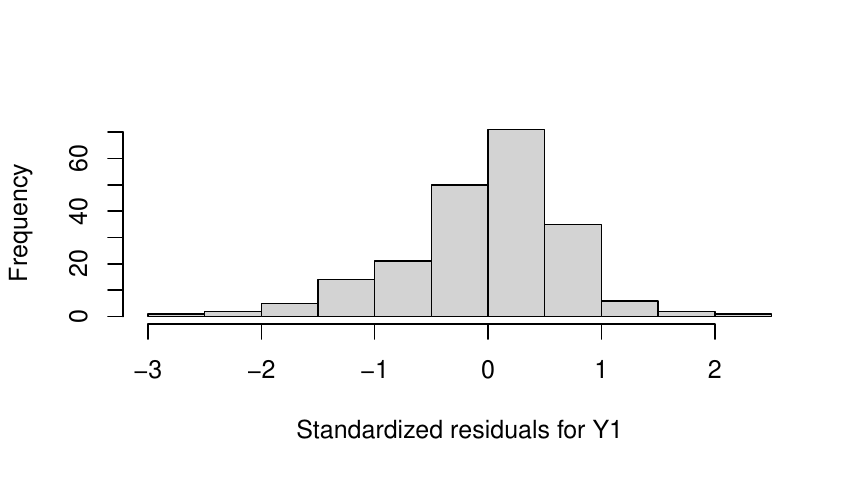} 
			\caption{MS-OH} 
		\end{subfigure}
  \begin{subfigure}[t]{0.4\textwidth}
			\centering
			\includegraphics[width=\linewidth]{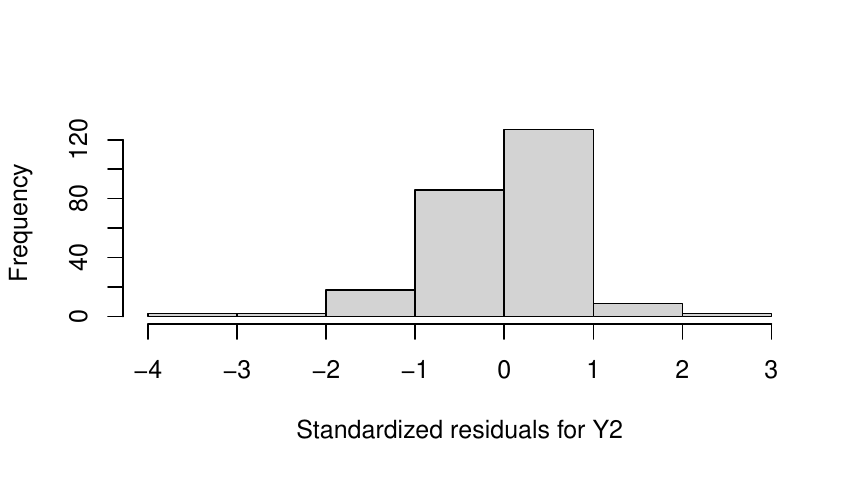} 
			\caption{MS-OH} 
		\end{subfigure}
		
		\caption{Histograms of standardized residuals for both variables from MS-SRE, MS-MCAR, and MS-OH.}
		\label{fig:histres}
	\end{figure}

	\end{appendices}


\begin{thebibliography}{11}
		
		\bibitem[Banerjee et al.(2015)]{Banerjee15} Banerjee, S., Carlin, B.P. \& Gelfand, A.E. (2015) \textit{Hierarchical Modeling and Analysis for Spatial Data}. London, UK: Chapman and Hall.
		
		\bibitem[Banerjee et al.(2008)]{Banerjee08} Banerjee, S., Gelfand, A.E., Finley, A.O. \& Sang, H. (2008) Gaussian predictive process models for large spatial data sets. \textit{Journal of the Royal Statistical Society: Series B}, 70(4), 825--848.
		
		\bibitem[Benedetti et al.(2022)]{Benedetti22} Benedetti, M.H., Berrocal, V.J. \& Little, R.J. (2022) Accounting for survey design in Bayesian disaggregation of survey-based areal estimates of proportions: An application to the American Community Survey. \textit{Annals of Applied Statistics}, 16(4), 2201--2230.

        \bibitem[Bradley et al. (2015a)]{bradley2015spatio} Bradley, J.R., Wikle, C.K., \& Holan, S.H.  (2015) Spatio-temporal change of support with application to American Community Survey multi-year period estimates. \textit{Stat}, 4(1), 255--270.
	
		\bibitem[Bradley et al.(2015b)]{bradley15} Bradley, J.R., Holan, S.H. \& Wikle, C.K. (2015) Multivariate Spatio-Temporal Models for High-Dimensional Areal Data with Application to Longitudinal Employer-Household Dynamics. \textit{The Annals of Applied Statistics}, 9(4), 1761--1791.
		
		\bibitem[Bradley et al.(2018)]{bradley18} Bradley, J.R., Holan, S.H. \& Wikle, C.K. (2018) Computationally Efficient Multivariate Spatio-Temporal Models for High-Dimensional Count-Valued Data (with Discussion). \textit{Bayesian Analysis}, 13(1), 253--310.
		
		
		\bibitem[Bradley et al.(2016)]{bradley16} Bradley, J.R., Wikle, C.K. \& Holan, S.H. (2016) Bayesian Spatial Change of Support for Count-Valued Survey Data with Application to the American Community Survey. \textit{Journal of the American Statistical Association}, 111(514), 472--487.
		
		\bibitem[Bradley et al.(2017)]{bradley17} Bradley, J.R., Wikle, C.K. \& Holan, S.H. (2017) Regionalization of multiscale spatial processes by using a criterion for spatial aggregation error. \textit{Journal of the Royal Statistical Society: Series B}, 79(3), 815--832.
		
		\bibitem[Carlin \& Banerjee(2003)]{CB03} Carlin, B.P. \& Banerjee, S. (2003) Hierarchical Multivariate CAR Models for Spatio-Temporally Correlated Survival Data. \textit{Bayesian Statistics}, 7, 45--63.

        \bibitem[Cressie(1993)]{cressie1993} Cressie, N. (1993) \textit{Statistics for Spatial Data, Rev'd Edn}. Hoboken, NJ: John Wiley \& Sons.
		
		\bibitem[Cressie \& Johannesson(2008)]{CJ08} Cressie, N. \& Johannesson G. (2008) Fixed Rank Kriging for Very Large Spatial Data Sets. \textit{Journal of the Royal Statistical Society: Series B}, 70(1), 209--226.

        \bibitem[Cressie \& Wikle(2011)]{cressie2011statistics} Cressie, N. \& Wikle, C.K. (2011) \textit{Statistics for spatio-temporal data}. Hoboken, NJ: Wiley.
		
		\bibitem[Dartmouth Atlas Data(2022)]{Data22} Dartmouth Atlas Data (2022) Selected Primary Care Access and Quality Measures. Available at \textit{https://data.dartmouthatlas.org/primary-care/}.
		
		\bibitem[Delamater(2006)]{Delamater06} Delamater, A.M. (2006) Clinical Use of Hemoglobin A1c to Improve Diabetes Management. \textit{Clinical Diabetes}, 24(1), 6--8.
		
		\bibitem[Gelfand \& Vounatsou(2003)]{Gelfand03} Gelfand, A.E. \& Vounatsou, P. (2003) Proper multivariate conditional autoregressive models for spatial data analysis. \textit{Biostatistics}, 4(1), 11--25.
		
		\bibitem[Gelfand et al.(2001)]{Gelfand01} Gelfand, A.E., Zhu, L. \& Carlin B.P. (2001) On the change of support problem for spatio-temporal data. \textit{Biostatistics}, 2(1), 31--45.
		
		\bibitem[Gelman \& Rubin(1992)]{GR92} Gelman, A. \& Rubin D.B. (1992) Inference from Iterative Simulation Using Multiple Sequences. \textit{Statistical Science}, 7(4), 457--472.

        \bibitem[Gneiting \& Raftery(2007)]{GR2007} Gneiting, T. \& Raftery A. (2007) Strictly proper scoring rules, prediction, and estimation. \textit{Journal of the American statistical Association}, 102(477), 359--378.
		
		\bibitem[Gotway \& Young(2002)]{GY02} Gotway, C.A. \& Young L.J. (2002) Combining Incompatible Spatial Data. \textit{Journal of the American Statistical Association}, 97(458), 632–-648.

        \bibitem[Guhaniyogi et al.(2011)]{Guhaniyogi11} Guhaniyogi, R., Finley, A.O., Banerjee, S. \& Gelfand, A.E. (2011) Adaptive Gaussian predictive process models for large spatial datasets. \textit{Environmetrics}, 22(8), 997--1007.
		
		\bibitem[Habib \& Aslam(2003)]{HA03} Habib, S.S. \& Aslam, M. (2003) Risk factors, knowledge and health status in diabetic patients. \textit{Saudi Medical Journal}, 24(11), 1219--1224.

        \bibitem[Hanks et al.(2015)]{hanks} Hanks, E. M., Schliep, E. M.,  Hooten, M. B. \& Hoeting, J. A. (2015) Restricted spatial regression in practice: geostatistical models, confounding, and robustness under model misspecification. \textit{Environmetrics}, 26(4), 243--254.

        \bibitem[Hodges \& Reich(2010)]{hodges} Hodges, J. S. \& Reich, B. J. (2010) Adding spatially-correlated errors can mess up the fixed effect you love. \textit{The American Statistician}, 64(4), 325--334.
		
		\bibitem[Hughes \& Haran(2013)]{HH13} Hughes, J. \& Haran, M. (2013) Dimension reduction and alleviation of confounding for spatial generalized linear mixed models. \textit{Journal of the Royal Statistical Society: Series B}, 75(1), 139--159.
		
		\bibitem[Khan \& Calder(2022)]{KC22} Khan, K. \& Calder, C.A. (2022) Restricted Spatial Regression Methods: Implications for Inference. \textit{Journal of the American Statistical Association}, 117(537), 482--494.
		
		\bibitem[Kirkman et al.(2012)]{Kirkman12} Kirkman, M.S., Briscoe, V.J., Clark, N., Florez, H., Haas, L.B., Halter, J.B., Huang, E.S., Korytkowski, M.T., Munshi, M.N., Odegard, P.S., Pratley, R.E. \& Swift, C.S. (2012) Diabetes in Older Adults. \textit{Diabetes Care}, 35(12), 2650--2664.
		
		\bibitem[Larsen et al.(1990)]{LHM90} Larsen, M.L., Horder, M. \& Mogensen E.F. (1990) Effect of Long-Term Monitoring of Glycosylated Hemoglobin Levels in Insulin-Dependent Diabetes Mellitus. \textit{The New England Journal of Medicine}, 323(15), 1021–-1025.
		
		\bibitem[Mardia \& Goodall(1993)]{MG93} Mardia, K.V. \& Goodall C.R. (1993) Spatial-Temporal Analysis of Multivariate Environmental Monitoring Data. \textit{Multivariate Environmental Statistics} (pp. 347-–386). Elsevier Science Publishers B.V.
		
		\bibitem[Messick et al.(2017)]{MHH17} Messick, R.M., Heaton, M.J. \& Hansen, N. (2017) Multivariate spatial mapping of soil water holding capacity with spatially varying cross-correlations. \textit{Annals of Applied Statistics}, 11(1), 69--92.
		
		\bibitem[Moran(1950)]{Moran50} Moran, P.A.P. (1950) Notes on Continuous Stochastic Phenomena. \textit{Biometrika}, 37(1), 17--23.
		
		\bibitem[Mugglin \& Carlin(1998)]{MC98} Mugglin, A.S. \& Carlin, B.P. (1998) Hierarchical Modeling in Geographic Information Systems: Population Interpolation over Incompatible Zones. \textit{Journal of Agricultural, Biological, and Environmental Statistics}, 3(2), 111-–130.
		
		\bibitem[Mugglin et al.(1999)]{MC99} Mugglin, A.S., Carlin, B.P., Zhu, L. \& Conlon, E. (1999) Bayesian Areal Interpolation, Estimation, and Smoothing: An Inferential Approach for Geographic Information Systems. \textit{Environment and Planning A}, 31(8), 1337-–1352.
		
		\bibitem[Obled \& Creutin(1986)]{OC86} Obled, C. \& Creutin, J. (1986) Some developments in the use of empirical orthogonal functions for mapping meteorological fields. \textit{Journal of Applied Meteorology}, 25(9), 1189–-1204.
		
		\bibitem[Qu et al.(2021)]{Qu21} Qu, K., Bradley, J.R. \& Niu, X. (2021) Boundary Detection Using a Bayesian Hierarchical Model for Multiscale Spatial Data. \textit{Technometrics}, 63(1), 64--76.
		
		\bibitem[Raim et al.(2021)]{Raim21} Raim, A.M., Holan, S.H., Bradley, J.R. \& Wikle, C.K. (2021) Spatio-temporal change of support modeling with R. \textit{Computational Statistics}, 36, 749--780.
		
		\bibitem[Royle \& Berliner(1999)]{RB99} Royle, J.A. \& Berliner, L.M. (1999) A Hierarchical Approach to Multivariate Spatial Modeling and Prediction. \textit{Journal of Agricultural, Biological, and Environmental Statistics}, 4(1), 29--56.

        \bibitem[Simpson et al.(2012)]{Simpson12} Simpson, D., Lindgren, F. \& Rue, H. (2012) Think continuous: Markovian Gaussian models in spatial statistics. \textit{Spatial Statistics}, 1, 16--29.]

        \bibitem[Tiefelsdorf \& Boots(2015)]{MI} Tiefelsdorf. M \& Boots, B. (2015) The exact distribution of Moran's I. \textit{Environment and Planning A}, 27(6), 985--999.

        \bibitem[Vehtari et al.(2017)]{VGG2017} Vehtari, A., Gelman, A. \& Gabry, J. (2017) Practical Bayesian model evaluation using leave-one-out cross-validation and WAIC. \textit{Statistics and Computing}, 27, 1413--1432.
		
			
		\bibitem[Waller \& Gotway(2004)]{WG04} Waller, L.A. \& Gotway, C.A. (2004) \textit{Applied Spatial Statistics for Public Health Data}. Hoboken, NJ: Wiley.
		
		\bibitem[Watanabe(2010)]{Watanabe10} Watanabe, S. (2010) Asymptotic Equivalence of Bayes Cross Validation and Widely Applicable Information Criterion in Singular Learning Theory. \textit{The Journal of Machine Learning Research}, 11, 3571-–3594.
		
		\bibitem[Wikle \& Berliner(2005)]{WB2005} Wikle, C.K. \& Berliner L.M. (2005) Combining Information across Spatial Scales. \textit{Technometrics}, 47(1), 80--91.
		

		
		
	\end{thebibliography}
\end{document}